\documentclass[opre,nonblindrev]{informs4}

\OneAndAHalfSpacedXI

\usepackage{amsmath,amssymb,amsfonts}

\usepackage{natbib}
 \bibpunct[, ]{(}{)}{,}{a}{}{,}%
 \def\bibfont{\small}%

\usepackage{rotating}
\usepackage{fancyvrb}

\TheoremsNumberedThrough     
\ECRepeatTheorems
\JOURNAL{Operations Research}

\EquationsNumberedThrough    

\usepackage{dutchcal}
\usepackage{hyperref}
\usepackage{algorithm}
\usepackage{algpseudocode}
\usepackage{tcolorbox}
\usepackage{booktabs}
\usepackage[shortlabels]{enumitem}
\usepackage{multirow}
\usepackage{makecell}
\newcommand{\kprelax}[1]{\textsc{kp-relax}(#1)}
\newcommand{\kprelaxeq}[1]{\textsc{kp-relax-eq}(#1)}

\def\subdual{\textsc{sub-dual}}
\def\subdualaux{\textsc{sub-dual-aux}}

\def\i{\textsf{i}}
\def\j{\textsf{j}}
\def\k{\textsf{k}}
\def\r{\mathcal{r}}
\def\c{\mathfrak{c}}

\def\th{W_{\textsc{TH}}}
\def\minwt{\textsc{min-wt}}
\def\minwtset{\textsc{set}}

\newcommand{\kpcard}[1]{\textsc{kp}(#1)}

\def\argmax{\mathop{\rm arg\,max}}

\def\rev{\textsc{rev}}
\def\revmod{\textsc{rev-cost}}
\def\revmodaux{\textsc{rev-cost-aux}}

\usepackage{bbm}
\usepackage{array}
\usepackage{float}
\usepackage{subfig}
\usepackage{wrapfig}
\usepackage{caption}
\usepackage{bm}

\renewenvironment{thebibliography}[1]
{
\begin{oldthebibliography}{#1}\baselineskip12.0pt
\setlength{\parskip}{0pt} 
\setlength{\itemsep}{0pt plus 0.3ex} 
}
{
\end{oldthebibliography}
}

\renewcommand{\footnotesize}{\fontsize{9pt}{12pt}\selectfont}

\begin{document}

\RUNAUTHOR{Chen, Golrezaei, and Susan}

\RUNTITLE{Fair Assortment Planning}

\TITLE{Fair Assortment Planning}

\ARTICLEAUTHORS{%
\AUTHOR{Qinyi Chen}
\AFF{Operations Research Center, Massachusetts Institute of Technology, Cambridge, MA 02139, \EMAIL{qinyic@mit.edu}}

\AUTHOR{Negin Golrezaei}
\AFF{Sloan School of Management, Massachusetts Institute of Technology, Cambridge, MA 02139, \EMAIL{golrezae@mit.edu}}

\AUTHOR{Fransisca Susan}
\AFF{Operations Research Center, Massachusetts Institute of Technology, Cambridge, MA 02139, \EMAIL{fsusan@mit.edu}}
}

\ABSTRACT{%
Many online platforms, ranging from online retail stores to social media platforms, employ algorithms to optimize their offered assortment of items (e.g., products and contents). 
These algorithms often focus exclusively on achieving the platforms' objectives, highlighting items with the highest popularity or revenue. This approach, however, can compromise the equality of opportunities for the rest of the items, in turn leading to less content diversity and increased regulatory scrutiny for the platform.
{Motivated by this, we introduce and study a \emph{fair} assortment planning problem that enforces equality of opportunities via pairwise fairness, which requires any two items to be offered similar outcomes.}
We show that the problem can be formulated as a linear program (LP), called \eqref{eq:problem:fair}, that optimizes over the distribution of all feasible assortments. To find a near-optimal solution to \eqref{eq:problem:fair}, we propose a framework based on the Ellipsoid method, which requires a polynomial-time separation oracle to the dual of the LP. We show that finding an optimal separation oracle to the dual problem is an NP-complete problem, and hence we propose a series of approximate separation oracles, which then result in a $1/2$-approx. algorithm and an FPTAS for Problem \eqref{eq:problem:fair}. The approximate separation oracles are designed by (i) showing the separation oracle to the dual of the LP is equivalent to solving an infinite series of parameterized knapsack problems, and (ii) leveraging the structure of  knapsack problems. {Finally, we perform numerical studies on both synthetic data and real-world MovieLens data, showcasing the effectiveness of our algorithms and providing insights into the platform's price of fairness.}}


\maketitle

\section{Introduction}
\label{sec:introduction}

Algorithms are widely used by modern digital platforms to optimize decisions ranging from assortment planning, pricing, ranking and resource allocation. However, these algorithms, while enhancing efficiency, can inadvertently compromise fairness, marginalizing certain items or creators. In assortment planning, fairness concerns are prevalent across online retail stores, social media, job search sites, etc. These platforms often favor top-selling or well-established entities over lesser-known ones in their offered assortments, thus creating disparities among their items. For instance, Amazon highlights best-sellers on its front page \citep{amazon}, and Instagram prioritizes posts with high engagement or advanced features, sidelining content from smaller businesses \citep{instagram}. Such disparities risk reduced platform loyalty, decreased content diversity, and could lead to regulatory scrutiny on fair competition principles (e.g., Digital Markets Act \citep{DMA}).

Consequently, platforms must prioritize not only efficiency but also \emph{equality of opportunity} for all entities. Some companies are beginning to address this issue. For instance, LinkedIn's initial recommendation algorithms favored users with more connections and activity, disadvantaging newer or less active members with relevant skills. To address this, LinkedIn recently introduced fairness toolkits to ``provide equal opportunities to equally qualified members" \citep{linkedin_2}. In this work, we present a universal framework to promote equality of opportunity in assortment planning, fostering industry-wide equitable practices.

\textbf{Fair assortment planning problem.} We consider an online platform hosting $n$ items, representing products, social media contents, job candidates in respective contexts. Each item $i$ has a popularity weight $w_i$ and generates revenue $r_i$ when chosen. The platform offers an assortment of up to 
$K$ items to maximize expected revenue, with user choices following a multinomial model based on popularity weights $w_i$.

{To ensure fairness, we introduce a parameterized notion based on \emph{pairwise parity of outcomes}, ensuring comparable opportunities across items. Our framework also seamlessly integrates \emph{normalized outcomes}, where an item's outcome is scaled by its quality score—determined by factors such as popularity, revenue, and relevance. This allows for merit-based considerations while maintaining fairness.}

Defining fairness in assortment planning involves balancing the platform's revenue goals with equitable treatment of items. Traditional fairness notions, such as alpha fairness \citep{Mo2000Fair, Lan2010axiomatic}, often prioritize social welfare but may neglect the platform's revenue (see Section \ref{sec:discussion-fairness-notions} for a related discussion). To address this, we formulate the fair assortment planning problem as a constrained optimization problem (Problem \eqref{eq:problem:fair}), which maximizes revenue while ensuring fairness through constraints.

\textbf{Fair Ellipsoid-Based Assortment Planning Algorithms.} In Section \ref{sec:near-opt-fair}, we present a framework for near-optimal algorithms to solve Problem \eqref{eq:problem:fair} using the Ellipsoid method and a (near-)optimal polynomial-time separation oracle for its dual, Problem \eqref{eq:problem:dual}. While the Ellipsoid method ideally requires an optimal separation oracle to verify feasibility or provide a separating hyperplane, designing such an oracle is generally NP-complete (Theorem \ref{thm:np-hard}). However, we show that a $\beta$-approx. separation oracle for Problem \eqref{eq:problem:dual} suffices to construct a $\beta$-approx. solution to Problem \eqref{eq:problem:fair}, where $\beta \in (0,1]$ (Theorem \ref{thm:approx}).

\textbf{Separation oracle via transformation to infinite knapsack problems.} 
In Section \ref{sec:near-opt-sub-dual}, we develop near-optimal separation oracles for Problem \eqref{eq:problem:dual}, which requires maximizing cost-adjusted revenue over sets of at most 
$K$ items, where costs depend on dual variables from fairness constraints. 
We show that this optimization is equivalent to solving an infinite series of parameterized, cardinality-constrained knapsack problems. Leveraging this transformation, we design a series of polynomial-time $\beta$-approx. algorithms to efficiently solve these infinite knapsack problems.
Here, our approximation algorithms for Problem \eqref{eq:problem:dual} more broadly apply to \emph{cardinality-constrained assortment planning problem with arbitrary fixed costs}, which can be of independent interest to our readers; see Section \ref{sec:related-work} for details.

\textbf{$1/2$-approx. and FPTAS separation oracles.} 
While solving one knapsack problem approximately is doable in polynomial time, we cannot possibly solve an infinite number of them. 
In Section \ref{sec:half-approx}, we first present an $1/2$-approx. algorithm that leverages the structure of the parameterized knapsack problems
to overcome the challenge.
As an important technical contribution, we note that for each knapsack problem, an $1/2$-approx. solution can be obtained using a concept called \emph{profile}, which records the set of items that are fully added, fractionally added, and not added to the knapsack at the optimal basic solution to the problem's LP relaxation.
We observe that the profiles associated with these solutions undergo at most $\mathcal{O}(nK)$ changes as the knapsack parameter $W$ varies. Inspired by this, our $1/2$-approx. algorithm (Algorithm \ref{alg:half-approx}) adopts a key procedure called \emph{adaptive partitioning}, which dynamically detects changes in the profile, updates it, and collects the corresponding $1/2$-approx. solution in polynomial time.

Building on this, Section \ref{sec:fptas} presents a fully polynomial-time approximation scheme (FPTAS) that further refines the approximation. 
Our approach combines dynamic programming with a dual-partition strategy over 
$[0, \infty)$ while leveraging both the 
1/2-approx. algorithm and the problem structure. 

\textbf{Numerical studies and managerial insights.} We evaluate our algorithms (1/2-approx. and FPTAS) against some benchmark algorithms on various synthetic real-world-inspired instances (Section \ref{sec:synthetic_data}), and a real-world case study on MovieLens data (Section \ref{sec:numerics}). 
Our findings offer key managerial insights for platforms implementing fair policies:
\begin{itemize}[leftmargin=*]
\item The 1/2-approx. algorithm is shown to be well-suited for real-world deployment, given its good solution quality and runtime efficiency. Specifically, its adaptability allows it to exploit problem structures, reducing the number of partitions needed and consistently outperforming other benchmarks.
\item In real-world scenarios such as the movie recommendation setting in Section \ref{sec:numerics}, it is possible to implement fair policies that come with minimal tradeoff for the platform. This is also a setting where fairness can create impact for a wide range of contents, especially given that item qualities are similar.
\item Finally, we propose that platforms consider the price of fairness, alongside key business metrics like visibility, when selecting fairness parameters. Our analysis also underscores how fairness parameters, problem settings, and market conditions—such as price sensitivity—affect the price of fairness.
\end{itemize}

\section{Related Work}
\label{sec:related-work}

Our research is relevant to or contributes to several lines of works: 

\textbf{Fairness in supervised learning.} The rise of machine learning algorithms has spurred extensive research on algorithmic fairness in supervised learning, particularly in binary classification and risk assessment \citep{calders2009building,dwork2012fairness, hardt2016equality, goel2018non,ustun2019fairness}. These works are typically categorized into individual fairness \citep{dwork2012fairness, kusner2017counterfactual}, group fairness \citep{calders2009building, hardt2016equality}, and subgroup fairness \citep{kearns2018preventing}. 
Our work focuses on achieving individual fairness for every item on a platform. However, instead of addressing prediction problems using labeled data, we solve an assortment optimization problem while integrating fairness constraints.

\textbf{Fairness in resource allocation.} Our problem involves an online platform allocating exposure (an intangible resource) to items fairly while optimizing its objective. Works on fair resource allocation study two settings: (i) a static setting where allocation occurs in a single shot and the trade-off between fairness and efficiency is investigated \citep{bertsimas2011price,bertsimas2012efficiency, hooker2012combining, donahue2020fairness}; see,  also,\citep{cohen2021price} and \citep{deng2022fairness} that study fairness in price discrimination and individual welfare in auctions, respectively; (ii) a dynamic setting where resources are allocated over time, including fairness in online resource allocation \citep{manshadi2021fair,balseiro2021regularized, bateni2021fair}, online matching \citep{ma2021fairness}, sequential allocation \citep{Sinclair2021}, scheduling \citep{mulvany2021fair}, bandit \citep{baek2021fair}, content spread \citep{Schoeffer2022}, search \citep{aminian2023fair} and recommendation \citep{chen2023interpolating}.
Our work aligns more closely with the first stream, focusing on fairness and efficiency trade-offs. However, it stands out as the first to explore fairness in assortment planning. 

\textbf{Fairness in ranking.} Our work also intersects with the literature on fairness in ranking, which focuses on optimizing permutations of items displayed to users \citep{patro2022fair}. We can view our assortment planning problem as a specialized ranking problem, where items in the top $K$ positions are chosen for the assortment. Several works, such as \citep{singh2018fairness, biega2018equity, singh2019policy}, emphasize the importance of ensuring equal opportunities for items and devise ranking methods that aim to achieve proportional exposure or visibility based on item relevance or quality. However, the aforementioned works lack theoretical guarantees. In contrast, our algorithms offer provable performance guarantees. We believe our fairness notion and mathematical framework can be easily extended to product ranking settings.

\textbf{Fairness in assortment planning.} 
{Fairness remains largely unexplored in the context of assortment planning. A potentially relevant work by \cite{chen2021assortment} briefly considers fairness but within a partitioned display policy, where sellers (items) are divided into disjoint groups assigned to different traffic segments. This approach can lead to significant disparities between
partitions in terms of the number of sellers, product quality, and customer volume. The authors incorporate fairness constraints to ensure that marketshare and attractiveness are similar across partitions. In contrast, our work focuses on the standard assortment planning framework, where items are not partitioned for different user segments. We impose item fairness
by similarly introducing fairness constraints but also by incorporating randomization over the entire family of assortments that can be shown to customers, which further distinguishes our approach from theirs.}

To the best of our knowledge, our work is the first to integrate fairness into assortment planning. Since our initial manuscript, follow-up studies have emerged. For example, \cite{lu2023simple} introduces a fairness constraint that ensures minimum item visibility. 
{As discussed in Section \ref{sec:discussion-fairness-notions}, under pairwise fairness, tuning the fairness parameters allows the platform to focus on high-quality items. In contrast, a uniform lower bound across all items could lead to oversaturation with low-quality items, ultimately reducing user engagement.} Another follow-up by \cite{barre2023assortment} considers a multi-period setting, where fairness is enforced by ensuring each item is displayed to a minimum number of customers, differing from our single-shot model. 

\textbf{Cardinality-Constrained Assortment Planning with Fixed Costs.}  
In our Ellipsoid-based framework, we design separation oracles to find a set \(S\) with cardinality at most \(K\) that maximizes cost-adjusted revenue. Our approximation schemes, including a \(1/2\)-approx. algorithm and an FPTAS, can be directly applied to \emph{general cardinality-constrained assortment planning problems with arbitrary fixed costs}, which is a technical contribution that extends beyond prior work and could hold independent interests for our readers. As shown in Table \ref{tab:comparison}, existing studies that address assortment planning with fixed costs often focus on specific cases, such as unconstrained problems or non-positive fixed costs, or have certain limitations. For example, \cite{Kunnumkal2010} handles unconstrained problems with positive fixed costs, \cite{kunnumkal2019tractable} only obtains solution upper bounds, while \cite{lu2023simple}, our follow-up work, only addresses non-positive fixed costs. In contrast, we uniquely address the challenge of cardinality constraints while accommodating both positive and non-positive fixed costs. Notably, we overcome the difficulties introduced by positive fixed costs, which directly offset item revenues. Compared to other works, our approach offers both stronger theoretical guarantees and broader applicability.

\vspace{-2mm}
\begin{table}[htbp]
\centering
\caption{Comparison of Works on Assortment Planning with Fixed Costs}
\label{tab:comparison}
\footnotesize 
\begin{tabular}{p{3.5cm}|p{4.5cm}|p{3.1cm}|p{4cm}}
\hline
\textbf{Work} & \textbf{Problem Setting} & \textbf{Main Results} & \textbf{Limitations} \\ 
\hline
\cite{Kunnumkal2010} & Unconstrained planning with positive fixed costs & \(1/2\)-approx. and PTAS &  Does not address cardinality constraints. \\ 
\hline
\cite{kunnumkal2019tractable} & Constrained and unconstrained planning with positive fixed costs & Solution upper bounds & No direct algorithms proposed. \\ 
\hline
\cite{lu2023simple}  (a follow-up work) & Cardinality-constrained planning with non-positive fixed costs & \(1/2\)-approx. and FPTAS & Does not handle positive fixed costs. \\ 
\hline
\textbf{This Work} & Cardinality-constrained planning with arbitrary fixed costs & \(1/2\)-approx. and FPTAS  \\ 
\hline
\end{tabular}
\end{table}

\section{Model}
\label{sec:model}

\subsection{User's Choice Models}
We consider a platform with $n$ items, indexed by $i\in [n]$. Each item $i$ has a \emph{weight} $w_i\ge 0$ that measures how popular the item is to the platform's users.
Upon offering an assortment/set $S$ of items, the users purchase/choose item $i \in S$ according to the MNL model \citep{train1986qualitative}, with probability $\frac{w_i}{1+w(S)}\,,$
where $w(S)= \sum_{i\in S} w_i$.
The users may also choose not to purchase any item with probability 
$\frac{1}{1+w(S)}$; that is, the no-purchase option (item $0$) is always included in set $S$. 
If item $i$ is purchased by the users, it generates revenue $r_i \in (0, \bar{r}]$.
The action of purchasing an item can be viewed as ``taking the desired action'' under different contexts (e.g., purchasing a product in an online retail store, clicking on a post on a social media platform, or watching a movie on a movie recommendation site). The main objective of the platform is to optimize over its offered assortment that can contain at most $K$ items in order to maximize its expected revenue, which we denote as 
$
\textsc{rev}(S) = \frac{\sum_{i \in S}  r_i w_i}{1 + w(S)}\,.
$

\subsection{Fairness Notions}
\label{sec:fairness-notions}

The platform aims to create a fair marketplace for all items, ensuring equitable allocation of outcomes among items. Let $p(S)$ represent the probability that the platform offers set $S$ with $S \subseteq [n]$ and $|S| \leq K$. 
{We let $O_i(S)$ be the \emph{outcome} received by item $i$ upon offering a set $S$, and, with a slight abuse of notation, let $O_i = \sum_{i \in S} O_i(S)p(S)$ be the expected outcome received by item $i$ given the platform's assortment planning decision $\{p(S)\}_{S:|S|\leq K}$. Given set $S$, let the outcome of item $i \in S$ take the following generic form: 
\begin{equation}
\label{eqn:outcome}
O_i(S) = \Big( a_i \cdot \frac{w_i}{1+w(S)} +  b_i \Big) \cdot \mathbbm{1}\{i \in S\}
\end{equation}
for some $a_i, b_i \geq 0$. Note that an item not in the offered set always receives zero outcome.} 

Our generic definition of item outcome can encompass various metrics that an item may care about, including: (1) \textbf{visibility} ($a_i=0, b_i=1$), where $O_i = \sum_{S:i \in S}p(S)$ is the probability that item $i$ gets shown to users;
(2) \textbf{revenue} ($a_i=r_i, b_i=0$), where $O_i = \sum_{S:i \in S} \frac{r_iw_i}{1+w(S)} \cdot p(S)$ is the expected revenue generated by item $i$; (3) \textbf{marketshare} ($a_i=1, b_i=0$), where $O_i = \sum_{S:i\in S}\frac{w_i}{1+w(S)} \cdot p(S)$ is the expected marketshare of item $i$. Here, the preferred interpretation of the outcome received by item $i$ varies by the context. For instance, content creators on streaming platforms may prioritize visibility, while sellers on online marketplaces may focus on revenues or profits. Under Eq. \eqref{eqn:outcome}, any weighted combination of visibility, revenue, and marketshare can serve as the outcome metric for items, accommodating different contexts.

To ensure a fair ecosystem, the platform aims to ensure the parity of \emph{pairwise outcome} across different items. 
We say that the platform is \emph{$\delta$-fair} if the pairwise outcome between any two items $(i,j)$ satisfies:
\begin{align}
O_i - O_j \le \delta\quad 
\Longleftrightarrow \quad 
\sum_{S: i\in S} p(S) \cdot O_i(S) - \sum_{S: j\in S} p(S)\cdot O_j(S) \le \delta~~\quad 
\text{$\forall ~~ i, j\in [n]$}\,, 
\label{eq:fair_relaxed} 
\end{align}
Here, the pairwise fairness constraints can be imposed with respect to any interpretation of item outcomes as defined in Eq. \eqref{eqn:outcome}.\footnote{Multiple sets of pairwise fairness constraints with respect to different outcome metrics can also be imposed  simultaneously, such as visibility and marketshare. In general, any fairness constraint can be incorporated as long as the dual counterpart of Problem \eqref{eq:problem:fair} resembles Problem \eqref{eq:problem:dual} that we will present in Section \ref{sec:dual}, which involves solving a maximization problem similar to Problem \eqref{eq:subdual}. All subsequent analysis and results remain consistent.}
A larger $\delta$ in Eq. \eqref{eq:fair_relaxed} generally allows for greater outcome disparity.\footnote{It is also possible to consider a different fairness parameter $\delta_{ij}$ for each pair of items $(i,j)$. To incorporate this, one simply needs to replace $\delta$ in Problems \eqref{eq:problem:fair} and \eqref{eq:problem:dual} with $\delta_{ij}$. All of our results would remain valid.}  When $\delta=0$ (i.e., when the platform is $0$-fair), the above constraint can be written as ${O_i} = {O_j}$, indicating that all items receive equal outcomes. For any $\delta \ge \max_{i\in [n]} O_i$, the fairness constraint is simply satisfied at the optimal solution without the fairness constraint, achieved by always offering the single assortment with the largest expected revenue.
Nonetheless, the optimal solution without fairness constraints, while maximizing platform revenue, can lead to very unfair outcomes for some of the items that are not included in the assortment. 

Pairwise notions of fairness, such as the one defined in Eq. \eqref{eq:fair_relaxed}, are widely studied and have been used in recommendation systems \citep{beutel2019fairness}, ranking and regression models \citep{narasimhan2020pairwise, kuhlman2019fare, singh2021fairness}, and predictive risk scores \citep{kallus2019fairness}. Our approach to achieving pairwise fairness is closely related to the fairness notion employed in \cite{biega2018equity, singh2019policy, Schoeffer2022} for ranking problems, which, like our setting, aims to allocate visibility in proportion to item quality. However, our formulation of pairwise fairness in the assortment planning setting is not explored in prior work. 
See Section \ref{sec:discussion-fairness-notions} for an expanded discussion on the benefits of enforcing fairness via constraints and the advantages of pairwise fairness.

\textbf{Normalized outcome scaled by quality.} 
Our fairness notions can be flexibly imposed based on each item's normalized outcome, scaled by its \emph{quality score}. These quality scores can be estimated by platforms using metrics like ratings, purchase frequency, or engagement. Let $q_i$ be the quality of item $i$. We can then define the following fairness constraints:
$
\frac{O_i}{q_i} - \frac{O_j}{q_j} \leq \delta
$
for all $i, j \in [n]$.
Note that the normalized outcome, scaled by quality, maintains the same affine structure:
$\frac{O_i(S)}{q_i} = \frac{a_i}{q_i} \cdot \frac{w_i}{1 + w(S)} + \frac{b_i}{q_i}$. Hence, to simplify the exposition, we will, without loss of generality, assume $q_i = 1$ for all $i$ throughout this paper.\footnote{We will revisit normalized outcome scaled by quality scores in our numerical experiments in Section \ref{sec:experiments}.}

\subsection{Fair Assortment Planning Problem}
Having defined the user's choice model and the fairness notion, we define an instance of our fair assortment planning problem using  
$\mathbf{\Theta} = (\textbf{a}, \textbf{b}, \textbf{r}, \textbf{w}) \in \mathbb{R}_{+}^{4n}$, where $\textbf{a} = \{a_i\}_{i \in [n]}$ and $\textbf{b} = \{b_i\}_{i \in [n]}$ determine outcomes of items (see Eq. \eqref{eqn:outcome}), $ \textbf{r} = \{r_i\}_{i \in [n]}$ represents item revenues, and $\textbf{w} = \{w_i\}_{i \in [n]}$ represents item weights. We now define two specific instances of our problem.
\begin{itemize}[leftmargin=*]
    \item \textbf{Revenue-fair instance.} We call an instance with the form $\mathbf{\Theta} = (\textbf{a}, \mathbf{0}, \textbf{r}, \textbf{w})$ a \emph{revenue-fair} instance.
    Here, item $i$'s outcome when it gets offered in set $S$ is $O_i(S) =a_i \cdot \frac{w_i}{1+w(S)}$, which can be its revenue if $a_i= r_i$. In the special case when $a_i= 1$, the outcome would be item's marketshare. 
    \item \textbf{Visibility-fair instance.} We call an instance with the form $\mathbf{\Theta} = (\mathbf{0}, \textbf{b}, \textbf{r}, \textbf{w})$ a \emph{visibility-fair instance}. Here, the expected outcome received by item $i$ is $O_i = b_i \cdot \sum_{S:i \in S}p(S)$, which is proportional to the probability that item $i$ is shown to a user (i.e., the visibility of item $i$).  
\end{itemize}
While our algorithms can be applied to Problem \eqref{eq:problem:fair} for any instance $\mathbf{\Theta} \in \mathbb{R}_{+}^{4n}$, better results can be achieved for revenue-fair instances. Specifically, we will show that there exists a polynomial-time optimal algorithm for Problem \eqref{eq:problem:fair} for any revenue-fair instance.

Given any instance $\mathbf{\Theta} \in \mathbb{R}_{+}^{4n}$, the platform wishes to maximize its expected revenue by optimizing over its offered assortment with cardinality of at most $K$, subject to the fairness constraints in Eq. \eqref{eq:fair_relaxed}. This gives the platform's 
optimization problem:
\begin{align}
    \notag \textsc{fair}~=~\max_{p(S)\ge 0: |S|\le K} & \sum_{S: |S|\le K} p(S)\cdot \rev(S)\\   \notag
    \text{s.t.} \qquad & ~~\sum_{S: i\in S} p(S) \cdot O_i(S) - \sum_{S: j\in S} p(S) \cdot O_j(S) \le \delta \quad\quad~~~~~ \forall~ i,j\in [n], i \neq j\\   \notag
    \quad\quad\quad\quad\quad 
    & \sum_{S:|S|\leq K} p(S) \leq 1 
    \label{eq:problem:fair} \tag{\textsc{fair}}\,,
\end{align}
where 
$
\rev(S) =\frac{\sum_{i \in S} r_i w_i}{1+w(S)}
$
is the platform's expected revenue upon offering assortment $S\subseteq [n]$. Here, we slightly abuse the notation and denote both the platform's problem and the optimal expected revenue with the term  \ref{eq:problem:fair}. In Problem \eqref{eq:problem:fair}, the objective is the platform's expected revenue. {The first set of constraints is the pairwise fairness constraints}, while the second constraint enforces $\{p(S)\}_{S:|S|\leq K}$ to be a probability distribution. 
Our goal is to develop computationally efficient algorithms with provable performance guarantees for Problem \eqref{eq:problem:fair} for any $\delta \geq 0$. {Our algorithm should identify a family of assortments for randomization, along with their corresponding probabilities.}\footnote{In this work, we focus on the platform's problem in a static/offline setting, justified when the model primitives are well-estimated. An online version, where the platform maximizes revenue under fairness constraints while learning model primitives, could leverage an efficient offline algorithm (e.g., \cite{niazadeh2021online, kakade2007playing}). See Section \ref{sec:conclusion} for future directions on such settings.}

\section{(Near) Optimal Algorithms for Problem (FAIR)}
\label{sec:near-opt-fair}
In this section, we present a general framework for developing (near-)optimal algorithms using the dual counterpart of Problem \eqref{eq:problem:fair} (Section \ref{sec:dual}) and the Ellipsoid method (Section \ref{sec:elipsoid}). 

\subsection{A Property of Problem (FAIR)}
\label{sec:main-property}
Before proceeding, we introduce a key property of Problem \eqref{eq:problem:fair} that informs our algorithm design. The following proposition ensures that an optimal solution always randomizes over at most 
$\mathcal{O}(n^2)$ sets.
\begin{proposition}
   [Randomization over at most $\mathcal{O}(n^2)$ sets]  
\label{lem:n_2}
For any $\delta \ge 0$, there exists an optimal solution $p^\star(.)$ to  Problem \eqref{eq:problem:fair} such that
$\big|\{S : p^\star(S) > 0\}\big| \leq n(n-1) + 1.$
\end{proposition} 
This result motivates our development of (near-)optimal algorithms that also randomize over a polynomial number of sets, easing practical implementation.

\subsection{Dual Counterpart of Problem (FAIR)}
\label{sec:dual}
Our framework relies on the dual counterpart of Problem \eqref{eq:problem:fair}, which can be written as
\vspace{-2mm}
\begin{align}
\begin{aligned}
\textsc{fair-dual}~=~ \min_{\rho\ge 0, \mathbf{z}\ge \mathbf{0}}&   ~\, \rho+\sum_{i=1}^{n} \sum_{j = 1, j \neq i}^{n} \delta \cdot  z_{i j} \\
\text { s.t.} ~~~ & ~ \sum_{i \in S} O_i(S)\cdot \left(\sum_{j=1, j \neq i}^{n} (z_{i j}- z_{ji})\right)+\rho \geq \rev(S), ~ \forall S:|S| \leq K \,.
\end{aligned}
\tag{\textsc{fair-dual}} 
\label{eq:problem:dual}
\end{align}
Here, $\mathbf z=(z_{ij})_{i,j \in [n], i\ne j} \ge {\bf 0}$ and $\rho\ge { 0}$ are respectively the dual variables associated with constraints in Problem (\ref{eq:problem:fair}). 
We call each constraint in Problem \eqref{eq:problem:dual} \emph{a dual fairness constraint} on set $S$. Note that the constraints in Problem \eqref{eq:problem:dual} can be expressed as a single constraint: 
\begin{align} 
\label{eq:inner_dual}
    \rho \ge \max_{S: |S|\le K}\Big\{ \rev(S) - \sum_{i \in S} O_i(S) \cdot \Big(\sum_{j=1, j \neq i}^{n} (z_{i j}- z_{ji}) \Big)\Big\} := \max_{S: |S|\le K}~ \revmod(S, \mathbf{z}) \,.
\end{align}
Here, we refer to $\revmod(S, \mathbf{z})$ as the \emph{cost-adjusted revenue}. Given that $O_i(S) = \frac{a_i w_i}{1+w(S)} + b_i$, we can thus reformulate the cost-adjusted revenue as:
\begin{align} 
\label{eqn:revcost-form-1}
\revmod(S, \mathbf{z}) & = \rev(S) - \sum_{i \in S} \left(\frac{a_i w_i}{1+w(S)} + b_i \right) \cdot c_i(\mathbf{z}) = \sum_{i \in S}\frac{(r_i-a_i c_i(\mathbf{z}))w_i}{1+w(S)} - \sum_{i \in S} b_i c_i(\mathbf{z})\,,
\end{align}
where we define $c_i(\mathbf{z}) = \sum_{j=1, j \neq i}^{n} (z_{i j}- z_{ji})$ as the 
\emph{cost} of item $i$,
Let us additionally define
$
\r_i(\mathbf{z}) = r_i - a_i c_i(\mathbf{z})$ as the \emph{post-fairness revenue} of item $i$ and 
$\c_i(\mathbf{z}) = b_i c_i(\mathbf{z})$ as the \emph{post-fairness cost} of item $i$. The cost-adjusted revenue in Eq. \eqref{eqn:revcost-form-1} can be further simplified as:
\begin{align} 
\label{eq:welmod} 
\revmod(S, \mathbf{z}) 
& = \sum_{i \in S}\frac{ \r_i(\mathbf{z}) w_i}{1+w(S)} - \sum_{i \in S} \c_i(\mathbf{z}) \,.
\end{align}
Determining the feasibility of any solution for Problem \eqref{eq:problem:dual} thus boils down to maximizing the cost-adjusted revenue over all possible assortments.
However, this problem, which we denote as \eqref{eq:subdual}, is in general NP-complete:
\begin{theorem} [NP-Completeness of Problem \eqref{eq:subdual}] 
\label{thm:np-hard} 
Consider the following problem that represents the first set of constraints in Problem \eqref{eq:problem:dual}:
\begin{align}\subdual(\mathbf{z}, K) = \max_{S: |S|\le K}~\revmod(S, \mathbf{z})\,,
 \label{eq:subdual} 
 \tag{\subdual($\mathbf{z}, K$)}
 \end{align}
\begin{enumerate}[leftmargin=*]
    \item For revenue-fair instances where $\mathbf{\Theta} = (\textbf{a}, \textbf{0}, \textbf{r}, \textbf{w})$, there exists an polynomial-time optimal algorithm for Problem \eqref{eq:subdual}.
    \item For visibility-fair and general instances where $\mathbf{\Theta} = (\textbf{a}, \textbf{b}, \textbf{r}, \textbf{w})$, with $b_i \ne 0$ for some $i \in [n]$, Problem \eqref{eq:subdual} is NP-complete. 
\end{enumerate}
\end{theorem}

\subsection{Fair Ellipsoid-based Framework} 
\label{sec:elipsoid}
We now present our framework for obtaining (near-)optimal solutions to Problem \eqref{eq:problem:fair}, leveraging the Ellipsoid method and the dual problem \eqref{eq:problem:dual}. We show that to obtain a $\beta$-approx. solution to Problem \eqref{eq:problem:fair}, for $\beta \in (0, 1]$, it suffices to have a $\beta$-approx. solution to Problem \eqref{eq:subdual}.

Assume that $\mathcal{A}$ is a polynomial-time, $\beta$-approx. algorithm for \eqref{eq:subdual}, where $\beta \in (0, 1]$. Our fair Ellipsoid-based algorithm has the following two steps:

\textit{Step 1.}
At each iteration, check the feasibility of the current solution using an approximate separation oracle as follows:
(1) First, use $\mathcal{A}$ to find $S_\mathcal{A}$ with $|S_\mathcal{A}| \leq K$ such that $\revmod(S_\mathcal{A}, \mathbf{z}) \ge \beta \cdot \subdual(\mathbf{z}, K)$.
(2) If $\revmod(S_\mathcal{A}, \mathbf{z}) > \rho$, the set $S_\mathcal{A}$ violates the dual fairness constraint. If not, $(\mathbf{z}, \rho)$ is feasible.
See Section \ref{sec:ellipsoid-details} for details of the Ellipsoid method for Problem \eqref{eq:problem:dual}.

\textit{Step 2.}
Throughout the Ellipsoid method, we collect all sets violating the dual fairness constraint in $\mathcal{V}$, the size of which is polynomial in $n$. 
We then solve primal problem \eqref{eq:problem:fair} by setting $p(S) = 0$ for $S \notin \mathcal{V}$, reducing non-zero variables to a polynomial size, and obtain an optimal basic feasible solution.

{The following theorem shows that this fair Ellipsoid-based algorithm (i) gives a $\beta$-approx. solution for Problem \eqref{eq:problem:fair} that randomizes over $\mathcal{O}(n^2)$ assortments, and (ii) runs in polynomial time.}

\begin{theorem} [$\beta$-Approx. Algorithm for Problem (FAIR)]\label{thm:approx} 
Suppose that for any $\mathbf{z} \geq \mathbf{0}$ and $K \in [n]$, we have a polynomial-time $\beta$-approx. algorithm $\mathcal A$ for Problem \eqref{eq:subdual}, for some $\beta \in (0, 1]$. 
Then, the fair Ellipsoid-based algorithm returns a $\beta$-approx., feasible solution $\widehat p(S)$, $S\subseteq [n]$ to Problem \eqref{eq:problem:fair} in polynomial time. In addition, the number of sets $S$ such that $\widehat{p}(S) > 0$ is $\mathcal{O}(n^2)$. 
\end{theorem}

\section{(Near) Optimal Algorithms for Problem (SUB-DUAL)}
\label{sec:near-opt-sub-dual}
We now design a polynomial-time \(\beta\)-approx. algorithm for Problem \eqref{eq:subdual}, which acts as a (near-)optimal separation oracle for Problem \eqref{eq:problem:dual} within the Ellipsoid-based framework in Section \ref{sec:elipsoid}. Since the dual variables \(\mathbf{z}\) are fixed at each Ellipsoid iteration, in the rest of this paper, we will simplify notation by suppressing their dependency and denote \(\r_i = \r_i(\mathbf{z})\) and \(\c_i = \c_i(\mathbf{z})\).

\subsection{Optimal Algorithm for Revenue-Fair Instances}
\label{sec:rev-ms-fair}

In revenue-fair instances, Problem \eqref{eq:subdual} simplifies to a cardinality-constrained assortment optimization problem, with $\r_i$ being the revenue of item $i$ and $\c_i = 0$ for all $i \in [n]$. This problem can be solved optimally in polynomial time using the \textsc{StaticMNL} algorithm \citep{rusmevichientong2010dynamic}.
\begin{proposition}
\label{prop:staticMNL}
Given any revenue-fair instance $\mathbf{\Theta} = (\textbf{a}, \mathbf{0}, \textbf{r}, \textbf{w})$, the \textsc{StaticMNL} algorithm finds the optimal assortment $S^\star = \argmax_{S:|S|\le K}\revmod(S, \mathbf{z})$ in $\mathcal{O}(n^2)$ time. 
\end{proposition}

In this case, we  have access to an exact separation oracle for Problem \eqref{eq:problem:dual}, solving Problem \eqref{eq:problem:fair} optimally in polynomial-time using the Ellipsoid-based algorithm in Section \ref{sec:elipsoid}.

\subsection{Near Optimal Algorithm for General Instances}
\label{sec:general-instance}

In visibility-fair and general instances, Problem \eqref{eq:subdual} is NP-complete and requires us to design approximation algorithms. Given any instance $\mathbf{\Theta}$, solving Problem \eqref{eq:subdual} is equivalent to solving a \emph{capacitated assortment optimization problem with fixed costs}, with $\r_i$ being the revenue of item $i$ and $\c_i$ (which can be positive or negative) being the fixed cost of item $i$. Prior works have recognized assortment planning with fixed costs as a challenging problem (see our detailed discussion in Section \ref{sec:related-work}). However, no existing work has addressed the added complexity of cardinality constraints with general fixed costs, making this problem non-trivial to solve.

We start by first establishing an important equivalence that transforms Problem \eqref{eq:subdual} to an infinite series of parameterized knapsack problems.
Consider the following knapsack problem with capacity $W\ge 0$ and cardinality upper bound $K$:
\begin{equation}
\label{eqn:knapsack-with-size}\tag{$\kpcard{W, K}$}
\begin{aligned}
\kpcard{W, K}= \max_{S} & 
\sum_{i\in S} \left(\frac{\r_iw_i}{1+W}-\c_i\right) 
\text { s.t. } & \sum_{i\in S}w_{i} \leq W \quad \text{and} \quad |S| \leq K\,,
\end{aligned}
\end{equation}
where for a given $W$, we denote the \emph{utility} of item $i$ as $u_i(W):= \frac{\r_iw_i}{1+W}-\c_i$. 
The following theorem characterizes the relationship between Problems \eqref{eqn:knapsack-with-size}, $W \ge 0$, and Problem \eqref{eq:subdual}.
\begin{theorem}[Problem \eqref{eq:subdual} as an infinite series of knapsack problems] \label{thm:kp-su-dual}
For any $\mathbf z \geq \mathbf{0}$ and $K \in [n]$, we have 
$\subdual(\mathbf{z}, K) = \max_{W\ge 0}~\kpcard{W, K}\,.$
\end{theorem}

\setlength{\intextsep}{-4mm}
\begin{wrapfigure}[6]{r}{0.35\textwidth}
\centering
\includegraphics[width=0.35\textwidth]{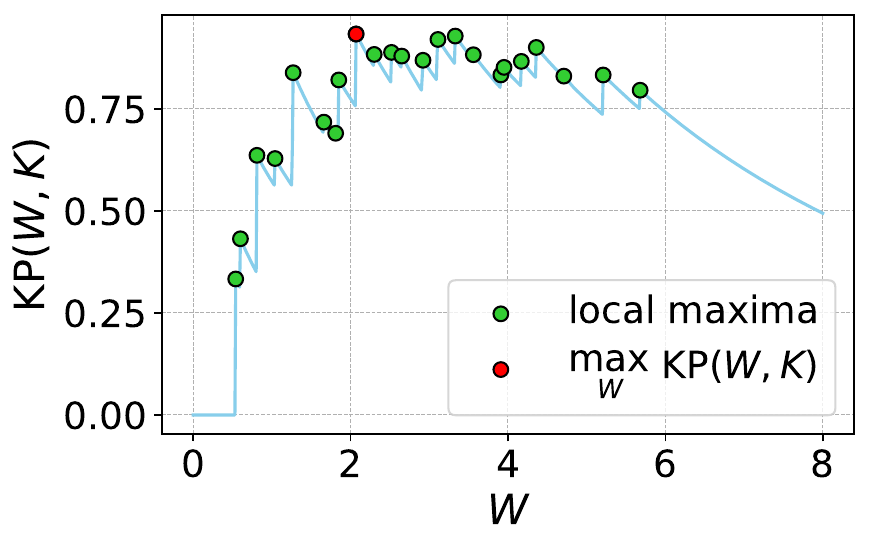}
\vspace{-0.8cm} 
\caption{\footnotesize{Evolution of $\kpcard{W,K}$ with $W$.}}
\label{fig:kpcard}
\end{wrapfigure}
Despite this equivalence, solving $\max_{W\ge 0}~\kpcard{W, K}$ is also challenging as it
involves solving infinite cardinality-constrained knapsack problems.
Further note that in Problem \eqref{eqn:knapsack-with-size}, increasing 
$W$ enlarges the knapsack capacity but reduces item utility $u_i(W) = \frac{\r_i w_i}{1+W} - \c_i$, introducing further complexity.  
As illustrated in Figure \ref{fig:kpcard},\footnote{For illustration in Figure \ref{fig:kpcard}, we consider a random instance where $n = 10, K = 5, w_i\sim\text{i.i.d.} \textsc{Unif}([0.5, 1.5]), \r_i\sim\text{i.i.d. } \textsc{Unif}([1, 2]), \c_i\sim\text{i.i.d. } \textsc{Unif}([0, 1])$, for all $i \in [n]$. We then solve Problem \eqref{eqn:knapsack-with-size} for $W \in [0, 8]$.}
the values of $\kpcard{W, K}$ are not monotonic with respect to $W$, but rather can have many local maxima as $W$ evolves. Hence, it can be difficult to tell which $W$ maximizes $\kpcard{W, K}$, or if our transformation has even simplified the problem. 

A potential approach to solving $\max_{W \geq 0} \kpcard{W, K}$ is to discretize $[0, \infty)$ using geometric grids and approximate $\kpcard{W, K}$ at each grid point. This grid-based method achieves an $1/(2+2\epsilon')$ approximation ratio, for $\epsilon' > 0$, as shown in Section \ref{sec:enumeration}. 
However, we will later show in Section \ref{sec:experiments} that this static grid-based enumeration fails to fully exploit the problem structure, resulting in suboptimal performance—particularly in practical instances where structural properties can be more effectively leveraged.

In the next two sections, we introduce more adaptive approximation algorithms that not only leverage the problem's structure but also attain better theoretical guarantees. Specifically, we present a $1/2$-approx. algorithm (Section \ref{sec:half-approx}) and an FPTAS (Section \ref{sec:fptas}) for solving Problem \eqref{eq:subdual} by relying on the key transformation to an infinite series of knapsack problems.

\section{1/2-Approx. Algorithm for Problem (SUB-DUAL)}
\label{sec:half-approx}

Leveraging the transformation to $\max_{W\ge 0} \kpcard{W, K}$, we first devise a $1/2$-approx. algorithm for general instances of Problem \eqref{eq:subdual}. This, along with the fair Ellipsoid-based framework from Section \ref{sec:elipsoid}, yields our \emph{$1/2$-approx. fair Ellipsoid-based algorithm}.

\subsection{Relaxed Knapsack Problems and the Notion of Profile}
\label{sec:profile-notion}
Our algorithm crucially uses the LP relaxation of the original knapsack problem \eqref{eqn:knapsack-with-size}, defined as
\begin{equation}
\label{eqn:relaxed-knapsack}\tag{$\textsc{kp-relax}(W, K)$}
\begin{aligned}
\kprelax{W, K} =\max_{\mathbf{x} \in [0,1]^n} & 
\sum_{i \in [n]} (\frac{\r_i w_i}{1+W} - \c_i) x_{i} \quad 
\mathrm{s.t.}  \sum_{i \in [n]} w_{i} x_{i} \leq W {\text { and }} 
 \sum_{i \in [n]} {x_i} \leq K \,.
\end{aligned}
\end{equation} 
Here, the variable $x_i$ can be viewed as the fraction of item $i$ that we fit in the knapsack. For integer solutions $\mathbf x\in \{0,1\}^n$ to  Problem \eqref{eqn:relaxed-knapsack}, we sometimes use the term ``sets" where  the set associated with $\mathbf{x} \in\{0,1\}^n$ is $\{i\in [n]: x_i = 1\}$.

For any fixed $W\in [0, \infty)$, the relaxed problem \eqref{eqn:relaxed-knapsack} always has an \emph{optimal basic feasible solution}\footnote{See Section~\ref{sec:LP} for the definition of a basic (feasible) solution. In Problem \eqref{eqn:relaxed-knapsack}, the feasibility region is nonempty (as $\mathbf{x} = \mathbf{0}$ is feasible) and bounded (since $\mathbf{x} \in [0, 1]^{n}$). By Corollary 2.2 in \cite{bertsimas-LPbook}, the problem has at least one basic feasible solution. Furthermore, because its optimal objective is also bounded, by Theorem 2.8 in \cite{bertsimas-LPbook}, it admits at least one optimal basic feasible solution.}, which can have at most two fractional variables. We denote an optimal basic solution to \eqref{eqn:relaxed-knapsack} as $\mathbf{x}^\star(W)$.\footnote{The optimal basic solution to Problem \eqref{eqn:relaxed-knapsack}, denoted as $x^{\star}(W)$, may not be unique. 
While one could represent profile $\mathcal{P}(W)$ as $\mathcal{P}(x^{\star}(W))$, we opt to avoid this notation for simplicity.} 
Let $P_1(W) = \{i\in [n]: x^\star_i(W) = 1\}$ be the set of items that are fully added to the knapsack. Similarly, let $P_f(W) = \{i\in [n] : 0 < x^\star_i(W) < 1\}$ (respectively $P_0(W) = \{i\in [n] : x^\star_i(W) = 0\}$) as the set of items that have  a fractional value (respectively value of zero) in $\mathbf{x}^\star(W)$. 
We denote $P_f(W)$ by an ordered pair of items $(\i,\j)$. When $P_f(W)$ contains one (zero) item, we set $\j = 0$ ($\i, \j = 0$) where $0$ represents a dummy item; otherwise, we ensure $w_{\i} \leq w_{\j}$. We denote $\mathcal P(W) =\{P_1(W), P_f(W), P_0(W)\}$ as the \emph{profile} associated with the optimal basic solution ${\bf x}^{\star}(W)$.

The following lemma from \cite{caprara2000approximation} shows that an optimal basic solution to the relaxed Problem \eqref{eqn:relaxed-knapsack} can be used to construct an integer \(1/2\)-approx. solution for Problem \eqref{eqn:knapsack-with-size}.

\begin{lemma}[$1/2$-Approx. Solutions to Problem \eqref{eqn:relaxed-knapsack} \citep{caprara2000approximation}]
\label{lemma:1/2-approx}
For a fixed $W \in [0, \infty)$, let $\mathbf{x}^\star(W)$ be an optimal basic solution to Problem \eqref{eqn:relaxed-knapsack} and let $\mathcal{P}(W)= \left\{P_1(W), P_f(W), P_0(W)\right\}$ be its profile, where $P_f(W)=(\i, \j)$. We have one of the following:
\begin{enumerate}[leftmargin =*]
    \item  If $P_f(W) = (0,0)$,  $P_1(W)$ is an integer optimal solution to Problem  \eqref{eqn:relaxed-knapsack}.
    \item   If $P_f(W) = (\i, 0)$ for some $\i\in [n]$, either $P_1(W)$ or $\{\i\}$ is an integer $1/2$-approx. solution to Problem \eqref{eqn:relaxed-knapsack}.
    \item If $P_f(W) = (\i, \j)$ for some $\i, \j \in [n]$ such that $w_{\i} \le w_{\j}$, either $P_1(W) \cup \{\i\}$ or $\{\j\}$ is an integer $1/2$-approx. solution to Problem \eqref{eqn:relaxed-knapsack}.
\end{enumerate}
\end{lemma}
Lemma \ref{lemma:1/2-approx} shows  that to find a \(1/2\)-approx. solution for Problems \eqref{eqn:knapsack-with-size} or \eqref{eqn:relaxed-knapsack}, only the profile \(\mathcal{P}(W)\) is needed, not the full optimal solution \(\mathbf{x}^\star(W)\). While \(\mathbf{x}^\star(W)\) varies across different knapsack problems \(\kpcard{W, K}\), as we will see later in Theorem~\ref{thm:half-approx-ratio},  \(\mathcal{P}(W)\) changes only a polynomial number of times. This enables partitioning the range of $W$ (i.e., \([0, \infty)\)) into polynomially many sub-intervals where \(\mathcal{P}(W)\) is fixed, allowing us to construct a \(1/2\)-approx. solution for \(\max_{W \geq 0}~\kpcard{W, K}\) in polynomial time.

\subsection{Description of the 1/2-Approx. Algorithm}
\label{sec:half-approx-description}
As alluded above, a main challenge for solving $\max_{W\ge 0}~\kpcard{W, K}$ is that the key components of our knapsack problem (e.g., utilities and capacity constraint) keep evolving as $W$ changes. To address this, we first partition the interval \([0, \infty)\) into a number sub-intervals that is \emph{well-behaving}, as defined below.

\begin{definition}[Well-behaving interval]
Given a collection of $n$ items, indexed by $i \in [n]$, we say that an interval $I = [W_{\min}, W_{\max}) \subset [0, \infty)$ is \emph{well-behaving} if the following conditions are satisfied on $I$:
\begin{enumerate}[(i)]
\item {For any given item $i \in [n]$, it satisfies either $w_i \leq W_{\min}$ or $w_i \geq W_{\max}$. }
\item The signs of the utilities of items are the same for any $W\in I$.
\item The ordering of the utilities of items is the same for $W\in I$. 
\item The ordering of the utility-to-weight ratios of items is the same for any $W\in I$.
Here, the utility-to-weight ratio of item $i \in [n]$ is $\frac{u_i(W)}{w_i} = \frac{\frac{\r_i w_i }{1+W} - \c_i}{w_i} = \frac{\r_i}{1+W} - \frac{\c_i}{w_i}$. 
\item For any given item $i \in [n]$ and two distinct items $j \neq i, k \neq i, k \neq j$, the ordering between $\frac{u_j(W) - u_i(W)}{w_j - w_i}$ and $\frac{u_k(W) - u_i(W)}{w_k - w_i}$ is the same for any $W\in I$. 
\end{enumerate}
\label{def:well-behaving}
\end{definition}

We formally show in Lemma \ref{lemma:well-behaving-subinterval} that there are at most $\mathcal{O}(n^3)$ sub-intervals to be considered.
\begin{lemma}
\label{lemma:well-behaving-subinterval}
There exists a partition of $[0, \infty)$ with at most $\mathcal{O}(n^3)$ sub-intervals such that each sub-interval $I \subset [0, \infty)$ in this partition is well-behaving, per Definition \ref{def:well-behaving}.
\end{lemma}

Note that for each well-behaving interval \( I = [W_{\min}, W_{\max}) \), the set of items that can both fit into the knapsack (i.e., \( w_i \le W \)) and have positive utility (i.e., \( u_i(W) > 0 \)) remains unchanged for all \( W \in I \). We call this set of items the \emph{eligible items} of $I$. 
Given each well-behaving interval $I$, we will focus exclusively on the set of eligible items.\footnote{{\label{footnote:1/2}If the set of eligible items for $I$ is empty, our \( 1/2 \)-approx. algorithm will directly return the trivial solution \( \emptyset \).}} To ease the expositions and avoid reindexing, we will assume $[n]$ refers to the ``eligible items'' of $I$ when presenting the algorithms, with a slight abuse of notation.

Our $1/2$-approx. algorithm is formally presented in Algorithm \ref{alg:half-approx}, which returns a $1/2$-approx. solution $S_I$ for $\max_{W \in I}~\kpcard{W, K}$ for any well-behaving interval $I$.

\setlength{\textfloatsep}{4mm}
\begin{algorithm}
\caption{$\mathcal A_{1/2}(\mathbf{w}, \mathbf{\r}, \mathbf{\c}, K, I)$: ~$1/2$-approx. algorithm for $\max_{W\in I}~\kpcard{W, K}$ }
\footnotesize{
\textbf{Input:} 
weights $\mathbf{w} = \{w_i\}_{i\in [n]}$, post-fairness revenues $\mathbf{\r} = \{\r_i\}_{i\in [n]}$, post-fairness costs $\mathbf{\c} = \{\c_i\}_{i\in [n]}$, cardinality upper bound $K$,  a well-behaving interval $I = [W_{\min}, W_{\max}) \subset [0, \infty)$ per Definition \ref{def:well-behaving}.\\
\textbf{Output:} assortment $S_{I}$. 
\begin{enumerate} 
  \item 
  \textbf{Initialization.} 
  \label{step:initialization}   
      Rank the items by their utility-to-weight ratios, where the utility and weight of an item $i$ are $u_i(W) = \frac{\r_i w_i}{1+W} - \c_i$ and $w_i$, respectively.  
      Let $h_j$ be the index of the item with the $j$th highest utility-to-weight ratio, for $j \in [n]$, and define $H_j = \{h_1, \dots, h_j\}$, for $j \in [K]$ and $\th = w(H_K)$, and initialize $\mathcal{C}_I = \emptyset$. For $j \in [n]$, add $\{j\}$ to $\mathcal{C}_I$.
  \item  \label{step:I_low}{\textbf{Interval $I_{\text{low}} = I \cap [0, \th)$}. If $I_{\text{low}}$ is non-empty:} 
  \begin{enumerate}
      \item For $j = 1, \dots, K$, add $H_j$ to $\mathcal{C}_I$.
      \item \label{step:2b}
      \textit{Stopping rule.} 
      If $u_{h_K}(\th) \geq 0$, go to Step 3; otherwise, go to Step 4 (i.e., the termination step).
  \end{enumerate}
  \item \label{step:I_high}
  \textbf{Interval $I_{\text{high}} = I \cap [\th, \infty)$.} If $I_{\text{high}}$ is non-empty:
  \begin{enumerate} 
      \item \label{step:3a}
      \textbf{Initialize the profile.}
      \begin{itemize}
          \item [i.] If $W_{\min} < \th$, set $P_1 = H_K, P_0 = [n] \setminus H_K$, and $W_\text{next} = \th$. 
          \item [ii.] If $W_{\min} \geq \th$, compute an optimal basic solution for Problem $(\kprelax{W_{\min}, K})$ with profile $\mathcal{P}(W_{\min}) = \{P_1(W_{\min}), (\i, \j), P_0(W_{\min})\}$, where $w_i < w_j$. 
          Add $P_1(W_{\min}) \cup \{\i\}$ to $\mathcal{C}_I$. Set $P_1 = P_1(W_{\min}) \cup \{\j\}$, $P_0 = [n] \setminus P_1$, and $W_\text{next} = w(P_1)$.
      \end{itemize}
      \item \label{step:3b}
      \textbf{Adaptively partitioning $I_{\text{high}}$.}
      While there exist $i \in P_1, j \in P_0$ such that $w_i < w_j$:
      \begin{enumerate}
        \item \label{step:3bi} 
        Update indices $\i^\star, \j^\star$ as follows: 
        \begin{align}\label{eq:opt}
        (\i^\star, \j^\star) \leftarrow \argmax_{i \in P_1, j \in P_0 \atop {w_i < w_j}} \left[ \frac{u_j(W_{\text{next}}) - u_i(W_{\text{next}})}{w_j - w_i} \right]\,.
        \end{align}
        \item \label{step:3bii} \textit{Stopping rule.} If
        $
        W_\text{next} \geq W_\text{max}
        $ or
        $
        u_{\i^\star}(W_{\text{next}}) > u_{\j^\star}(W_{\text{next}}) \,,
        $
        go to Step 4 (i.e., the termination step).
        \item \label{step:3b_iii}
        \textit{Swapping the two items.} 
        Update $P_1, P_0, W_{\text{next}}$ as follows:
        $$
        P_1 \leftarrow P_1 \cup \{\j^\star\} \setminus \{\i^\star\} 
        \quad \text{and} \quad  
        P_0 \leftarrow P_0 \cup \{\i^\star\} \setminus \{\j^\star\}
        \quad \text{and} \quad  
        W_{\text{next}} \leftarrow  W_{\text{next}} - w_{\i^\star} + w_{\j^\star}\,.
        $$
        Add $P_1$ to $\mathcal{C}_I$. 
      \end{enumerate}
   \end{enumerate}
     \item \textbf{Termination Step.} Return the set $S_{I} = \argmax_{S \in \mathcal{C}_I} \revmod(S, \mathbf{z})$. 
\end{enumerate}
}
\label{alg:half-approx}
\end{algorithm}

We define $H_j= \{h_1, \dots, h_j\} \subset [n]$ as the $j$ items in $[n]$ with the highest utility-to-weight ratios. Recall that on a well-behaving interval $I$, the ordering of utility-to-weight ratios of items remains constant, so the sets $\{H_j\}_{j = 1}^K$ are well-defined.
Algorithm \ref{alg:half-approx} partitions $I = [W_{\min}, W_{\max})$ into two intervals: $I_{\text{low}} = I \cap [0, \th)$ and $I_{\text{high}} = I \cap [\th, \infty)$, where $\th = w(H_K)$ is the total weight of the $K$ items with the highest utility-to-weight ratios. 
We now discuss $I_\text{low}$ and $I_\text{high}$ separately.

\textbf{Interval $I_{\text{low}} = I \cap [0, \th)$.} Solving Problem \eqref{eqn:relaxed-knapsack} for any $W \in I_{\text{low}}$ is straightforward because the cardinality constraint is not binding. The solution involves filling the knapsack with items with the highest utility-to-weight ratios until reaching capacity $W$. Consequently, the profile $\mathcal{P}(W) = \{P_1(W), P_f(W), P_0(W)\}$ of our optimal basic solution would be $P_1(W) = H_{j-1}$ and $P_f(W) = (h_{j}, 0)$ for some $j \in [K]$. This explains why we consider sets $H_j, j \in [K]$ in Step \ref{step:I_low} of Algorithm \ref{alg:half-approx}.

\textbf{Interval $I_{\text{high}} = I \cap [\th, \infty)$.} Solving Problem \eqref{eqn:relaxed-knapsack} for $W \in I_{\text{high}}$ is more challenging because profile of the optimal basic solution no longer maintains the structured form seen in interval $I_{\text{low}}$.
As capacity $W$ exceeds $\th$, the cardinality constraint becomes binding, making it suboptimal to simply fill the knapsack with items having the highest utility-to-weight ratios. Inspired by Lemma \ref{lemma:1/2-approx}, we partition
$I_{\text{high}}$ into a polynomial number of sub-intervals where the profile $\mathcal{P}(W)$ remains unchanged, allowing us to apply the same $1/2$-approx. solution. However, these sub-intervals must be determined \emph{adaptively}.

\textbf{\textit{Adaptive partitioning.}} 
The algorithm iteratively updates: (i) capacity \emph{change points} $W_{\text{next}}$, and (ii) profile sets $P_1$, which contains all items completely added to the knapsack in Problem $(\kprelax{W_{\text{next}}, K})$, and $P_0$, which contains all items not added to the knapsack. 
The next change point 
$W_{\text{next}}$ is chosen such that $P_f(W_{\text{next}}) = (0, 0)$. To update $P_1, P_0$, and $W_{\text{next}}$, the algorithm identifies two items $\i^{\star}\in P_1$ and $\j^{\star}\in P_0$ by solving a simple optimization problem stated in Step \ref{step:3bi}, aiming to swap the pair that maximizes the marginal utility increase. The swap---removing $\i^{\star}$ from $P_1$ and adding it to $P_0$, while removing $\j^{\star}$ from $P_0$ and adding it to $P_1$---is a key innovation enabling a $1/2$-approx. solution in polynomial time.

The following theorem is the main result of this section:
\begin{theorem}[$1/2$-Approx. Algorithm]
\label{thm:half-approx-ratio}
Consider a well-behaving interval $I \subset [0, \infty)$ per Definition \ref{def:well-behaving}. For any $\mathbf{z}\ge \mathbf{0}$, set $S_{I}$ returned by Algorithm \ref{alg:half-approx} satisfies
$
\revmod(S_{I}, \mathbf{z})\ge \frac{1}{2} \max_{W\in I}~ \kpcard{W,K}\,. 
$
The runtime of Algorithm~\ref{alg:half-approx} is in the order of $\mathcal{O}(n^2 \log n + n K^2)$.
\end{theorem}
We apply Algorithm \ref{alg:half-approx} to at most $\mathcal{O}(n^3)$ well-behaving intervals and select the $1/2$-approx. set $S_I$ with the highest cost-adjusted revenue, which is the $1/2$-approx. solution for $\max_{W \in [0, \infty)} \kpcard{W, K}$.
The overall runtime of our $1/2$-approx. algorithm is thus at most $\mathcal{O}(n^5 \log n + n^4 K^2)$.

\section{FPTAS for Problem (SUB-DUAL)}
\label{sec:fptas}

In this section, we further propose an FPTAS for general instances of Problem \eqref{eq:subdual}.
Similar to what we did in the design of the $1/2$-approx. algorithm, we start by pre-partitioning $[0,\infty)$ into $\mathcal{O}(n^3)$ well-behaving intervals based on Definition \ref{def:well-behaving} and Lemma \ref{lemma:well-behaving-subinterval}. Then, for each of these well-behaving interval $I$, we design an FPTAS for $\max_{W \in I} \kpcard{W, K}$, which is formally presented in Algorithm \ref{alg:fptas}.
Similar to Section \ref{sec:half-approx}, here we also assume that the items in $[n]$, fed into Algorithm \ref{alg:fptas}, refers to the eligible items of $I$.

\vspace{6mm}
\begin{algorithm}[htbp]
\caption{$\text{FPTAS}(\mathbf{w}, r, \mathbf{\c}, K, I)$: \texttt{FPTAS} for $\max_{W\in I}~\kpcard{W, K}$ }
\footnotesize{
\textbf{Input:} 
weights $\mathbf{w} = \{w_i\}_{i\in [n]}$, post-fairness revenues $\mathbf{\r} = \{\r_i\}_{i\in [n]}$, post-fairness costs $\mathbf{\c} = \{\c_i\}_{i\in [n]}$, cardinality upper bound $K$, a well-behaving interval $I \subset [0, \infty)$ per Definition \ref{def:well-behaving}, accuracy parameter $\epsilon\in (0,1)$.
\\
\textbf{Output:} assortment $S_I$

\begin{enumerate}
  \item Initialize the collection of assortments $\mathcal{C}_I = \emptyset$.
  \item \textbf{Adaptive partitioning via the $1/2$-approx. algorithm.} 
  Apply the modified $1/2$-approx. algorithm (Algorithm \ref{alg:half-approx-modified}):
  $$
  \Pi_I, \mathcal{D} \leftarrow \mathcal A^\text{m}_{1/2}(\mathbf{w}, \mathbf{\r}, \mathbf{\c}, K, I)\,,
  $$
  where $\Pi_I = \{I_\ell\}_{\ell \in [L]}$ is a partition of $I$ and $\mathcal{D}(I_\ell)$ is the $1/2$-approx. solution corresponding to interval $I_\ell$. 
  \item \textbf{Further partitioning via monotonicity.} For each $I_\ell \in \Pi_I$, partition $I_\ell$ into at most $\mathcal{O}(nK/\epsilon)$ sub-intervals $I'_\ell$ such that the rescaled utility $\tilde{u}_i(W)$, defined as follows, takes the same value for all $i \in [N]$:
  \begin{equation}
  \label{eqn:rescaled-util}
    \tilde{u}_i(W) = \Big\lceil \dfrac{u_{i}(W)}{\sum_{i' \in \mathcal{D}(I_\ell)} u_{i'}(W)} \cdot \frac{K}{\epsilon} \Big\rceil \,.
    \end{equation}
  \item \textbf{Dynamic Programming (DP).} For each $I'_\ell = [W_{\min}, W_{\max}) \subseteq I_\ell$, perform the following DP scheme.
  \begin{enumerate}
  \item \textbf{Re-scale the utilities.} 
  Let $\tilde{u}_i$ denote the rescaled utility of item $i$ on interval $I'_\ell$, based on Eq. \eqref{eqn:rescaled-util}.
  \item \textbf{Initialization.} 
  Let $\chi_{\max}= \lceil 2 \cdot \frac{K}{\epsilon} + K \rceil$. 
  Let $\minwt_0(0, 0) = 0$, $\minwtset_0(0, 0) = \emptyset$. Let  $\minwt_0(\chi, \kappa) = \infty$ for all $\chi \in \{0, \dots, \chi_{\max}\}$, and $\kappa \in \{0, \dots, K\}$. 
 Finally, for any $i\in \{0, 1, \ldots, n\}$,  set $\minwt_i(\chi, \kappa) = \infty$ and $\minwtset_i(\chi, \kappa) = \emptyset$ when either $\chi$ or $\kappa$ are negative. 
  \item \textbf{Recursion.} For $i = 1, \dots, n$, 
  \begin{itemize}
      \item For any $\chi \in \{0, \dots, \chi_{\max}\}$ and $\kappa \in \{0, \dots, K\}$, compute the entries of $\minwt_i(.)$ as follows:
      \begin{equation*}
      \minwt_{i}(\chi, \kappa) = \min \big\{\minwt_{i-1}(\chi, \kappa),~ \minwt_{i-1}\left(\chi - \tilde{u}_i, \kappa - 1 \right) + w_i\big\}
      \end{equation*}
      \item If $\minwt_i(\chi, \kappa) < \infty$, update its corresponding assortment:
        \begin{equation}\label{eq:update}
        \minwtset_{i}(\chi, \kappa) = 
        \left\{\begin{array}{ll}
        \begin{aligned}
        & \minwtset_{i-1}(\chi, \kappa) & \quad & \text{if $\minwt_i(\chi, \kappa) = \minwt_{i-1}(\chi, \kappa)$} \\
        & \minwtset_{i-1}\left(\chi - \tilde{u}_i, \kappa - 1 \right) \cup \{i\} & \quad & \text {if $\minwt_i(\chi, \kappa) = \minwt_{i-1}(\chi-\tilde{u}_i, \kappa-1) + w_i$}
        \end{aligned}
        \end{array}\right.
        \end{equation}
     \end{itemize}
  \item \textbf{Collect assortments.}  For any $\chi \leq \chi_{\max}$ and $\kappa \leq K$ such that $\minwt_n(\chi, \kappa) \leq W_{\max}$, add $\minwtset_n(\chi, \kappa)$ to $\mathcal{C}_I$.
  \end{enumerate}
\item Return the assortment $S_I = \argmax_{S \in \mathcal{C}_I} \revmod(S, \mathbf{z})$.
\end{enumerate}
}
\label{alg:fptas}
\end{algorithm}
\vspace{3mm}

\textbf{Adaptive partitioning via the $1/2$-approx. algorithm.} Given a well-behaving interval $I$, our FPTAS starts by applying a slightly modified version of our $1/2$-approx. algorithm, detailed in Algorithm \ref{alg:half-approx-modified} of Section \ref{sec:half-mod-for-fptas}, to handling $I$. Algorithm \ref{alg:half-approx-modified} remains largely consistent with Algorithm \ref{alg:half-approx}, with the key difference being its keeping track of the partition generated by our adaptive partitioning process. Algorithm \ref{alg:half-approx-modified} returns a partition of $I$ containing at most $\mathcal{O}(nK)$ sub-intervals, denoted by $\Pi_I = \{I_\ell : \ell = 0, 1, \dots, L \}$, where $I_\ell = [W_\ell, W_{\ell+1})$ with $I = [W_0, W_L)$ and $W_0 < W_1 < \dots < W_L$. Moreover, it produces a mapping $\mathcal{D}$ assigning each sub-interval $I_\ell$ a set $\mathcal{D}(I_\ell)$, which constitutes a feasible $1/2$-approx. solution for any $W \in I_\ell$. In other words, for any $W \in I_\ell$, we have 
$\sum_{i \in \mathcal{D}(I_\ell)} u_i(W) \geq \frac{1}{2} \kpcard{W, K}$ and $\sum_{i \in \mathcal{D}(I_\ell)} w_i \leq W$.

\textbf{Partitioning via monotonicity.}
We next focus on each sub-interval $I_\ell$ that is well-behaving and admits a common $1/2$-approx. solution $\mathcal{D}(I_\ell)$.
Our FPTAS adopts a DP scheme that resembles the one used for a single knapsack problem with cardinality constraint \citep{caprara2000approximation}. 
However, the DP scheme necessitates rescaling each item's utility, which is not straightforward for us.

One idea is to follow the approach in \cite{caprara2000approximation} and re-scale the utility of each item $i$ using a factor that depends on $K, \epsilon$ and the total utility of the $1/2$-approx. solution $\mathcal{D}(I_\ell)$, as follows:
\begin{equation}
\tilde{u}_i(W) = \Big\lceil \dfrac{u_{i}(W)}{\sum_{i' \in \mathcal{D}(I_\ell)} u_{i'}(W)} \cdot \frac{K}{\epsilon} \Big\rceil \,.
\end{equation}
Nonetheless, the hurdle here is that the utility of our $1/2$-approx. solution $\sum_{i \in \mathcal{D}(I_\ell)} u_i(W)$ evolves with $W$ and the scaled utility $\tilde{u}_i(W)$ might change its value within interval $I_\ell$. To tackle this, we show that $\tilde{u}_i(W)$ only changes monotonically within interval $I_\ell$, as stated in the following lemma:
\begin{lemma}
\label{lemma:monotonic}
If an interval $I_{\ell}$ is well-behaving 
per Definition \ref{def:well-behaving} 
and admits a common $1/2$-approx. solution $\mathcal{D}(I_{\ell})$, the re-scaled utility $\tilde{u}_i(W)$ defined in Eq. \eqref{eqn:rescaled-util} changes monotonically with respect to $W$. 
\end{lemma}
A similar monotonicity property was also adopted by \cite{holzhauser2017fptas}, who designed an FPTAS for a single parametric knapsack problem.  
Given Lemma \ref{lemma:monotonic}, and given that $\tilde{u}_i(W) \in [0, 2K/\epsilon]$ since we rescale the utility using the $1/2$-approx. solution, we know that for any $i \in [n]$, the value of $\tilde{u}_i(W)$ changes at most $O(K/\epsilon)$ times within the interval $I_\ell$.
This allows us to further divide $I_\ell$ into at most $O(nK/\epsilon)$ intervals, each denoted by $I'_\ell$, such that on each interval $I'_\ell$, the re-scaled utility $\tilde{u}_i(W)$ defined in Eq. \eqref{eqn:rescaled-util} takes the same value for all items $i \in [N]$.

\textbf{The DP scheme.}
After two rounds of partitioning described above, we focus on each interval $I'_\ell =[W_{\min}, W_{\max}) \subseteq I_\ell$ such that $I'_\ell$ is well-behaving, admits a common $1/2$-approx. solution $\mathcal{D}(I_\ell)$, and the rescaled utility $\tilde{u}_i(W)$ of each item $i$ takes the same integer value for all $W \in I'_\ell$.
This allows us to simply write the rescaled utility of each item $i$ as 
$
\tilde{u}_i (W) = \tilde{u}_i 
$
for all $W\in I_{\ell}'$. 

\textbf{\textit{Recursion in DP.}} 
For any $W \in I'_\ell$, since the optimal value to our original knapsack problem $\kpcard{W, K}$ is bounded by twice our $1/2$-approx. solution, $2 \sum_{i' \in \mathcal{D}(I_\ell)} u_{i'}(W)$, the optimal value with re-scaled utilities is thus bounded by $ 2 \cdot \frac{K}{\epsilon} + K := \chi_{\max}$. 
For any $\chi \in \{0,1, \ldots, \chi_{\max}\}$, $\kappa \in \{0,1, \ldots, K\}$, and $i\in [n]$, let $\minwt_i(\chi, \kappa)$ denote the minimum total weight of $\kappa$ items from $\{1, \dots, i\}$ that can yield total utility $\chi$, and let $\minwtset_i(\chi, \kappa)$ be the corresponding assortment consisting of these $\kappa$ items. 
If we cannot find exactly $\kappa$ items from $\{1, \dots, i\}$ whose total utility is $\chi$, we set $\minwt_i(\chi, \kappa) = \infty$. 
We compute the values for $\minwt_i(.)$ using values from $\minwt_{i-1}(.)$, and construct the assortment $\minwtset_i(.)$ from $\minwtset_{i-1}(.)$ in a recursive manner. For any $i = 1, \dots, n$, we let:
\begin{align}
\minwt_{i}(\chi, \kappa) = \min \big\{\minwt_{i-1}(\chi, \kappa), \minwt_{i-1}\left(\chi - \tilde{u}_i, \kappa - 1 \right) + w_i\big\}\,. 
\end{align}
In the first term, we do not include item $i$ in the set, while in the second term, we include item $i$ in the set. 
If $\minwt_{i}(\chi, \kappa) < \infty$, we update its corresponding assortment according to Eq. \eqref{eq:update}.
\footnote{When $\minwt_{i}(\chi, \kappa) = \infty$, its corresponding set is not used by our algorithm, and hence we do not define it.} 
This then gives us access to the matrix $\minwt_{n}(\chi, \kappa)$. Set $\minwtset_n(\chi, \kappa)$, if $\minwt_{n}(\chi, \kappa) < \infty$, is the assortment with minimum weight that satisfies 
$\sum_{i \in \minwtset_n(\chi, \kappa)} \tilde{u}_i = \chi$ and $|\minwtset_n(\chi, \kappa)| = \kappa$.

If we were to solve a single knapsack problem $\kpcard{W, K}$ for a fixed $W$, a near-optimal solution value would be given by 
$\max_{\chi, \kappa} \{ \chi : \minwt_n(\chi, \kappa) \leq W\}\,.$
Here, since our interval $I_{\ell}$ involves infinite knapsack problems, we instead consider all $\minwtset_n(\chi, \kappa)$ that can fit into a knapsack of capacity $W_{\max}$, i.e., $\minwt_n(\chi, \kappa) \leq W_{\max}$. 
One of these assortments is guaranteed to be an $(1-\epsilon)$-approx. solution to $\max_{W \in I'_\ell} \kpcard{W, K}$.

In the following theorem, which is the main result of Section~\ref{sec:fptas}, we show that Algorithm \ref{alg:fptas} achieves $(1-\epsilon)$ approximation ratio with runtime polynomial in $n, K,1/\epsilon$.

\begin{theorem}[Near-optimality of Algorithm~\ref{alg:fptas}]
\label{thm:fptas-optimality}
Consider a well-behaving interval $I \subset [0, \infty)$ per Definition \ref{def:well-behaving}. For any $\epsilon > 0$ and $\mathbf {z}\ge \mathbf{0}$, set $S_I$ returned by Algorithm~\ref{alg:fptas} satisfies
$\revmod(S_I, \mathbf{z})\ge (1-\epsilon) \max_{W\in I}~ \kpcard{W,K}\,.$
The runtime of Algorithm~\ref{alg:fptas} is in the order of $\mathcal{O}(n^3 K^4/\epsilon^2)$. 
\end{theorem}

Similar to Section \ref{sec:half-approx}, we apply Algorithm~\ref{alg:fptas} to each well-behaving sub-interval $I$. The $S_I$ with the highest cost-adjusted revenue is an $(1-\epsilon)$-approx. solution to Problem \eqref{eq:subdual}. The overall runtime of our FPTAS is $\mathcal{O}(n^6 K^4/\epsilon^2)$
since Algorithm \ref{alg:fptas} is applied to $\mathcal{O}(n^3)$ well-behaving sub-intervals. 

\section{Numerical Experiments}
\label{sec:experiments}

In this section, we numerically evaluate our method's efficacy and investigate the impact of fair policies in assortment planning. Section \ref{sec:synthetic_data} compares the $1/2$-approx. algorithm, FPTAS, and benchmark algorithms  on synthetic data, highlighting $1/2$-approx. algorithm's practical performance. Section \ref{sec:numerics} further presents a real-world MovieLens case study, showcasing the impact of imposing fairness in practical settings.

\subsection{Experiments on Synthetic Data}
\label{sec:synthetic_data}

\subsubsection{Setup.}
We consider an assortment planning problem with $n = 10$ items, where the offered assortment can include up to $K = 5$ items, resulting in 637 possible assortments.\footnote{We have also experimented with synthetic instances with larger $n$, where $n = 10, 15, \dots, 40$, and the results remain consistent. See Section \ref{sec:comparison-enumeration} for related discussions.}
We generate 100 instances, drawing $r_i$ (item $i$'s revenue) independently from $\text{Uniform}[0, 1]$ and $\theta_i$ (other item features) from $\text{Uniform}[0, 0.5]$. Each item's popularity score is defined as $w_i = \exp(\beta \cdot r_i + \theta_i)$, correlating popularity with revenue and items' features \citep{train2009discrete}. The price sensitivity parameter $\beta$ is set to $-1$ (high sensitivity) or $-0.1$ (low sensitivity), and $q_i = w_i$. The platform aims to maximize expected revenue while ensuring $\delta$-fairness, where $\delta \in \{0, 0.2, 0.4, 0.6, 0.8, 1.0\}$ and an item's outcome is its visibility. 

We compare our \(1/2\)-approx. and FPTAS (with $1-\epsilon=0.75$)  algorithms with the following benchmark algorithms. In all algorithms, to enhance computational efficiency, we use a column generation method  instead of the theoretically polynomial-time Ellipsoid-based method.  See Section \ref{sec:column-generation} for details.

\begin{itemize}[leftmargin=*]
    \item \textbf{Randomized greedy algorithm.}  
    As a benchmark, we include the randomized greedy algorithm from \cite{Buchbinder2014}, which iteratively selects items that maximize the objective function until no further improvement is possible. While this algorithm achieves a \(1/e\) approximation ratio for non-monotone submodular functions, the objective function in Problem \eqref{eq:subdual} is generally not submodular, except in the special case when \(r_i = r\), $i\in [n]$.  
    \item \textbf{Grid-based enumeration algorithm.}  
    We also consider a grid-based enumeration method, which discretizes the interval \(I = [0, \infty]\) using a geometric grid with ratio \((1+\epsilon')\) for some $\epsilon' > 0$. It solves the LP relaxation and applies the rounding scheme from Lemma \ref{lemma:1/2-approx} to achieve a \(1/2\)-approx. solution at each grid point. This method guarantees a \(1/(2+2\epsilon')\)-approx. for Problem \eqref{eq:subdual}; see Section \ref{sec:enumeration} for details. In our experiments, we set \(\epsilon' = 1/49\), resulting in an approximation ratio of 0.49.
    
    \noindent While this method offers theoretical guarantee close to our 1/2-approx. algorithm, we will see that the 1/2-approx. algorithm outperforms in both solution quality and runtime. Unlike the enumeration method that relies on a fixed partitioning scheme, the $1/2$-approx. algorithm efficiently exploits and adapts to problem structure, especially in cases with large $\delta$ values or homogeneous revenues (see Tables \ref{tbl:numerical_results_synthetic} and \ref{tbl:comparison}).
\end{itemize}

\subsubsection{Performance and Runtime Comparison.}
\label{sec:synthetic-comparison}
\begin{figure}[htb]
\centering
\subfloat[The setting of high price sensitivity $(\beta = -1)$.]{
\includegraphics[width=0.47
\textwidth]{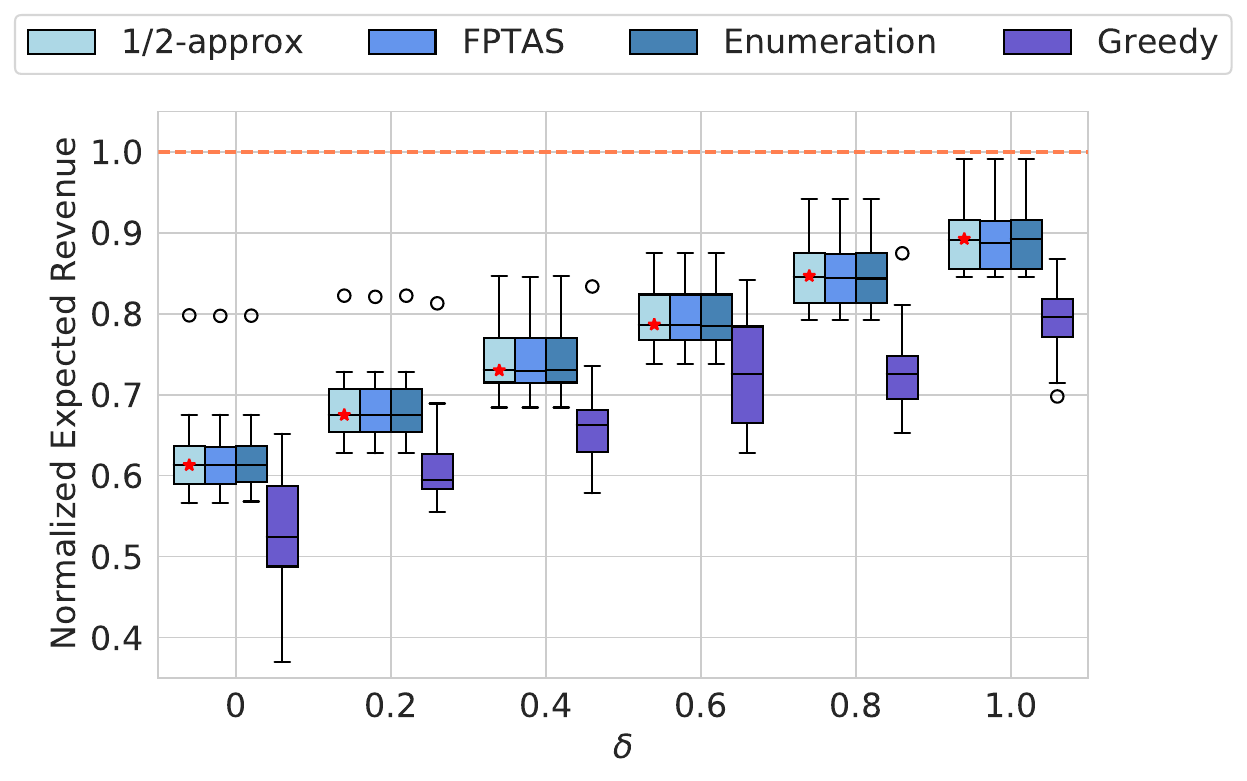}
\label{subfig:synthetic-rev-high}
}
\subfloat[The setting of low price sensitivity $(\beta = -0.1)$.]
{
\includegraphics[width=0.47\textwidth]{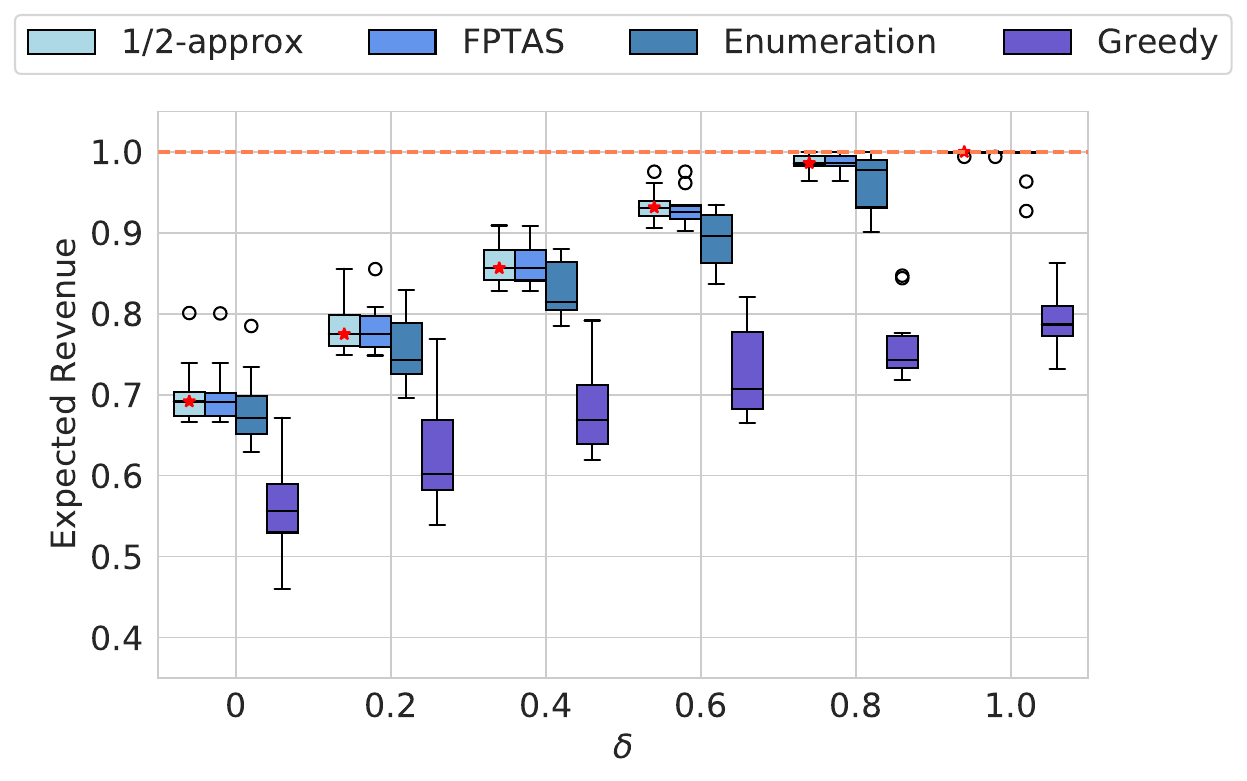}
\label{subfig:synthetic-rev-low}
}
\vspace{2mm}
\caption[PDF]{\footnotesize{Boxplots of expected revenue across algorithms under different fairness levels ($\delta$). The box spans the first to third quartile, with a median line. The dashed line represents average optimal revenue without fairness constraints, and the red star marks the median optimal solution to Problem \eqref{eq:problem:fair}.
} 
}
\label{fig:synthetic-rev}
\end{figure}

In Figure \ref{fig:synthetic-rev}, we show the normalized expected revenue obtained by different algorithms under varying fairness levels \(\delta\). The normalized revenue is calculated as the expected revenue divided by the maximum attainable revenue without fairness constraints. Our \(1/2\)-approx. algorithm and FPTAS produce results very close to the optimal solution to Problem \eqref{eq:problem:fair} across all settings and fairness levels. The greedy algorithm performs poorly, as expected, due to the lack of theoretical guarantees. The enumeration method achieves comparable revenue to our methods in high price sensitivity settings but underperforms in low price sensitivity scenarios. This is because the enumeration method is non-adaptive, where the geometric grid partitioning is determined solely by the item weights, without accounting for the underlying structure of the problem (see Section \ref{sec:enumeration}). In the low price sensitivity setting, insufficient partitioning of \([0, \infty)\) results in reduced performance.

Table \ref{tbl:numerical_results_synthetic} records the runtime of all algorithms on the synthetic instances. As expected, the greedy algorithm is the fastest but it underperforms in terms of solution quality. 
The FPTAS is comparatively slow, as expected from its runtime complexity, making it less practical for real-world deployment. However, its key value lies in being the first algorithm to provide arbitrary accuracy guarantees, serving as a strong benchmark for near-optimality.
The enumeration method is also slow, as it relies on a static partitioning scheme, repeatedly solving LPs regardless of problem structure.
Our 1/2-approx. algorithm outperforms both the FPTAS and the enumeration method in runtime while maintaining near-optimal solutions. Notably, as $\delta$ increases, the runtime gap between $1/2$-approx. and enumeration widens.
The adaptive partitioning in the 1/2-approx. algorithm efficiently captures the expanded decision space, while the enumeration method remains static. 

To highlight the adaptability of the $1/2$-approx. algorithm, we provide a detailed comparison with the enumeration method in Section \ref{sec:comparison-enumeration}. There, we show as $n$ increases, the $1/2$-approx. algorithm remains runtime efficient by exploiting problem structure. 

\vspace{6mm}
\setlength{\belowcaptionskip}{0pt}
\begin{table}[h]
\footnotesize
    \centering
    \caption[XXX]{\footnotesize{Average runtime comparison in the synthetic experiments.}
    }
    \label{tbl:numerical_results_synthetic}
    \footnotesize
    \begin{tabular}{ l| c | cccccc } 
    \toprule
     & \multicolumn{1}{c|}{$\delta$} & 0 & 0.2 & 0.4 & 0.6 & 0.8 & 1.0 \\
    \midrule
   \multirow{2}{*}[-5mm]{\makecell{Time taken by a\\ single execution (s)}} &  {$1/2$-approx} 
   & $9.9 \times 10^{-2}$
   & $9.3 \times 10^{-2}$
   & $8.7 \times 10^{-2}$
   & $7.2 \times 10^{-2}$
   & $5.0 \times 10^{-2}$
   & $4.6 \times 10^{-2}$
    \\
    & {FPTAS} & 1.4 & 1.3 & 1.2 & 1.1 & $8.8 \times 10^{-1}$ & 
    $8.0 \times 10^{-1}$ 
    \\
    & {Greedy} 
    & $4.5 \times 10^{-4}$ 
    & $4.5 \times 10^{-4}$ 
    & $4.4 \times 10^{-4}$ 
    & $4.5 \times 10^{-4}$ 
    & $4.7 \times 10^{-4}$ 
    & $4.5 \times 10^{-4}$ 
    \\
    & {Enumeration} 
    & 1.4 & 1.3 & 1.4 & 1.4 & 1.3 & 1.3
    \\
    \midrule
     \multirow{2}{*}[-5mm]{\makecell{Total time taken in\\ solving \eqref{eq:problem:fair} (s)}}
    & \multirow{1}{*}[0ex]{\makecell{$1/2$-approx}}
    & 2.7 & 2.5 & 2.2 & 1.6 & 1.0 & $7.8 \times 10^{-1}$
    \\
    & {\makecell{FPTAS}} & 
    28.9 & 28.7 & 24.9 & 20.0 & 13.3 & 10.1
    \\
    & {Greedy} 
    & $1.3 \times 10^{-1}$ 
    & $1.4 \times 10^{-1}$ 
    & $1.3 \times 10^{-1}$ 
    & $1.3 \times 10^{-1}$ 
    & $9.2 \times 10^{-2}$ 
    & $1.2 \times 10^{-1}$ 
    \\
    & {Enumeration} 
    & 26.4 & 24.6 & 24.4 & 22.5 & 19.6 & 16.3
    \\
    \bottomrule \addlinespace 
    \end{tabular}
\end{table}

\subsubsection{Price of Fairness.} 
\label{sec:pof-synthetic}
We briefly examine the tradeoff between fairness and the platform's expected revenue, known as the price of fairness. This serves as a key indicator for selecting the appropriate fairness level to balance fairness considerations with platform's revenue goals. See, also, our expanded analysis of price of fairness in our real-world case study in Section \ref{sec:movielens-insights}.

Figure \ref{fig:synthetic-rev} shows the clear tradeoff that as fairness constraints increase, the platform's expected revenue decreases, with this tradeoff varying between high and low price sensitivity settings.  It is noteworthy that in high price sensitivity markets, fairness constraints can ($\delta \in [0, 1]$) lead to revenue losses from 36.8\% to 10.6\%, whereas in low price sensitivity markets, losses range from 29.8\% to just 0.08\%. This highlights the need for caution in high-sensitivity markets, where fairness enforcement has a greater revenue impact.

\subsection{A Real-World Case Study on MovieLens Data}
\label{sec:numerics}
In this section, we conduct a real-world case study on the MovieLens data \citep{Harper2015} to illustrate the impact of fair policies in a realistic setting. Given our analysis in Section \ref{sec:synthetic_data}, in our case study, we primarily applied the $1/2$-approx. algorithm, given its practicality and effectiveness in yielding near-optimal solutions.

\subsubsection{Setup.}The MovieLens 100K dataset contains 100,000 ratings (1–5) from 943 users on 1,682 movies. We focus on \(n = 20\) drama movies with at least five ratings and an average rating of at least 3.
Movies are ordered by average rating $\rho_i$, ranging from $\rho_1 = 4.09$ to $\rho_{20} = 3.14$.
The popularity weight is set as \(w_i = s \cdot \rho_i\) with \(s = 1/20\), and we use \(q_i = w_i\) as the quality measure for each movie.\footnote{The scaling factor $s$ determines how movies compare to the no-purchase option. Results remain consistent with \(s = 1/10\), and with \(s = 1/20\), the weights are close, making fairness constraints crucial. Our experiment results are also consistent across genres.} 
We consider a platform that recommends at most 5 of the 20 drama movies, yielding \( \sum_{k=1}^5 {20 \choose k} = 21,699 \) possible assortments. Since movies lack associated revenues, the platform aims to maximize marketshare---the likelihood that a user selects a recommended movie from the assortment---while enforcing \(\delta\)-fairness with respect to movie visibilities, with \(\delta\) ranging from 0 to 5.

\subsubsection{Managerial insights.} 
\label{sec:movielens-insights}
Our case study provides key managerial insights in a real-world setting. 

\begin{figure}[htbp]
\centering
{
\includegraphics[width=0.38
\textwidth]{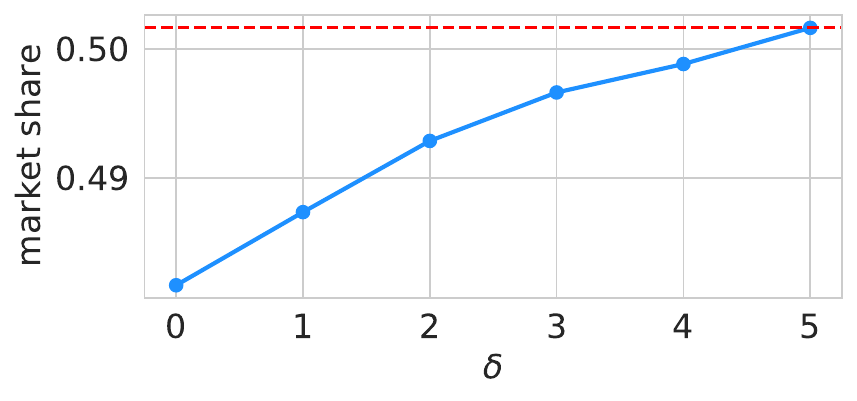}
\label{subfig:real-world-price-fairness-MS}
}\qquad
{
\includegraphics[width=0.35\textwidth]{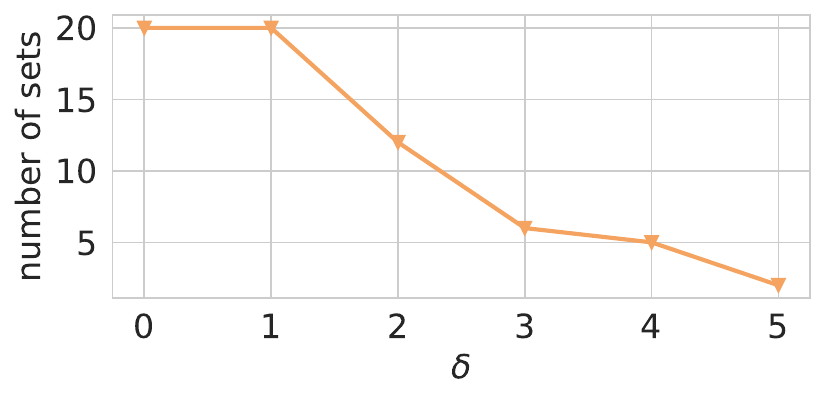}
\label{subfig:real-world-price-fairness-numsets}
}
\caption[PDF]{\footnotesize{Price of fairness in the MovieLens case study. Left: platform's marketshare obtained under different fairness parameters $\delta$. The dashed line is the optimal marketshare attainable in the absence of fairness. Right: Number of sets to randomize over under different fairness parameters $\delta$.} 
}
\label{fig:real-world-price-fairness}
\end{figure}

We first note that fairness can come with minimal tradeoff for the platform in a setting such as movie recommendation, where item qualities do not differ by that much and maintaining fairness is not in contrast with platform's goal of providing high-quality contents. Figure \ref{fig:real-world-price-fairness} illustrates the price of fairness in terms of both the loss in marketshare and the number of sets to randomize over. In terms of the marketshare, a clear tradeoff between fairness level and platform marketshare still exists: as fairness constraints relax (\(\delta\) increases), marketshare approaches the maximum attainable in the absence of fairness. However, even under strict fairness (\(\delta = 0\)), marketshare drops by only ~4\%, showing fairness can be achieved with minimal impact. On the other hand, the number of sets required for randomization ranges from 1 to 20, with the number of sets needed decreasing as $\delta$ increases. This suggests that a platform needs to randomize over a small number of sets even when strict fairness is enforced, reinforcing the practicality of our approach. 

\vspace{5mm}

Here, the minimal tradeoff is because the MovieLens setting comes with homogeneous revenues, which naturally aligns fairness with marketshare maximization, making fairness constraints highly cost-effective. When determining fairness levels $\delta$ in real-world settings beyond that of movie recommendation, it is also important to note that fairness constraints are more costly in heterogeneous revenue settings, particularly in high price sensitivity markets, as observed in Section \ref{sec:synthetic_data}. This occurs when high-quality items might generate lower revenue, whereas fairness considerations could result in the platforms allocating more visibility to them in the expense of its own revenue. This suggests that in such cases, the platform may need to adopt more conservative fairness parameters to balance fairness with its revenue goals.

\begin{figure}[htbp]
\centering
\includegraphics[width=0.7\textwidth]{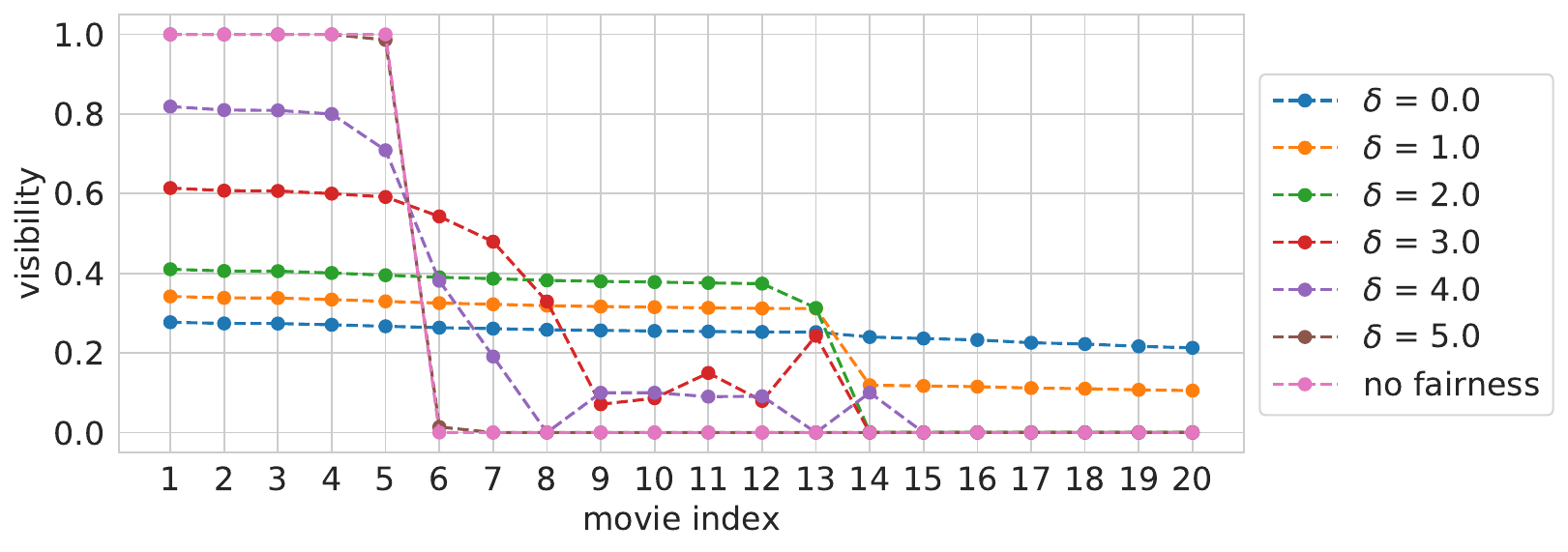}
\caption{\footnotesize Visibility received by each movie under different $\delta$. The movie indices are decreasing based on qualities.}
\label{fig:visibility-real-world}
\end{figure}

Beyond the price of fairness, a platform can also rely on other key performance metrics to guide the selection of $\delta$. Figure \ref{fig:visibility-real-world} illustrates how visibility changes under different fairness levels given our $1/2$-approx. solution. Generally, lower-quality movies receive less visibility, and for each $\delta$, there is a threshold quality score beyond which movies receive nonzero visibility. As $\delta$ increases, this threshold decreases, allowing lower-quality movies to gain more exposure. The observed behavior provides a practical approach for dynamically adjusting fairness: by analyzing the visibility distribution, platforms can identify when certain items drop out of the recommended set and fine-tune $\delta$ accordingly.

\vspace{6mm}

\section{Conclusion and Future Directions}
\label{sec:conclusion} 
Our work develops a novel framework and viable approaches for fair assortment planning, which contributes to promoting equal opportunities for items on digital platforms. We also provide managerial insights on the impact of incorporating fairness in practical assortment planning settings and how platforms can assess the tradeoff between revenue and fairness to inform its fair policies. This work has opened several exciting future directions. One is exploring how our framework adapts to different settings, such as alternative choice models or more constraints on the feasible set of assortments. Another is studying fair assortment planning in a dynamic environment, where item quality or popularity evolves based on past decisions. Ensuring fairness in such settings is crucial, as platform choices can shape item attributes, potentially reinforcing the ``winner-takes-all'' effect. While we do not address the dynamic case here, incorporating fairness constraints could promote long-term equity.
Overall, fairness remains an underexplored aspect of operational decisions on online platforms, and we hope this work inspires further research in this area.

%
%

\bibliographystyle{plainnat}
\bibliography{ref}

\newpage 

\ECHead{Appendices}

\begin{APPENDICES}

\renewcommand*{\theHsection}{\thesection}
 \renewcommand*{\theHsubsection}{\thesubsection}

\section{Discussions on Our Pairwise Fairness Notion}
\label{sec:discussion-fairness-notions}
 
In this section, we discuss the motivations and justifications for our fairness notions in Section \ref{sec:fairness-notions}. 

\textbf{Fairness through constraints.}
In existing literature, fairness is often addressed through social welfare functions like alpha fairness \citep{Mo2000Fair, Lan2010axiomatic}, which combine fairness and welfare metrics. 
However, these approaches either overlook the platform's revenue, which is crucial in assortment planning, or improperly merge it into the social welfare function, mistreating the platform as an equivalent to its items. Such a conflict of interests, where the platform's revenue goals clash with the maximization of social welfare, leads to fairness being sidelined by many platforms. The concern that dominant online platforms may leverage their power to funnel users towards the most profitable services or products has drawn regulatory attention, notably with initiatives like the proposed Digital Markets Act (DMA) by the European Union \citep{DMA}.

To reconcile the conflict of interests, we prioritize revenue maximization as the platform's primary goal, while introducing fairness in the form of constraints. Similar fairness-through-constraints approaches are seen in \cite{singh2018fairness, singh2019policy} for fair rankings, \cite{cohen2021price, cohen2021dynamic} for fair pricing, \cite{zafar2019fairness} for fair classification, \cite{kasy2021fairness} for fair algorithmic decision-making, \cite{Hogsgaard2023optimally} for fairness-welfare trade-offs, and \cite{chen2023interpolating} for fair recommendations.

\textbf{Advantages of pairwise fairness.}
In our work, we focus on pairwise fairness due to its intuitive comparison of fairness across individuals or items, aligning well with our goal of ensuring ``equality of opportunity''. As we remarked in Section \ref{sec:fairness-notions}, this notion is widely used in applications like recommendation systems and ranking models (e.g., \cite{beutel2019fairness, singh2019policy}). To the best of our knowledge, we are the first that incorporates pairwise fairness in assortment planning. 

Our pairwise fairness also offers a more nuanced and practical approach to alternative fairness notions, such as applying a uniform lower bound to all item outcomes. By ensuring that items are treated equitably in a relative rather than absolute manner, pairwise fairness strikes a balance between equitable treatment and maintaining content quality. As seen in our MovieLens case study (Figure \ref{fig:visibility-real-world}, Section \ref{sec:numerics}), under pairwise fairness, the platform can adaptively selects a subset of high-quality items that receive nonzero visibility (exposure). In contrast, a uniform lower bound on all item outcomes could potentially oversaturate the platform with too many low-quality items, hence diluting user engagement.

\textbf{Generality of our fairness notion.} 
We also highlight the rationale for considering a broad fairness framework that accommodates multiple outcome metrics, such as visibility, revenue, and market share, as done in Section \ref{sec:fairness-notions}. From the platform’s perspective, no single metric should be universally prioritized for fairness, as the impact of fairness constraints is influenced by factors such as market price sensitivity, purchasing power, and the distribution of item popularity and revenue. As shown in Figure \ref{fig:visibility-vs-sellability}, being fair with respect to different outcome metrics can yield varying effects depending on market conditions, with no single approach consistently outperforming the others. Our generalized fairness formulation enables platforms to flexibly adapt fairness constraints to their specific objectives and operational considerations.

\vspace{2mm}
\begin{figure}[hbt]
\centering
\subfloat[Low price sensitivity ($\beta = -0.1$).]{
\includegraphics[width=0.4
\textwidth]{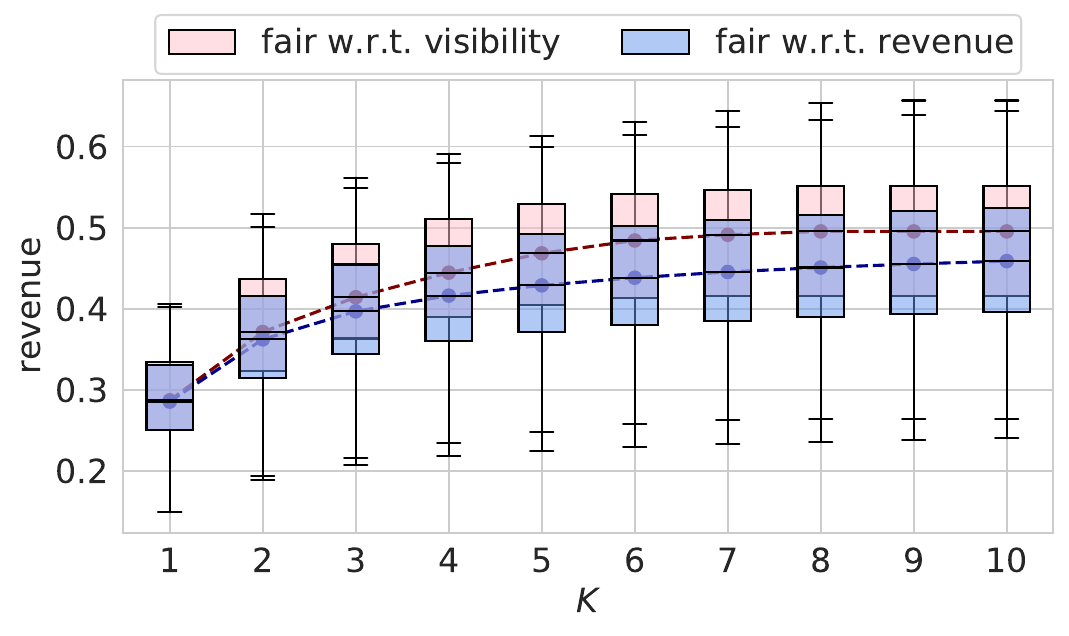}
\label{fig:visibility-vs-sellability-low}
}\qquad
\subfloat[High price sensitivity ($\beta = -1$).]{
\includegraphics[width=0.4\textwidth]{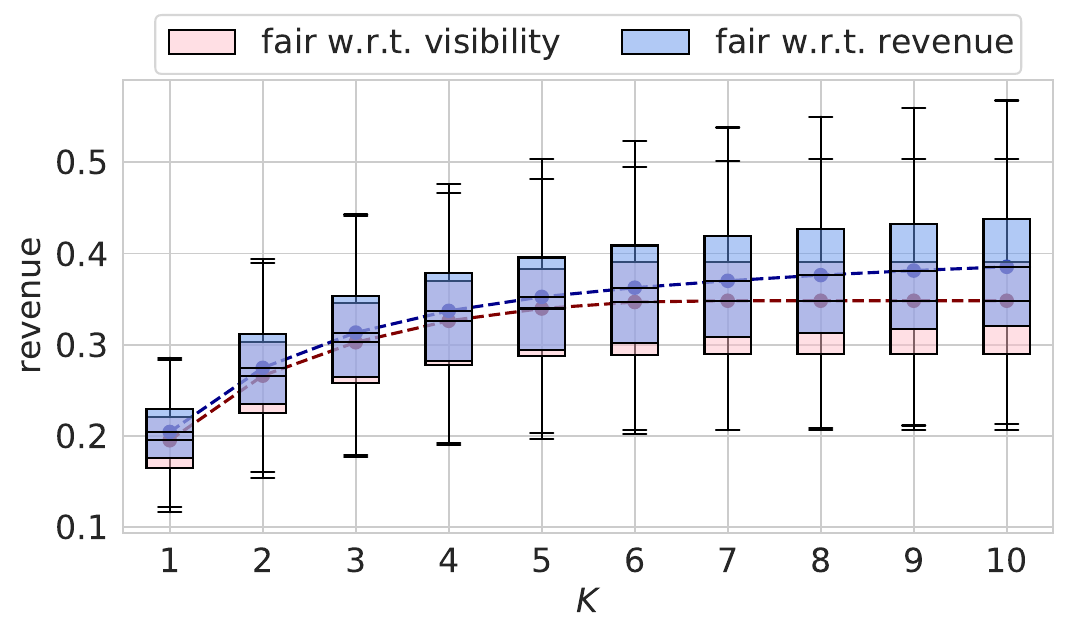}
\label{fig:visibility-vs-sellability-high}
}
\caption[PDF]{\footnotesize{
We solve Problem \eqref{eq:problem:fair} for $n=10$ and $K \in [n]$. Each item $i$ has $r_i {\sim} \text{ i.i.d. } \textsc{Unif}([0,1])$, and $w_i = \exp(\beta \cdot r_i + \theta_i)$, where $\beta$ is price sensitivity, $\theta_i \sim \text{ i.i.d. } \textsc{Unif}([0.4,1])$ for $i\leq m$ and $\textsc{Unif}([0.2,0.4])$ for $i>m$, where $1\leq m\leq \lceil n/4 \rceil$ are highly attractive items.  We set $q_i = w_i$ for all $i$. For each $(K, m, \beta)$, we generate $50$ instances and compute Problem \eqref{eq:problem:fair}'s objective when being $0$-fair w.r.t. different item outcomes. In markets with lower price sensitivity (Figure \ref{fig:visibility-vs-sellability-low}), fairness based on item visibility leads to higher platform revenue, while in markets with higher price sensitivity, fairness based on item revenue is more profitable (Figure \ref{fig:visibility-vs-sellability-high}).
} 
}
\label{fig:visibility-vs-sellability}
\end{figure}

\section{Details on the Ellipsoid Method}
\label{sec:ellipsoid-details}
In this section, we provide more details on the Ellipsoid method used in Step 1 of the fair Ellipsoid-based algorithm, proposed in Section~\ref{sec:elipsoid}. Recall that in Section~\ref{sec:elipsoid}, we assume that we have access to a polynomial-time algorithm $\mathcal{A}$ that gives a $\beta$-approx. solution for Problem \eqref{eq:subdual}. We will use algorithm $\mathcal{A}$ within the Ellipsoid method, as part of our approximate separation oracle.

In Step 1 of the fair Ellipsoid-based algorithm, we apply the Ellipsoid method, outlined in Algorithm~\ref{alg:ellipsoid-method}, to solve Problem \eqref{eq:problem:dual}. Here, we restate Problem \eqref{eq:problem:dual}:
\begin{align*}
\begin{aligned}
\textsc{fair-dual}~=~ \min_{\rho\ge 0, \mathbf{z}\ge \mathbf{0}}&   ~\, \rho+\sum_{i=1}^{n} \sum_{j = 1, j \neq i}^{n} \delta \cdot  z_{i j} 
\\
\text { s.t.} ~~~ & ~ \sum_{i \in S} O_i(S)\cdot \left(\sum_{j=1, j \neq i}^{n} (z_{i j}- z_{ji})\right)+ \rho \geq \rev(S), ~ \forall S:|S| \leq K \,.
\end{aligned}
\tag{\textsc{fair-dual}} 
\end{align*}
For simplicity of notation, we can rewrite Problem \eqref{eq:problem:dual} in the form of $\min\{\mathbf{d}^\top \mathbf{s} : \mathbf{A}\mathbf{s} \geq \mathbf{b}, \mathbf{s} \geq 0\}$. Here, $\mathbf{s} = (\mathbf{z}, \rho) \in \mathbb{R}^{n^2 +1}$ is the vector of decision variables. $\mathbf{d}$ is a vector of size $n^2 +1$ chosen such that $\mathbf{d}^\top \mathbf{s} =  \rho+\sum_{i=1}^{n} \sum_{j=1, j \neq i}^n \delta \cdot  z_{i j}$. $\mathbf{A}$ is a $N \times (n^2+1)$ matrix and $\mathbf{b}$ is a vector of size $N$, where $N = |\{ S \subseteq [n] : |S| \leq K\}|$. 
Let each row of $\mathbf{A}$ and $\mathbf{b}$ be indexed by a set $S$ with $|S| \leq K$. 
The matrix $\mathbf{A}$ is chosen such that $\mathbf{a}_S^\top \mathbf{s} = \sum_{i \in S} O_i(S) \cdot \left(\sum_{j\in [n], j \neq i} (z_{i j}-z_{j i})\right)+\rho$, where $\mathbf{a}_S$ is the row vector in $\mathbf{A}$ indexed by set $S$. The vector $\mathbf{b}$ is chosen such that $b_S = \rev(S)$, where $b_S$ is component of $\mathbf{b}$ indexed by set $S$.

Within the Ellipsoid method, we keep track of the following quantities: 
(i) $(\mathbf{z}, \rho) \in \mathbb{R}^{n^2 + 1}$, the center of the current ellipsoid, which is also the current solution to the dual problem; note that this solution might not be feasible for the dual problem. 
(ii) $(\mathbf{z}^\star, \rho^\star)$, the best feasible solution to the dual problem we have found so far. We initialize $(\mathbf{z}^\star, \rho^\star)$ to be $(\mathbf{0}_{n^2}, 1)$, where $\mathbf{0}_{n^2}$ is a zero vector of length $n^2$; this is always a feasible solution to Problem \eqref{eq:problem:dual}. 
(iii) the current best objective $\textsc{obj}$. We initialize it to be $1$, which is the objective of $(\mathbf{0}_{n^2}, 1)$. 
(iv) a positive-definite matrix $\mathbf{D} \in \mathbb{R}^{(n^2 + 1) \times (n^2+1)}$, which represents the shape of the ellipsoid.
(v) a collection $\mathcal{V}$ of sets that have violated the dual fairness constraint during the execution of the Ellipsoid method. 

\vspace{6mm}
\begin{algorithm}
\caption{The Ellipsoid method for Problem \eqref{eq:problem:dual}}
\footnotesize{
\textbf{Input:} Starting solution $(\mathbf{z}_0, \rho_0)$, starting matrix $\mathbf{D}_0$, maximum number of iterations $t_{\max}$, a $\beta$-approx. algorithm $\mathcal{A}$ for Problem \eqref{eq:subdual}. Here,  we  rewrite Problem \eqref{eq:problem:dual} in the form of $\min\{\mathbf{d}^\top \mathbf{s} : \mathbf{A}\mathbf{s} \geq \mathbf{b}, \mathbf{s} \geq 0\}$. 
 The matrix $\mathbf{A}$ is chosen such that $\mathbf{a}_S^\top \mathbf{s} = \sum_{i \in S} O_i(S) \cdot \left(\sum_{j\in [n], j \neq i} (z_{i j}-z_{j i})\right)+\rho$, where $\mathbf{a}_S$ is the row vector in $\mathbf{A}$ indexed by set $S$. The vector $\mathbf{b}$ is chosen such that $b_S = \rev(S)$, where $b_S$ is component of $\mathbf{b}$ indexed by set $S$. \\
\textbf{Output:} (i) A collection $\mathcal{V}$ of sets that have violated constraints. (ii) An optimal, feasible solution $(\mathbf{z}^\star, \rho^\star) \geq 0$ (iii) optimal objective $\textsc{obj}$.
\begin{enumerate}
  \item \textbf{Initialization.} $(\mathbf{z} ,\rho) = (\mathbf{z}_0, \rho_0),  (\mathbf{z}^\star, \rho^\star) = (0, 1),  \textsc{obj} = 1, \mathbf{D} = \mathbf{D}_0, \mathcal{V} = \emptyset,$ and  $t = 0$.
  \item \label{step:ellipsoid-while} While $t \leq t_{\max}$:
  \begin{enumerate}
      \item \textbf{Find a violated constraint.} 
      \begin{itemize}
          \item Check if we can reduce the objective further. If $\rho+\sum_{i=1}^{n} \sum_{j=1, j \neq i} \delta \cdot z_{i j}^n  \geq \textsc{obj}$, 
          set $\mathbf{a} = -\mathbf{d}$ and go to Step 2(b).           \item Check if the non-negativity constraints hold. If $\rho < 0$, set $\mathbf{a} = \mathbf{e}_{n^2+1}$; else if $z_{ij} < 0$ for some $(i, j)$, set $\mathbf{a} = \mathbf{e}_{i \cdot n + j}$, where $\mathbf{e}_k$ is the unit vector with a $1$ in the $k$th coordinate. Go to Step 2(b).
          \item Check if the dual fairness constraints holds using the approximate separation oracle. \begin{itemize}
          \item Apply the $\beta$-approx. algorithm $\mathcal{A}$ to Problem \eqref{eq:subdual} that returns $S_\mathcal{A}$ with $|S_\mathcal{A}| \leq K$ such that
          $\revmod(S_\mathcal{A}, \mathbf{z}) \ge \beta \cdot \subdual(\mathbf{z}, K)$. 
          \item
          If $\revmod(S_\mathcal{A}, \mathbf{z}) > \rho$, then set $S_\mathcal{A}$ is violating the constraint. Set $\mathbf{a} = \mathbf{a}_S$ and add $S_\mathcal{A}$ to $\mathcal{V}$.
          \end{itemize}
          \item If we have found no violated constraint, update our best feasible solution and its objective:
          $$
          (\mathbf{z}^\star, \rho^\star) \leftarrow (\mathbf{z}, \rho) \quad \text{and} \quad 
          \textsc{obj} \leftarrow \rho^\star+\sum_{i=1}^{n} \sum_{j=1, j \neq i}^n \delta \cdot  z^\star_{i j}\,.
          $$
          Then, go back to the start of Step~\ref{step:ellipsoid-while} and re-enter the while loop. 
      \end{itemize}
      \item \textbf{Use the violated constraint to decrease the volume of the ellipsoid and find a new solution.}
      $$
      \begin{aligned}
      (\mathbf{z}, \rho) & \leftarrow (\mathbf{z}, \rho) + \frac{1}{n^2 + 2}\frac{\mathbf{D}\mathbf{a}}{\sqrt{\mathbf{a}^\top \mathbf{D}\mathbf{a}}} \,;\\
      \mathbf{D} & \leftarrow \frac{(n^2+1)^2}{(n^2+1)^2 -1} \Big(\mathbf{D} - \frac{2}{n^2+2}\frac{\mathbf{D}\mathbf{a}\mathbf{a}^\top\mathbf{D}}{\mathbf{a}^\top\mathbf{D}\mathbf{a}}\Big) \,.
      \end{aligned}
      $$
  \item $t \leftarrow t + 1\,.$
  \end{enumerate}
\item \textbf{The ellipsoid is sufficiently small.} Return $\mathcal{V}, (\mathbf{z}^\star, \rho^\star), \textsc{obj}$. 
\end{enumerate}
}
\label{alg:ellipsoid-method}
\end{algorithm}
\vspace{2mm}

At a high level, the Ellipsoid method for Problem \eqref{eq:problem:dual} (Algorithm~\ref{alg:ellipsoid-method}) works as follows. At each iteration, it generates an ellipsoid $E$ centered at the current solution $\mathbf{s} = (\mathbf{z}, \rho)$, which is defined as:
$$
E = \{\mathbf{x} : (\mathbf{x} - \mathbf{s})^\top \mathbf{D}^{-1} (\mathbf{x} - \mathbf{s}) \leq 1\} \,.
$$
By design of the Ellipsoid method, the ellipsoid $E$ always contains the intersection of the feasibility region of Problem \eqref{eq:problem:dual} and the half space $\{\mathbf{s} : \mathbf{d}^\top \mathbf{s} < \textsc{obj}\}$. 

In Step 2(a), the algorithm first attempts to check if the current solution is feasible and improves our objective, via finding a violated constraint. There are three types of violated constraints that we consider: (i) the \emph{objective constraint} is violated if the current solution $(\mathbf{z}, \rho)$ does not yield an objective less than $\textsc{obj}$; (ii) the \emph{non-negativity constraint} is violated if any of the entries in $(\mathbf{z}, \rho)$ is negative; (iii) the \emph{dual fairness constraint} is violated if we can find a set $S$ with $|S| \leq K$ such that $\revmod(S, \mathbf{z}) > \rho$. Note that when examining the dual fairness constraint, instead of examining the dual fairness constraint for every $S$ such that $|S| \leq K$, we instead use an approximate separation oracle that relies on algorithm $\mathcal{A}$. First, it applies the $\beta$-approx. algorithm $\mathcal{A}$ to Problem \eqref{eq:subdual}, and gets $S_\mathcal{A}$ with $|S_\mathcal{A}| \leq K$ such that
$\revmod(S_\mathcal{A}, \mathbf{z}) \ge \beta \cdot \subdual(\mathbf{z}, K)$. Then, if $\revmod(S_\mathcal{A}, \mathbf{z}) > \rho$, set $S_\mathcal{A}$ has violated the dual fairness constraint. Otherwise, the solution $(\mathbf{z}, \rho)$ is declared feasible. In the proof of Theorem~\ref{thm:approx}, we will show that the use of the approximate separation oracle might cause the  solution $(\mathbf{z}^\star, \rho^\star)$ returned by the Ellipsoid method to be infeasible for Problem \eqref{eq:problem:dual}; however, the objective $\textsc{obj}$ returned by the Ellipsoid method would stay close to the true optimal objective $\ref{eq:problem:dual}$.

If we have found one of the constraints violated in Step 2(a), which can be written as $\mathbf{a}^\top \mathbf{s} < b$ for some $\mathbf{a} \in \mathbb{R}^{n^2+1}$ and $b \in \mathbb{R}$, we then go to Step 2(b) and construct a new ellipsoid $E'$. We do so by updating the center of the ellipsoid $(\mathbf{z}, \rho)$ and the positive-definite matrix $\mathbf{D}$ that defines our ellipsoid; see Step 2(b) (with slight differences in notations).
$$
(\mathbf{z}', \rho') \leftarrow (\mathbf{z}, \rho) + \frac{1}{n^2 + 2}\frac{\mathbf{D}\mathbf{a}}{\sqrt{\mathbf{a}^\top \mathbf{D}\mathbf{a}}} \quad \text{and} \quad
\mathbf{D}' \leftarrow \frac{(n^2+1)^2}{(n^2+1)^2 -1} \Big(\mathbf{D} - \frac{2}{n^2+2}\frac{\mathbf{D}\mathbf{a}\mathbf{a}^\top\mathbf{D}}{\mathbf{a}^\top\mathbf{D}\mathbf{a}}\Big) \,.
$$
The volume of the new ellipsoid $E'$ defined by $(\mathbf{z}', \rho')$ and $\mathbf{D}'$ is only a fraction of the volume of the previous ellipsoid $E$. In addition, this new ellipsoid has a new center $\mathbf{s}' = (\mathbf{z}', \rho')$ that satisfies $\mathbf{a}^\top \mathbf{s}' > \mathbf{a}^\top \mathbf{s}$. We then start a new iteration, in which we check whether the new solution $(\mathbf{z}', \rho')$ is feasible and improves our objective. 

On the other hand, if we have found no violated constraint, this means that our current solution $(\mathbf{z}, \rho)$ is (approximately) feasible and also improves the objective $\textsc{obj}$ obtained by the previous best feasible solution $(\mathbf{z}^\star, \rho^\star)$. If this happens, we then set the current solution to be our current best feasible solution, i.e., $(\mathbf{z}^\star, \rho^\star) \leftarrow (\mathbf{z}, \rho)$, and update the current optimal objective $\textsc{obj}$ to be the objective obtained by the current solution. Then, we start a new iteration by re-entering Step 2, and seeks to find a feasible solution that can further reduce our objective function.

The Ellipsoid method for Problem \eqref{eq:problem:dual} terminates when the ellipsoid $E$ is sufficiently small, by selecting a sufficiently large $t_{\max}$. 
Recall that the ellipsoid $E$ always contains the intersection of the feasibility region of Problem \eqref{eq:problem:dual} and the half space $\{\mathbf{s} : \mathbf{d}^\top \mathbf{s} < \textsc{obj}\}$. Intuitively, when the ellipsoid gets reduced to a sufficiently small volume, it is unlikely that there exists a feasible solution that can further reduce our objective. It is shown, in \cite{bertsimas-LPbook}, that given a separation oracle, the Ellipsoid method is guaranteed to terminate in $t_{\max} = \mathcal{O}(n^{12}\log(n))$ iterations. When the Ellipsoid method terminates, it returns the following: (i) the collection $\mathcal{V}$ of sets that have violated the dual fairness constraint; (ii) the optimal solution $(\mathbf{z}^\star, \rho^\star)$ and (iii) the optimal objective $\textsc{obj}$. Note that this collection $\mathcal{V}$ would then play an important role in Step 2 of the fair Ellipsoid-based algorithm. As discussed in Section~\ref{sec:elipsoid}, we use the collection $\mathcal{V}$ to reduce the number of variables to be considered in the primal problem \eqref{eq:problem:fair} to a polynomial size. 

\section{Details on the Column Generation Method}
\label{sec:column-generation}
In this section, we provide a brief description of the column generation method for Problem \eqref{eq:problem:fair} (see \cite{bertsimas-LPbook} for more information on this method). We have employed this method, in combination with our approximation algorithms (1/2-approx. algorithm and FPTAS) for Problem \eqref{eq:subdual}, to numerically solve the fair assortment planning problems in Sections \ref{sec:experiments}. While there is no polynomial-time guarantee for the column generation method, it is often computationally efficient in practice and serves as a viable alternative to the fair Ellipsoid-based framework in Section \ref{sec:elipsoid}, which theoretically establishes that Problem \eqref{eq:problem:fair} can be solved near-optimally in polynomial time. 

In Algorithm \ref{alg:column-generation}, we outline the column generation method for Problem \eqref{eq:problem:fair}. At a high level, the column generation considers two optimization problems---a master problem and a subproblem. The master problem is a simplified version of the original optimization problem, which, in our case, is Problem \eqref{eq:problem:fair}, restricted to a subset of variables. The subproblem is a new problem that identifies one additional variable that we can consider in the master problem, which would further improve the objective function. In our case, the subproblem reduces to solving Problem \eqref{eq:subdual}.

\vspace{5mm}
\begin{algorithm}
\caption{Column generation method for Problem \eqref{eq:problem:fair}}
\footnotesize{
\textbf{Input:} Problem instance $\mathbf{\Theta} = (\mathbf{a}, \mathbf{b}, \mathbf{r}, \mathbf{w})$, 
a $\beta$-approx. algorithm $\mathcal{A}$ for Problem \eqref{eq:subdual}. \\
\textbf{Output:} A $\beta$-approx. solution $\widehat{p}(S)$ to Problem \eqref{eq:problem:fair}.
\begin{enumerate}
  \item \textbf{Initialization.} 
  Initialize the collection of sets to include only sets with size one: $\mathcal{V} = \Big\{\{1\}, \dots, \{n\}\Big\}$. 
  \item Repeatedly solve the master and the subproblem until the stopping rule is satisified.
  \begin{enumerate}
      \item \textbf{Solve the master problem.} Solve Problem \eqref{eq:problem:fair} with $p(S) = 0$ for all $S \notin \mathcal{V}$, which returns primal solution $\widehat{p}(S)$ and dual solution $(\mathbf{{z}}, {\rho})$.
      \item \textbf{Solve the subproblem.} Apply the $\beta$-approx. algorithm $\mathcal{A}$ to solving Problem \eqref{eq:subdual}, which returns $S_\mathcal{A}$ with $|S_\mathcal{A}| \leq K, S_\mathcal{A} \notin \mathcal{V}$ such that $\revmod(S_\mathcal{A}, \mathbf{z}) \ge \beta \cdot \subdual(\mathbf{z}, K)$.
      \item \textbf{Check whether $S_\mathcal{A}$ can improve the objective of Problem \eqref{eq:problem:fair}}
      \begin{itemize}
          \item If $\revmod(S_\mathcal{A}, \mathbf{z}) > \rho$, then set $S_\mathcal{A}$ is an improving variable. Add $S_\mathcal{A}$ to $\mathcal{V}$ and go back to Step 2(a).
          \item \emph{Stopping rule.} If $\revmod(S_\mathcal{A}, \mathbf{z}) \leq \rho$, return the solution $\widehat{p}(S)$.
      \end{itemize}
  \end{enumerate}
\end{enumerate}
}
\label{alg:column-generation}
\end{algorithm}
\vspace{5mm}

At each iteration of the column generation method, we first solve the master problem, i.e., Problem \eqref{eq:problem:fair}, with only a subset of variables under consideration. That is, we only consider a restricted collection of sets $\mathcal{V} \subset \{S \subset [n] : |S| \le K \}$ when solving Problem \eqref{eq:problem:fair}, and let $p(S) = 0$ for all $S \notin \mathcal{V}$. We then need to solve a subproblem to determine if there is an additional set $S \notin \mathcal{V}$ that we wish to consider, in order to improve the objective of Problem \eqref{eq:problem:fair}. To do that, we can simply compute the \emph{reduced cost} associated with each $S$ such that $S \notin \mathcal{V}$ (see Section \ref{sec:LP} for a formal definition of the reduced cost) and pick the set $S$ with the highest reduced cost. Given problem instance $\mathbf{\Theta} = (\mathbf{a}, \mathbf{b}, \mathbf{r}, \mathbf{w})$ and the dual variables $(\mathbf{z}, \rho)$, the reduced cost $C_S$ associated with each set $S \notin \mathcal{V}$ is:
$$
\bar{C}_S = \revmod(S, \mathbf{z}) - \rho\,.
$$
Hence, if we found $S \notin \mathcal{V}$ such that $\revmod(S, \mathbf{z}) > \rho$, adding it to the collection $\mathcal{V}$ and resolving Problem \eqref{eq:problem:fair} would allow us to obtain a better objective. On the other hand, if there exists no set $S \notin \mathcal{V}$ with a positive reduced cost (i.e., $\revmod(S, \mathbf{z}) \le \rho$ for all $S \notin \mathcal{V}$, the optimality conditions of Problem \eqref{eq:problem:fair} are satisfied (see Section \ref{sec:LP}) and the current solution of Problem \eqref{eq:problem:fair} is optimal. In Algorithm \ref{alg:column-generation}, in order to solve the subproblem and find the set $S$ with the highest reduced cost, we apply a $\beta$-approx. algorithm $\mathcal{A}$ for Problem \eqref{eq:subdual}, which returns a $\beta$-approx. solution $S_\mathcal{A}$ for $\max_{S:|S| \le K} \revmod(S, \mathbf{z})$. We then compare $ \revmod(S_\mathcal{A}, \mathbf{z})$ with the dual variable $\rho$: (i) if $\revmod(S_\mathcal{A}, \mathbf{z}) > \rho$, adding $S_\mathcal{A}$ to the collection $\mathcal{V}$ and resolving Problem \eqref{eq:problem:fair} would improve the objective; (ii) otherwise, we simply terminate our algorithm and return the current solution. 

Using the same proof presented in Segment 1 of proof for Theorem \ref{thm:approx}, it can be shown that Algorithm \ref{alg:column-generation}, utilizing the column generation method, returns a $\beta$-approx. feasible solution $\widehat{p}(S), S \subset [n]$, to Problem \eqref{eq:problem:fair}. Since the number of columns (variables) is finite, the column generation method is also guaranteed to terminate in finite, but not necessarily polynomial, number of iterations.

\section{Details on a Modified 1/2-Approx. Algorithm as a Subroutine of FPTAS}
\label{sec:half-mod-for-fptas}

In this section, we present a slightly modified version of the $1/2$-approx. algorithm (Algorithm \ref{alg:half-approx}), from Section \ref{sec:half-approx}, which we use as a subroutine of our FPTAS (Algorithm \ref{alg:fptas}) from Section \ref{sec:fptas}. The algorithm is detailed in Algorithm \ref{alg:half-approx-modified}.
Similar to what we did in Sections \ref{sec:half-approx} and \ref{sec:fptas}, given the well-behaving interval $I$, the collection of item $[n]$ that we feed into Algorithm \ref{alg:half-approx-modified} refers to the \emph{eligible items} of $I$.

\vspace{6mm}
\setlength{\textfloatsep}{2pt}
\begin{algorithm}
\caption{$\mathcal A^\text{m}_{1/2}(\mathbf{w}, \mathbf{\r}, \mathbf{\c}, K, I)$: Modified~$1/2$-approx. algorithm for $\max_{W\in I}~\kpcard{W, K}$ }
\footnotesize{
\textbf{Input:} 
Same input as Algorithm \ref{alg:half-approx} \\
\textbf{Output:} \underline{In addition} to assortment $S_{I}$, \underline{return} partition $\Pi_I = \{I_\ell : \ell = 0, 1, \dots, L \}$,
where $I_\ell = [W_\ell, W_{\ell+1})$ with $I = [W_0, W_L)$ and $W_0 < W_1 < \dots < W_L$, and a mapping $\mathcal{D}$ that maps each $I_\ell$ to its $1/2$-approx. solution.
\begin{enumerate}
  \item 
  \textbf{Initialization.} \underline{In addition} to step (1a) and (1b) in Algorithm \ref{alg:half-approx}, \underline{initialize} partition $\Pi_I = \emptyset$, and mapping $\mathcal{D}$. 
  
  \item  {\textbf{Interval $I_{\text{low}} = I \cap [0, \th)$}. If $I_{\text{low}}$ is non-empty:}
  \begin{enumerate}
      \item For $j = 1, \dots, K-1$, \underline{in addition} to step (2a) in Algorithm \ref{alg:half-approx}, 
      if $w(H_j) \in I$, let $I' = [w(H_j), w(H_{j+1})) \cap I$ and update:
      $$
      (\Pi_I, \mathcal{D}) \leftarrow \texttt{update}(I', H_j, \{h_{j+1}\}, \Pi_I, \mathcal{D})
      $$
    \item Same stopping rule as step (2b) in Algorithm \ref{alg:half-approx}
  \end{enumerate}
  
  \item 
  \textbf{Interval $I_{\text{high}} = I \cap [\th, \infty)$.} If $I_{\text{high}}$ is non-empty:
  \begin{enumerate} 
      \item 
      \textbf{Initialize the profile.} \underline{In addition} to step (3a)-i and (3a)-ii, of Algorithm \ref{alg:half-approx} when  $W_{\min} \geq \th$,  with the parameters obtained in step (3a)-ii, let $I' = [W_{\min}, w(P_1(W_{\min})) + w_{\j}) \cap I$ and update
              $$
              (\Pi_I, \mathcal{D}) \leftarrow \texttt{update}(I', P_1(W_{\min}) \cup \{\i\}, \{\j\}, \Pi_I,  \mathcal{D})\,.
              $$
      \item 
      \textbf{Adaptively partitioning $I_{\text{high}}$.}
      While there exist $i \in P_1, j \in P_0$ such that $w_i < w_j$:\\ 
      \indent \underline{In addition to and after}  steps (3b)-i, (3b)-ii, and \underline{before} step (3b)-iii in Algorithm \ref{alg:half-approx}, 
      let $I' = [W_{\text{next}},  W_{\text{next}} - w_{\i^\star} + w_{\j^\star}) \cap I$ and update:
        $$
        (\Pi_I, \mathcal{D}) \leftarrow \texttt{update}(I', P_1, \{\j^\star\}, \Pi_I, \mathcal{D})\,.
        $$

   \end{enumerate}
\item \textbf{Termination Step.} 
Return $\Pi_I$ and mapping $\mathcal D$. 
     
\end{enumerate}
}
\label{alg:half-approx-modified}
\end{algorithm}

\vspace{2mm}

Algorithm \ref{alg:half-approx-modified} follows the same procedures as in Algorithm \ref{alg:half-approx}. The only difference is that it additionally keeps track of the partition of the well-behaving interval $I$ that we obtained in Step 2 and 3. Instead of directly returning the $1/2$-approx. solution to $\max_{W \in I} \kpcard{W,K}$, it returns the following:
\begin{enumerate}
    \item Partition of $I$, denoted by $\Pi_I = \{I_\ell : \ell = 0, 1, \dots, L \}$ where $I_\ell = [W_\ell, W_{\ell+1})$ with $I = [W_0, W_L)$ and $W_0 < W_1 < \dots < W_L$. 
    \item A mapping $\mathcal{D}$ that maps each sub-interval $I_\ell$ to its $1/2$-approx. solution $\mathcal{D}(I_\ell)$. That is, for any $W \in I_\ell$,
    $$
    \sum_{i \in \mathcal{D}(I_\ell)} u_i(W) \geq \frac{1}{2} \kpcard{W, K} \quad \text{and} \quad \sum_{i \in \mathcal{D}(I_\ell)} w_i \leq W\,.
    $$
\end{enumerate}
Both the partition $\Pi_I$ and the mapping $\mathcal{D}$ would be used in the design of our FPTAS (Algorithm \ref{alg:fptas}) when we rescale the utility for each item. 

Whenever we find a well-behaving sub-interval $I' \in I$ on which the profile $\mathcal{P}(W) = (P_1, \{\i, \j\}, P_0)$ is the same for all $W \in I'$, we know from Lemma \ref{lemma:1/2-approx} that there are at most two candidate sets, denoted by $S_0$ and $S_1$, that might serve as the $1/2$-approx. solution for $W \in I'$. For instance, if $\i, \j \neq 0$ and $w_{\i} \leq w_{\j}$, then $S_0 = P_1 \cup \{\i\}$ and $S_1 = \{\j\}$ would be the two candidates. Given this, the following procedure differentiates our original $1/2$-approx. algorithm (Algorithm \ref{alg:half-approx}) and the modified algorithm (Algorithm \ref{alg:half-approx-modified}):
\begin{enumerate}
    \item In our original $1/2$-approx. algorithm (Algorithm \ref{alg:half-approx}), we simply add both $S_0$ and $S_1$ to the collection of assortments $\mathcal{C}_I$ without determining which set is the exact $1/2$-approx. set.
    \item In the slightly modified $1/2$-approx. algorithm (Algorithm \ref{alg:half-approx-modified}), we instead try to determine the exact $1/2$-approx. set for every $W \in I'$. Since the $1/2$-approx. solution is limited to at most 2 candidates $S_0$, and $S_1$ and the ordering between $\sum_{i \in S_0} u_i(W)$ and $\sum_{i \in S_1} u_i(W)$ only changes at $\widehat{W}$ such that $\sum_{i \in S_0} u_i(\widehat{W}) = \sum_{i \in S_1} u_i(\widehat{W})$ (defined in Eq. \eqref{eqn:dividing-pt}), we would need to divide $I'$ at most once to create a one-to-one mapping $\mathcal{D}$ between a sub-interval and its $1/2$-approx. solution. 
\end{enumerate} 
Hence, in Algorithm \ref{alg:half-approx-modified}, when we find an interval $I'$ on which the profile stays the same, we feed it and its candidate $1/2$-approx. sets $S_0, S_1$ into our \texttt{update} subroutine (detailed in Algorithm \ref{alg:subroutine}), which determines the exact $1/2$-approx. solution for $W \in I'$ using the idea above, and update our partition $\Pi_I$ and mapping $\mathcal{D}$ accordingly.

\setlength{\textfloatsep}{2pt}
\begin{algorithm}
\caption{\texttt{update}$(I', S_0, S_1, \Pi_I,\mathcal{D}_I)$}
\footnotesize{
\textbf{Input:} a well-behaving interval $I' = [\underline{W}, \overline{W})$ on which the profile does not change, the two candidate $1/2$-approx. solutions $S_0, S_1$, partition $\Pi_I$, mapping $\mathcal{D}$
\begin{enumerate}
\item 
Compute the utility generated by $S_0, S_1$ at two end points of $I'$:
$$
\underline{U}_0 = \sum_{i \in S_0} u_i(\underline{W}), \quad 
\underline{U}_1 = \sum_{i \in S_1} u_i(\underline{W}), \quad  
\overline{U}_0 = \sum_{i \in S_0} u_i(\overline{W}), \quad
\overline{U}_1 = \sum_{i \in S_1} u_i(\overline{W})\,.
$$
Compute the dividing point:
\begin{equation}
\label{eqn:dividing-pt}
\widehat{W} = \dfrac{\sum_{i \in S_0} \r_i w_i - \sum_{i \in S_1} \r_i w_i}{\sum_{i \in S_0} \c_i - \sum_{i \in S_1} \c_i}  - 1\,.
\end{equation}
and let $I'_0 = [\underline{W}, \widehat{W})$ and $I'_1 = [\widehat{W}, \overline{W})$.
\item Update the partition $\Pi_I$, the collection $\mathcal{C}_I$ and the mapping $\mathcal{D}$:
\begin{itemize}
    \item \textbf{Case 1:} If $\underline{U}_0 \geq \underline{U}_1$ and $\overline{U}_0 \geq \overline{U}_1$, $S_0$ is the $1/2$-approx. solution for all $W \in I'$. Add $I'$ to $\Pi_I$. Set $\mathcal{D}(I') \leftarrow S_0$.
    \item \textbf{Case 2:} If $\underline{U}_1 \geq \underline{U}_0$ and $\overline{U}_1 \geq \overline{U}_0$, $S_1$ is the $1/2$-approx. solution for all $W \in I'$. Add $I'$ to $\Pi_I$. Set $\mathcal{D}(I') \leftarrow S_1$.
    \item \textbf{Case 3:} If $\underline{U}_0 \geq \underline{U}_1$ and $\overline{U}_1 \geq \overline{U}_0$. $S_0$ is the $1/2$-approx. solution for $W \in I'_0$ and $S_1$ is the $1/2$-approx. solution for $W \in I'_1$. Add $I'_0, I'_1$ to $\Pi_I$. Set $\mathcal{D}(I'_0) \leftarrow S_0, \mathcal{D}(I'_1) \leftarrow S_1$.
    \item \textbf{Case 4:} If $\underline{U}_1 \geq \underline{U}_0$ and $\overline{U}_0 \geq \overline{U}_1$. $S_1$ is the $1/2$-approx. solution for $W \in I'_0$ and $S_0$ is the $1/2$-approx. solution for $W \in I'_1$. Add $I'_0, I'_1$ to $\Pi_I$. Set $\mathcal{D}(I'_0) \leftarrow S_1, \mathcal{D}(I'_1) \leftarrow S_0$.
\end{itemize}
\item Return the updated partition $\Pi_I$ and mapping $\mathcal{D}$.
\end{enumerate}
}
\label{alg:subroutine}
\end{algorithm}

Using the same proof for Theorem \ref{thm:half-approx-ratio}, we have the following corollary:
\begin{corollary}[Modified $1/2$-Approx. Algorithm]
\label{corollary:half-approx-ratio}
Consider a well-behaving interval $I \subset [0, \infty)$ per Definition \ref{def:well-behaving}. 
For any $\mathbf{z}\ge \mathbf{0}$, 
Algorithm \ref{alg:half-approx-modified} returns a partition $\Pi_I = \{I_\ell\}_{\ell \in [L]}$ of interval $I$ and a mapping $\mathcal{D}$ such that $\mathcal{D}(I_\ell)$ is a feasible $1/2$-approx. set for any $W \in I_\ell \subseteq \Pi_I$. That is, for any $W \in I_\ell$, 
\[
\sum_{i \in \mathcal{D}(I_\ell)} u_i(W)  \ge \frac{1}{2}~ \kpcard{W,K} \quad \text{and} \quad \sum_{i \in \mathcal{D}(I_\ell)} w_i \leq W\,.
\]
The overall complexity of Algorithm~\ref{alg:half-approx-modified} is in the order of $\mathcal{O}(n^2 \log n + n K^2)$.
\end{corollary}
The runtime of Algorithm \ref{alg:half-approx-modified} remains identical to that of Algorithm \ref{alg:half-approx}, as identifying the $1/2$-approx. set for each well-behaving interval 
$I'$ with a shared profile using Algorithm \ref{alg:subroutine} is an $\mathcal{O}(1)$ operation.

\section{Details on the Grid-Based Enumeration Algorithm}
\label{sec:enumeration}
In this section, we introduce a grid-based enumeration algorithm as an alternative approach for solving Problem \eqref{eq:subdual} via the transformation to an infinite series of knapsack problems $\max_{W \geq 0} \kpcard{W, K}$ (Theorem \ref{thm:kp-su-dual}). 

\subsection{Description of the Grid-Based Enumeration Method}

The grid-based enumeration algorithm is presented in Algorithm \ref{alg:enumeration}. 

\vspace{6mm}
\begin{algorithm}
\caption{A grid-based enumeration algorithm for $\max_{W\geq 0}~\kpcard{W, K}$ }
\footnotesize{
\textbf{Input:} 
weights $\mathbf{w} = \{w_i\}_{i\in [n]}$, post-fairness revenues $\mathbf{\r} = \{\r_i\}_{i\in [n]}$, post-fairness costs $\mathbf{\c} = \{\c_i\}_{i\in [n]}$, cardinality upper bound $K$, parameter $\epsilon' > 0$ \\
\textbf{Output:} Assortment $S$.
\begin{enumerate}
  \item \textbf{Initialization.} Initialize the collection of assortments $\mathcal{C} = \emptyset$. 
  \item For $i = 1, 2, \dots, n$, 
  \begin{itemize}
      \item First, let $W = w_i$. 
      \item While $W < n \cdot w_i$,
      \begin{enumerate}
          \item Compute an optimal basic feasible solution for Problem $(\kprelax{W, K})$ with profile $\mathcal{P}(W) = \{P_1(W), (\i, \j), P_0(W)\}$, where $w_i < w_j$. Add $P_1(W) \cup \{\i\}$ and $\{j\}$ to $\mathcal{C}$. 
          \item \textbf{Proceed to the next geometric grid.} Let $W \leftarrow W \cdot (1+\epsilon')$
     \end{enumerate}
    \end{itemize}
\item \textbf{Termination Step.} 
Return $S = \argmax_{S \in \mathcal{C}} \revmod(S, \mathbf{z})$. 
\end{enumerate}
}
\label{alg:enumeration}
\end{algorithm}
\vspace{4mm}

The grid-based enumeration algorithm first enumerates over which item $j \in [n]$ is the item of the highest weight in the optimal assortment that maximizes the revenue-adjusted cost. For each $j \in [n]$, we consider the interval $[w_j, n \cdot w_j)$ and discretize it using a geometric grid with ratio $(1+\epsilon')$ for some $\epsilon' > 0$. We then solve the LP relaxation $\kprelax{W, K}$ at each grid point and use the rounding scheme from Lemma \ref{lemma:1/2-approx} to achieve a 1/2-approx. solution at each grid point. It can be shown that the grid-based enumeration algorithm attains a $1/(2+2\epsilon')$ approximation ratio, as established in the following proposition.

\begin{proposition}
\label{prop:enumeration-perf}
Given $\epsilon' > 0$, the set $S$ returned by Algorithm \ref{alg:enumeration} satisfies $
\revmod(S_{I}, \mathbf{z})\ge 1/(2+2\epsilon') \cdot \max_{W \geq 0}~ \kpcard{W,K} \,.$ The runtime of Algorithm \ref{alg:enumeration} is in the order of $\mathcal{O}(n^2 \log(n)^2 / \log(1+\epsilon'))$.
\end{proposition}

\emph{Proof of Proposition \ref{prop:enumeration-perf}: }
Fix $\epsilon' > 0$. Let $W^\star = \argmax_{W \in [0, \infty)} \kpcard{W, K}$ and let $S^\star$ be the corresponding set. Let $k$ be the maximum-weight item in $S^\star$. Then, consider the interval $I = [w_k, n \cdot w_k)$. Since we partition the interval using geometric grids, there exists grid point $W$ such that $\frac{1}{1+\epsilon'} W^\star \leq W \leq W^\star$. Consider the optimal basic feasible solution that we obtained at grid point $W$, with profile $\mathcal{P}(W) = \{P_1(W), (\i, \j), P_0(W)\}$. By Lemma \ref{lemma:1/2-approx}, we know that either $P_1(W) \cup \{\i\}$ or $\{\j\}$ is a $1/2$-approx. solution to $\kprelax{W, K}$. Let $S_W$ denote this $1/2$-approx. solution. We have that 
\begin{equation}
\label{eqn:grid-proof-0}
\sum_{i \in S_W} u_i(W) \geq \frac{1}{2} \cdot \kprelax{W, K}\,.
\end{equation}

Now, note that we have the following inequalities:
\begin{equation}
\label{eqn:grid-proof-1}
\kpcard{W^\star, K} \leq \kprelax{W^\star, K}\,,  
\end{equation}
We also have that 
\begin{equation}
\label{eqn:grid-proof-2}
\kprelax{W^\star, K} \leq (1+\epsilon') \cdot \kprelax{W, K}\,.
\end{equation}
To see why Equation \eqref{eqn:grid-proof-2} holds, one can first consider an auxiliary knapsack problem with utilities $u_i(W)$. As one increases knapsack capacity from $W$ to $W^\star$, the total utilities increase at most by $W^\star/W \geq 1+\epsilon$. Then, since we also have $u_i(W) \geq u_i(W^\star)$, Equation \eqref{eqn:grid-proof-2} thus holds.

Equations \eqref{eqn:grid-proof-0},\eqref{eqn:grid-proof-1} and \eqref{eqn:grid-proof-2} together imply that 
$$
\kpcard{W^\star, K} \leq \kprelax{W^\star, K} \leq (1+\epsilon') \cdot \kprelax{W, K} \leq (2+2\epsilon') \cdot \sum_{i \in S_W} u_i(W)  \,.
$$
This suggests that the grid-based enumeration method will yield a $1/(2+2\epsilon')$-approx. solution.

In terms of the runtime, for each interval $[w_i, n \cdot w_i)$, the grid-based enumeration method will need to consider $\mathcal{O}(\log(n)/(\log(1+\epsilon'))$ grid points. Solving the relaxed knapsack problem takes $\mathcal{O}(n\log(n))$. Since the above procedure is repeated $n$ times for each interval, the total runtime complexity is $\mathcal{O}\Big(n \cdot \log(n)/\log(1+\epsilon) \cdot n\log(n)\Big) = \mathcal{O}(n^2 \log(n)^2 / \log(1+\epsilon'))\,.$
$\blacksquare$

\subsection{Additional Runtime Comparisons with the 1/2-Approx. Algorithm}
\label{sec:comparison-enumeration}

Recall that in Section \ref{sec:synthetic_data}, the grid-based enumeration method is one of the benchmarks that we evaluated against our near-optimal algorithms (1/2-approx. algorithm and FPTAS). Our numerical experiments from Section \ref{sec:synthetic_data} indicate that the $1/2$-approx. algorithm excels in terms of efficiency. In particular, despite its simplicity, the grid-based enumeration method suffers from lengthy runtimes due to its lack of adaptivity. Unlike the 1/2-approx. algorithm, which dynamically adapts to problem instances, the enumeration method always enumerates over the same number of LPs, regardless of the problem structure.

In this section, we make a few additional remarks on the runtime differences between the grid-based enumeration method and the 1/2-approx. algorithm, reinforcing why the latter is a more practical and efficient choice. Our key observations are as follows:
\begin{enumerate}

\item \emph{Our 1/2-approx. algorithm excels in realistic instances with exploitable structure, as demonstrated in the MovieLens case study (Section \ref{sec:numerics}).} 
Specifically, when the platform optimizes for marketshare with item visibility as the main outcome in the MovieLens case study, the runtime of the $1/2$-approx. algorithm improves significantly due to the simplicity of instances with homogeneous revenues. Table \ref{tbl:comparison} shows that the time taken by the $1/2$-approx. algorithm by a single execution and in solving Problem \eqref{eq:problem:fair} both reduced by a fair amount compared to more complex instances with heterogeneous revenues presented in Table \ref{tbl:numerical_results_synthetic}. 
The runtime improvement is driven by the algorithm’s adaptability to simpler instances, allowing it to evaluate fewer partitions. As shown in Table \ref{tbl:numerical_results_synthetic}, the 1/2-approx. algorithm iterates over as few as 31–49 sub-intervals per execution, resulting in superior performance on the MovieLens data.

In contrast, the runtime gap between the $1/2$-approx. algorithm and the grid-based enumeration method widened even further in the MovieLens case study. 
Table \ref{tbl:comparison} shows that while the $1/2$-approx. algorithm manages to adapt to MovieLens data by drastically reducing the number of partitions it evaluates, the enumeration method remains static, solving 1600 LPs per execution that result in heavy  runtime. 

Many real-world problem instances, including the MovieLens case study, have exploitable structural properties. This is where the adaptivity of the 
1/2-approx. algorithm proves especially valuable.

\vspace{6mm}
\begin{table}[htbp]
\footnotesize
    \centering
    \caption[XXX]{
    \footnotesize{
    Comparison of our $1/2$-approx. algorithm and grid-based enumeration method (with $\epsilon' = 1/49$) on MovieLens data. 
    }
    }
    \label{tbl:comparison}
    \renewcommand{\arraystretch}{1.0} 
    \footnotesize
    \begin{tabular}{ c | c | cccccc } 
    \toprule
     & \multicolumn{1}{c|}{$\delta$} & 0 & 1 & 2 & 3 & 4 & 5 \\
    \midrule
   \multirow{2}{*}[0.21ex]{
   \makecell{Avg \# sub-intervals\\
   per instance}} 
   &  {$1/2$-approx} 
   &   49
   &   34
   &   31
   &   33
   &   33
   &   42
   \\
    & {enumeration} 
    & 1600
    & 1600
    & 1600
    & 1600
    & 1600
    & 1600
    \\
    \midrule
     \multirow{2}{*}[0.22ex]{\makecell{Time taken by a\\single execution (s)}} &  {$1/2$-approx} 
   &  {$2.27 \times 10^{-2}$}
   &  {$1.75 \times 10^{-2}$}
   &  {$1.78 \times 10^{-2}$}
   &  {$1.64 \times 10^{-2}$}
   &  {$1.55 \times 10^{-2}$} 
   &  {$1.69 \times 10^{-2}$}
    \\
    & {enumeration} 
    & 3.24
    & 3.37
    & 3.37
    & 3.11
    & 3.10
    & 2.78
    \\
    \midrule
     \multirow{2}{*}[0.2ex]{\makecell{Total time taken in\\solving Problem (\texttt{FAIR})(s)}}
     & 
    {\makecell{$1/2$-approx}}
    &  1.77 
    &  0.47 
    &  0.64 
    &  0.29 
    &  0.23 
    &  0.18
    \\
    & 
    {\makecell{enumeration}} 
    & 116
    & 111 
    & 118 
    & 52 
    & 52 
    & 20
    \\
    \bottomrule \addlinespace 
    \end{tabular}
\end{table}
\vspace{2mm}

\item 

\emph{The $1/2$-approx. algorithm demonstrates superior runtime efficiency for problems of practical sizes.} 
One might argue that the grid-based enumeration attains better runtime complexity in terms of $n$, given by Proposition \ref{prop:enumeration-perf}. However, we 
first note that real-world recommendation systems and assortment planning often narrow down items through lightweight pre-filtering stages based on criteria like keywords or price range. This allows fairness to be enforced within smaller, context-specific subsets, typically reducing the number of items $n$ to fewer than 50. 
We thus evaluated the runtime performance for $1/2$-approx. algorithm and the grid-based enumeration method under various number of items $n$, where $10 \leq n \leq 40$, and the $1/2$-approx. algorithm consistently achieves better runtime for the range of items of practical sizes, despite its runtime complexity, as indicated by Figure \ref{subfig:time-comparison-alg}.

\begin{figure}[htbp]
\centering
\includegraphics[width=0.5\textwidth]{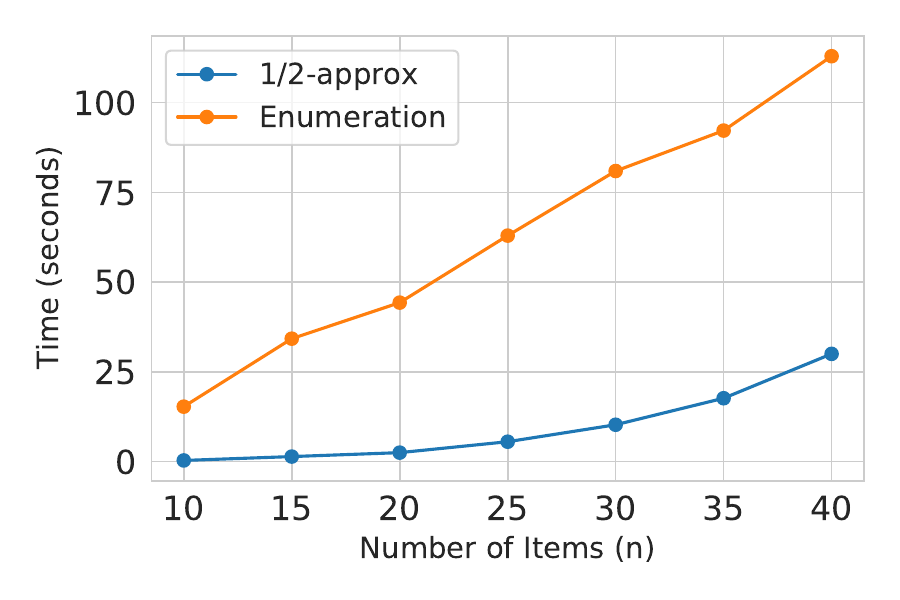}
\caption{\footnotesize 
Runtime comparison for solving Problem \eqref{eq:problem:fair} with 
$\delta = 0.8$. (Similar results are observed for other fairness parameters.) The problem instance is generated following the same approach as in Section \ref{sec:synthetic_data}.}
\label{subfig:time-comparison-alg}
\end{figure}

We also note that runtime complexity may be less critical in practical settings. For smaller platforms focused on specific item categories, assortment probabilities are often precomputed offline, eliminating the need for real-time computation. In such cases, actual runtime performance is more relevant than theoretical complexity.

\item 
\emph{Our algorithms can be parallelized for improved performance.} 
While parallelization has not been implemented in any of our approximation algorithms, both the $1/2$-approx. algorithm and the grid-based enumeration algorithm could benefit from parallel execution. For the $1/2$-approx. algorithm, a significant portion of the runtime is spent on the pre-partitioning procedure, rather than the adaptive partitioning steps. By parallelizing the execution of Algorithm \ref{alg:half-approx} across the well-behaving intervals identified during pre-partitioning, the algorithm's runtime could be further improved, while maintaining its adaptability to different problem instances. In this work, we do not include results from parallelized implementations, as they involve trade-offs between computational resources and runtime, and our primary focus is on establishing the strong theoretical guarantees and practicality of our methods. Nonetheless, parallelization could be beneficial for real-world deployments.
\end{enumerate}

\section{Proof of Proposition \ref{lem:n_2}}
Here, we restate Problem \eqref{eq:problem:fair} for reader's convenience.
\begin{align}
    \notag \textsc{fair}~=~\max_{p(S): S\subseteq [n]} & \sum_{S: |S|\le K} p(S)\cdot \rev(S)\\   \notag
    \text{s.t.} \qquad & ~~\sum_{S: i\in S} p(S) \cdot O_i(S) - \sum_{S: j\in S} p(S) \cdot O_j(S) \le \delta \quad\quad~~~~~ \forall~ i,j\in [n], i \neq j\\   \notag
    \quad\quad\quad ~~~~~~~~~& \sum_{S:|S|\leq K} p(S) \leq 1 \\
    & \quad p(S) \geq 0 \quad\quad\quad\quad\quad\quad\quad\quad\quad\quad\quad\quad\quad~~~~~~~~~~~~~~~~~~~ 
    S\subseteq [n], |S| \leq K\,, 
\end{align}
We start by showing that Problem \eqref{eq:problem:fair} admits an optimal basic feasible solution (for definition of basic feasible solution, see Definition 2.9 in \cite{bertsimas-LPbook} or Section~\ref{sec:LP}). Consider the polyhedron defined by the feasibility region of Problem \eqref{eq:problem:fair}. Note that this polyhedron is (i) nonempty, because $p(S) = 0$ for all $S \subseteq [n], |S| \leq K$ is a feasible solution for any $\delta \geq 0$; (ii) bounded, because $p(S) \in [0, 1]$ for all $S \subseteq [n], |S| \leq K$. Then, by Theorem 2.6 in \cite{bertsimas-LPbook}, there exists at least one extreme point in the polyhedron (for definition of extreme point, see Definition 2.6 in \cite{bertsimas-LPbook}). Since we have $\rev(S) = \sum_{i \in S} \frac{r_iw_i}{1+w(S)} \le \bar{r}$ for any $S$, this implies that $\sum_{S:|S| \leq K} p(S) \rev(S) \leq \bar{r}$; that is, the objective of Problem \eqref{eq:problem:fair} is also bounded. By Theorem 2.8 in \cite{bertsimas-LPbook}, since the polyhedron has at least one extreme point and Problem \eqref{eq:problem:fair} has a bounded optimal objective, there must exist one extreme point of the polyhedron that is optimal. By Theorem 2.3 in \cite{bertsimas-LPbook}, we have that an extreme point is equivalent to a basic feasible solution. Hence, we must have a basic feasible solution to Problem \eqref{eq:problem:fair} that is optimal, which we denote by $p^\star(.)$. 

Now, let $N = |\{ S \subseteq [n] : |S| \leq K\}|$ be the number of variables in Problem \eqref{eq:problem:fair}. There are $n(n-1) + 1 + N$ constraints in Problem \eqref{eq:problem:fair}, where $n(n-1)$ of them are in the first set of fairness constraints, and $N$ of them are the non-negativity constraints. By definition of a basic feasible solution, we must have that at least $N$ active (binding) constraints   at $p^\star(.)$. That implies that at most $n(n-1) + 1$ of the non-negativity constraints can be inactive, i.e. $p^\star(S) > 0$. We have thus showed that $\big|\{S : p^\star(S) > 0\}\big| \leq n(n-1) + 1$.  $\blacksquare$

\section{Proof of Theorem \ref{thm:np-hard}}
\indent 

\textbf{Revenue/marketshare-fair instance.}
    We first consider the revenue/marketshare-fair instance, where $\mathbf{\Theta} = (\mathbf{a}, 0, \mathbf{r}, \mathbf{w})$. By Proposition \ref{prop:staticMNL}, the \textsc{StaticMNL} algorithm introduced by \cite{rusmevichientong2010dynamic} solves Problem \eqref{eq:subdual} optimally and returns the optimal assortment $S^\star = \argmax_{S: |S|\le K}~\revmod(S, \mathbf{z})$. 
    The \textsc{StaticMNL} algorithm leverages the structural properties of the MNL model to generate a sequence of $\mathcal{O}(n^2)$ assortments that is guaranteed to include the optimal assortment. Hence, its running time is of order $\mathcal{O}(n^2)$.

\textbf{Visibility-fair and general instances.}
We now focus on visibility-fair and general instances, and establish the NP-completeness of Problem \eqref{eq:subdual}.
We first show that Problem 
\eqref{eq:subdual} is in NP. Then, we reduce an arbitrary instance of the partition problem, which is known to be NP-complete, to polynomially many visibility-fair instances of Problem 
\eqref{eq:subdual}.

To show that Problem 
\eqref{eq:subdual} is in NP, note that the decision version of Problem \eqref{eq:subdual}  is: given a set $S$ and some value $v$, we need to determine if $\revmod(S, \mathbf{z})$ is greater than $v$. We can compute $\textsc{rev-cost}(S,\mathbf{z})$ in polynomial time since $\mathbf{\Theta}$ and $\mathbf{z}$ are both known. This then allow us to tell if $\textsc{rev-cost}(S,\mathbf{z}) > v$ immediately. Hence, Problem \eqref{eq:subdual} is in NP.

We will now show the NP-completeness of Problem \eqref{eq:subdual} by considering an arbitrary instance of the \emph{partition problem}, and solving it using polynomially many \emph{visibility-fair} instances of Problem \eqref{eq:subdual}. The partition problem is known to be NP-complete \citep{hayes2002computing}, and is stated as follows: given a set of $n$ positive integers $\{\alpha_1, \alpha_2, \dots, \alpha_n\}$, we would like to determine if there exists a subset $S \subset [n]$ such that $\sum_{i \in S} \alpha_i = \frac{1}{2} \sum_{i \in [n]} \alpha_i := L$, and find the set $S$ if it exists. 

Before proceeding, let us first define an auxiliary optimization problem that we will work with throughout the proof. For some given vector $\mathbf{c}'$, we define Problem \eqref{eqn:auxiliary-max-problem} as:
\begin{equation}
\subdualaux(\mathbf{c}', K) = \max_{S:|S| \leq K} \frac{\sum_{i \in S} w_{i}}{1+\sum_{i \in S} w_{i}} -\sum_{i \in S} c'_i \,.
\tag{\subdualaux($\mathbf{c}', K$)}
\label{eqn:auxiliary-max-problem}
\end{equation}
We denote the objective function of Problem \eqref{eqn:auxiliary-max-problem} as 
$$
\revmodaux(S, \mathbf{c}') := \frac{\sum_{i \in S} w_{i}}{1+\sum_{i \in S} w_{i}} -\sum_{i \in S} c'_i \,.
$$
Our proof of Theorem~\ref{thm:np-hard} consists of two parts. In Part 1, we reduce an arbitrary instance of the partition problem to a particular instance of Problem \eqref{eqn:auxiliary-max-problem}. In Part 2, we show that this instance of Problem \eqref{eqn:auxiliary-max-problem} can be further reduced to a polynomial number of {visibility-fair} instances of Problem \eqref{eq:subdual}. 

\textbf{Part 1.} Let $\{\alpha_1, \alpha_2, \dots, \alpha_n\}$ be an instance of the partition problem. We first reduce it to the auxiliary optimization problem by creating an instance of Problem \eqref{eqn:auxiliary-max-problem} as follows. For each $i \in [n]$, let $w_i = \alpha_i/L$ and $c'_i = \alpha_i/4L$. 
Let $K = n/2$. Additionally, let $A(S) := \sum_{i \in S} \alpha_i$. We can rewrite the objective function of Problem \eqref{eqn:auxiliary-max-problem} to be a function dependent on $A(S)$:
$$
\frac{\sum_{i \in S} w_{i}}{1+\sum_{i \in S} w_{i}} -\sum_{i \in S} c'_i = \frac{A(S)}{L+A(S)} - \frac{A(S)}{4L} = \frac{3LA(S) - (A(S))^2}{4L^2 + 4LA(S)} : = f(A(S)), 
$$
where we let $f(x) = \dfrac{3Lx - x^2}{4L^2 + 4Lx}$ for $x \geq 0$. Differentiating $f(x)$ with respect to $x$ gives
$$
f'(x) = \frac{-x^2 - 2Lx + 3L^2}{4L(L+x)^2}\,.
$$
Taking $f'(x) = 0$ gives $x = L$. Since $f'(x) < 0$ for $x > L$ and $f'(x) > 0$ for $0 \leq x < L$, we have that for $x \geq 0$, $f(x)$ is uniquely maximized at $x = L$ and $\max_{x \geq 0} f(x) = f(L) = \frac{1}{4}$.

We now show that we can determine whether a partition $S$ exists and, if it exists, find $S$ 
if and only if 
we can find the optimal set $S^\star$ for instance of Problem \eqref{eqn:auxiliary-max-problem} and it attains optimal objective value of $1/4$. 
Suppose a partition $S^\star$ exists\footnote{Note that we can assume without loss of generality that $|S^\star| \leq n/2$. This is because if $S^\star$ satisfies $\sum_{i\in S} a_i = \frac{1}{2}\sum_{i \in [n]} a_i$, we must also have $\sum_{i\in [n]\setminus S^\star} a_i = \frac{1}{2}\sum_{i \in [n]} a_i$. That is, $S^\star$ and $[n]\setminus S^\star$ are both valid partitions.} such that $A(S^\star) = \sum_{i \in S^\star} a_i = L$, we have 
$$
\frac{1}{4} \geq \subdualaux(\mathbf{c}', K) = \max_{{S \subset [n]} \atop {|S| \leq K}} f(A(S)) \geq f(A(S^\star)) = f(L) = \frac{1}{4}.
$$
Hence, the inequalities above should be equalities. We have that $S^\star$ is the optimal set for Problem \eqref{eqn:auxiliary-max-problem} and the optimal objective value is $f(A(S^\star)) = 1/4$.
Reversely, if we find $S^\star = \argmax_{{S \subset [n]} \atop {|S| \leq K}} f(A(S))$ and $\max_{{S \subset [n]} \atop {|S| \leq K}} f(A(S)) = f(L) = \frac{1}{4}$, we must have that $A(S^\star) = \sum_{i \in S^\star} \alpha_i = L$, and $S^\star$ is thus a solution to the partition problem. The discussion above shows that we can reduce the partition problem to the instance of Problem \eqref{eqn:auxiliary-max-problem}. 

\textbf{Part 2.} 
Now, it suffices to show that the instance of Problem \eqref{eqn:auxiliary-max-problem} defined above can be solved using polynomially many instances of Problem \eqref{eq:subdual}. In particular, here we consider visibility-fair instances of Problem \eqref{eq:subdual} where $a_i = 0$ and $b_i = 1$ for all $i \in [n]$. We would also let all items have unit revenue, i.e., $r_i = 1$ for all $i \in [n]$. Such visibility-fair instances of Problem \eqref{eq:subdual} can be simplified into following form:
\begin{align}
\subdual(\mathbf{z}, K) 
= \max_{S: |S|\le K}~ \frac{\sum_{i \in S}  w_{i}}{1+\sum_{i \in S} w_{i}} -\sum_{i \in S} c_i(\mathbf{z}) = \max_{S: |S|\le K
}~\revmod(S, \mathbf{z}) \,,
\end{align}
where post-fairness revenue of item $i$ is $\r_i(\mathbf{z}) = 1$, post-fairness cost of item $i$ is $\c_i(\mathbf{z}) = c_i(\mathbf{z})$, and 
cost $c_i(\mathbf{z})$ is defined in Section \ref{sec:dual}. 
Note that Problem \eqref{eqn:auxiliary-max-problem} only differs from such visibility-fair instance of Problem \eqref{eq:subdual} in the second term of the objective function, where $\mathbf{c}'$ can be any vector and does not depend on $\mathbf{z}$.

Without loss of generality, assume that $\alpha_1 \geq \alpha_2 \geq \dots \geq \alpha_n$.  Before proceeding, let us first make the following definitions.
\begin{enumerate}[(1)]
    \item For every $k \in [n]$, let $(\subdualaux^{(k)})$ denote an instance of Problem \eqref{eqn:auxiliary-max-problem} with $k$ items, where $\mathbf{w} = (w_1, \dots, w_k) = (\alpha_1/L, \dots, \alpha_k/L)$, $\mathbf{c}' = (c'_1, \dots, c'_k) = (\alpha_1/4L, \dots, \alpha_k/L)$ and $K = n/2$. The instance of Problem \eqref{eqn:auxiliary-max-problem} that we considered in Part 1 of the proof is $(\subdualaux^{(n)})$.
    \item For every $k \in [n]$, let $(\subdual^{(k)})$ denote a visibility-fair instance of Problem \eqref{eq:subdual} with $k$ items, where $\mathbf{a} = \mathbf{0}, \mathbf{b} = \mathbf{1} = (1, \dots, 1), \mathbf{r} = \mathbf{1} = (1, \dots, 1), \mathbf{w} = (w_1, \dots, w_k) = (\alpha_1/L, \dots, \alpha_k/L)$, 
    and $K = n/2$. 
 We choose $\mathbf{z} = \mathbf{z}^{(k)}$ such that $c_i(\mathbf{z}^{(k)}) = c'_i = \alpha_i/4L$ for $i \in [k-1]$. Note that such $\mathbf{z}^{(k)}$ always exists, since the linear system 
    $$
    c_i(\mathbf{z}^{(k)}) 
    = \sum_{j \neq i}(z^{(k)}_{ij} - z^{(k)}_{ji})  
    = \frac{\alpha_i}{4L}, \quad  \forall i \in [k-1]
    $$
    can be written as 
    $$
    \sum_{j \neq i} \mathbbm{1}_{\{i < j\}} x_{ij} - \mathbbm{1}_{\{i > j\}} x_{ji} = \frac{\alpha_i}{4L},\quad  \forall i \in [k-1]
    $$
where $x_{ij} = z^{(k)}_{ij} - z^{(k)}_{ji}$ for all $i < j$. The rows of this $(k-1)\times(k-1)(k-2)$ linear system are linearly independent, and hence there exists a feasible solution. However, via row operations, one can check that the values of $c_1(\mathbf{z}^{(k)}), \dots, c_{k-1}(\mathbf{z}^{(k)})$ uniquely determine the value of $c_k(\mathbf{z}^{(k)})$ to be 
$$
c_k(\mathbf{z}^{(k)}) = - \frac{\sum_{i < k} \alpha_i}{4L }.
$$
\end{enumerate}
Note that the only difference between the instance $(\subdual^{(k)})$ and $(\subdualaux^{(k)})$ is in the difference between $c_k(\mathbf{z}^{(k)})$ and $c_k'$. 
Additionally, we have $c_k(\mathbf{z}^{(k)}) = -(\sum_{i < k} \alpha_i)/(4L ) \leq \alpha_k/4L = c_k'$ because $\alpha_i$'s are all positive integers.

Having defined the two types of instances, our proof of Part 2 proceeds in three steps: (i) first, we establish an important relationship between the two functions $\revmod$ and $\revmodaux$; (ii) next, we show that for any $k \in [n]$, we can solve the instance $(\subdualaux^{(k)})$ if we can solve $(\subdual^{(k)})$ and $(\subdualaux^{(k-1)})$; (iii) based on that, we use a recursive approach to show that the instance $(\subdualaux^{(n)})$ considered in Part 1 of the proof can be solved as long as we can solve $(\subdual^{(n)}), \dots, (\subdual^{(2)})$, which are $n-1$ instances of Problem \eqref{eq:subdual}.

\textbf{Step 1.}
Fix $k \in [n]$. Let us start by considering the two instances $(\subdual^{(k)})$ and $(\subdualaux^{(k)})$.
Recall that the objective function of $(\subdual^{(k)})$ is:
$$
\revmod(S, \mathbf{z}^{(k)}) = \frac{\sum_{i \in S} w_{i}}{1+\sum_{i \in S} w_{i}} -\sum_{i \in S} c_i(\mathbf{z}^{(k)}) 
$$ 
and the objective function of $(\subdualaux^{(k)})$ is 
$$
\revmodaux(S, \mathbf{c}) = \frac{\sum_{i \in S}  w_{i}}{1+\sum_{i \in S} w_{i}} -\sum_{i \in S} c'_i \,.
$$ 
Since $c_i(\mathbf{z}^{(k)}) = c'_i$ for all $i \in [k-1]$, we must have the following:
\begin{itemize}
    \item For any set $S$ such that $k \notin S$, we must have $\revmod(S, \mathbf{z}^{(k)}) = \revmodaux(S, \mathbf{c}')$. 
    \item For any set $S$ such that $k \in S$, we must have 
    \begin{equation*}
    \label{eqn:g-relations}
    \revmod(S, \mathbf{z}^{(k)}) = \revmodaux(S, \mathbf{c}') + \frac{\alpha_k}{4L} + \frac{\sum_{i < k} \alpha_i}{4L} \geq \revmodaux(S, \mathbf{c}').
    \end{equation*}
\end{itemize}

\textbf{Step 2.}
Now, suppose that we have solved $(\subdual^{(k)})$ with $S^\star_k$ as its optimal solution, and we have also solved $(\subdualaux^{(k-1)})$ with $S'_{k-1}$ as its optimal solution. We claim that the optimal solution to $(\subdualaux^{(k)})$, $S'_{k}$, would then be the best set among $S^\star_k$ and $S'_{k-1}$. 

To see why, we divide our arguments into the following cases:
\begin{itemize}[leftmargin =*]
    \item \textbf{Case 1: $k \notin S^\star_k$.} 
    In this case, we claim that $S^\star_k$ is an optimal solution to $(\subdualaux^{(k)})$. Let $S' \in [k], |S'| \leq K$ be any set that is different from $S^\star_k$. There are again two possibilities: 
    \begin{itemize}
        \item If $k \notin S'$, we have $$\revmodaux(S', \mathbf{c}') = \revmod(S', \mathbf{z}^{(k)}) \leq \revmod(S^\star_k, \mathbf{z}^{(k)}) = \revmodaux(S^\star_k, \mathbf{c}')\,,$$
        where the two equalities follow from Step 1.
        \item If $k \in S'$, 
        we have $$\revmodaux(S', \mathbf{c}') \leq \revmod(S', \mathbf{z}^{(k)}) \leq \revmod(S^\star_k, \mathbf{z}^{(k)}) = \revmodaux(S^\star_k, \mathbf{c}')\,,$$
        where the first inequality and the last equality follow from Step 1. 
    \end{itemize}
    Since we have $\revmodaux(S', \mathbf{c}') \leq \revmodaux(S^\star_k
    , \mathbf{c})$ for any set $S'$, 
    $S^\star_k$ must be an optimal solution to $(\subdualaux^{(k)})$.
    \item \textbf{Case 2}: $k \in S^\star_k$.
    In this case, we claim that the optimal solution to $(\subdualaux^{(k)})$, which we denote by $S'_k$, is either $S^\star_k$ or $S'_{k-1}$. Let us again consider the two possibilities:
    \begin{itemize}
        \item If $k \in S'_k$, we must have 
        $$
        \begin{aligned}
 \revmodaux(S'_k, \mathbf{c}') & = \revmod(S'_k, \mathbf{z}^{(k)}) - \big(\frac{\alpha_k}{4L} + \frac{\sum_{i < k} \alpha_i}{4L}\big) \\
        & \leq \revmod(S^\star_k, \mathbf{z}^{(k)}) - \big(\frac{\alpha_k}{4L} + \frac{\sum_{i < k} \alpha_i}{4L}\big) \\
        & = \revmodaux(S^\star_k, \mathbf{c}').
        \end{aligned}
        $$
    \end{itemize}
    where the equalities follow from Step 1. Hence, in this case, $S^\star_k$ is an optimal solution to $(\subdualaux^{(k)})$.
    \item If $k \notin S'_k$. In this case, an optimal solution to $(\subdualaux^{(k)})$ is simply the optimal solution to $(\subdualaux^{(k-1)})$; that is, $S'_{k-1}$. 
\end{itemize}
To summarize, our case discussion above shows that to solve $(\subdualaux^{(k)})$, it suffices to solve $(\subdual^{(k)})$ and $(\subdualaux^{(k-1)})$.

\textbf{Step 3.} Finally, we consider the instance $(\subdualaux^{(n)})$ from Part 1 of the proof. Since we have shown in Step 2 that for any $k \in [n]$, we can solve $(\subdualaux^{(k)})$ as long as we can solve $(\subdual^{(k)})$ and $(\subdualaux^{(k-1)})$, we can solve $(\subdualaux^{(n)})$ in a recursive manner. This requires us to solve at most $n-1$ visibility-fair instances of Problem \eqref{eq:subdual}, i.e., $(\subdual^{(n)}), (\subdual^{(n-1)}), \dots, (\subdual^{(2)})$ and one instance of Problem \eqref{eqn:auxiliary-max-problem}, i.e., 
$(\subdualaux^{(1)})$, for which the solution is trivial to obtain. 

In summary, in Part 2 we have shown that $(\subdualaux^{(n)})$ can be solved using a polynomial number of visibility-fair instances of Problem \eqref{eq:subdual}. Together, 
Part 1 and 2 of the proof show that an arbitrary instance of the partition problem can be solved using polynomially many visibility-fair instances of Problem \eqref{eq:subdual}. This establishes the NP-completeness of Problem \eqref{eq:subdual}. In particular, since we only consider visibility-fair instances of Problem \eqref{eq:subdual} with uniform revenues (i.e., $r_i = 1$ for all $i \in [n]$), our proof above in fact shows that even for visibility-fair instances with uniform revenues, Problem \eqref{eq:subdual} is also NP-complete. $\blacksquare$

\section{Proof of Theorem \ref{thm:approx}}
Our proof of Theorem \ref{thm:approx} consists of two segments. In Segment 1, we show that our fair Ellipsoid-based algorithm gives a $\beta$-approx. solution for Problem \eqref{eq:problem:fair}, and the returned solution requires us to randomize over a polynomial number of sets. In Segment 2, we show that the runtime of our fair Ellipsoid-based algorithm is polynomial in the size of our input. 

\textbf{Segment 1.} 
Before proceeding, recall that in Section~\ref{sec:elipsoid}, we define $\mathcal{V} = \{S : |S| \leq K\}$ as the collection of sets that violate the dual fairness constraint.
The first segment of our proof consists of three parts:
(i) We first show that if we apply the Ellipsoid method equipped with a $\beta$-approx. separation oracle to the dual problem \eqref{eq:problem:dual}, we will end up with a dual objective $r'$ that is not too far away from the optimal primal objective $\ref{eq:problem:fair}$. 
That is, it satisfies $r' \geq \beta \cdot \ref{eq:problem:fair}$.
(ii) We then construct a auxiliary dual problem \eqref{eq:problem:dual-aux}, in which we only keep the dual fairness constraints for $S \in \mathcal{V}$; and the corresponding auxiliary primal problem \eqref{eq:problem:fair-aux}, in which we enforce $p(S) = 0$ for all $S \notin \mathcal{V}$. The auxiliary primal problem is the one we solve in Step 2 of our algorithm. We will show that the optimal objectives of both Problem \eqref{eq:problem:dual-aux} and \eqref{eq:problem:fair-aux} exceeds $r'$, and hence exceeds $\beta \cdot \ref{eq:problem:fair}$. 
(iii) Finally, we show that it is possible to randomize over only $\mathcal{O}(n^2)$ sets.

\textbf{Part 1.} As described in Step 1 of the fair Ellipsoid-based algorithm, suppose that we apply the Ellipsoid method to the dual problem \eqref{eq:problem:dual}, and use algorithm $\mathcal{A}$ as the $\beta$-approx. separation oracle. Let $(\mathbf{z}', \rho')$ be the solution we get at the end of the Ellipsoid method, and let $r' = \rho' + \sum_{i=1}^{n} \sum_{j=1, j \neq i}^n \delta \cdot z'_{i j}$ denote its dual objective. We first note that the solution $(\mathbf{z}', \rho')$ is not necessarily feasible for Problem \eqref{eq:problem:dual}. This is because when we examine the feasibility of this solution, our separation oracle first solves Problem $(\subdual(\mathbf{z}', K))$, and obtained a set $S_\mathcal{A}$ such that $\revmod(S_\mathcal{A}, \mathbf{z}') \geq \beta \cdot \subdual(\mathbf{z}', K)$. The Ellipsoid method has declared solution $(\mathbf{z}', \rho')$ feasible because $\revmod(S_\mathcal{A}, \mathbf{z}') \leq \rho'$. However, it is still possible that
$\subdual(\mathbf{z}', K) =\max_{|S|\le K} \revmod(S, \mathbf{z}') > \rho'$. We will first show that the dual objective that we obtain here would still be close to the optimal primal objective 
$\ref{eq:problem:fair}$.

Suppose that we solve the dual problem \eqref{eq:problem:dual} by additionally fixing $\mathbf{z} = \mathbf{z}'$. This problem has a trivial solution because $\rho$ is the only decision variable, and it must be chosen such that the inequality in Eq. \eqref{eq:inner_dual} is tight. That is, we set $\rho = 
\subdual(\mathbf{z}', K)$ and the optimal objective is 
$$
r_{\mathbf{z'}} := \subdual(\mathbf{z}', K) + \sum_{i=1}^{n} \sum_{j=1, j \neq i}^n \delta \cdot z'_{i j}
$$
Clearly, we must have $\ref{eq:problem:dual} \leq r_{\mathbf{z'}}$ since fixing $\mathbf{z} = \mathbf{z'}$ is equivalent to introducing more constraints into Problem \eqref{eq:problem:dual}. Note that we also have
$$
r' = \rho' + \sum_{i=1}^{n}\sum_{j =1, j \neq i}^n \delta \cdot  z'_{i j} 
\geq \beta \cdot \subdual(\mathbf{z}', K) + \beta \cdot \sum_{i=1}^{n} \sum_{j =1, j \neq i}^n \delta \cdot  z'_{i j} 
= \beta \cdot r_{\mathbf{z'}}\,.
$$
where the inequality follows from $\rho' \geq \revmod(S_\mathcal{A}, \mathbf{z}') \geq \beta \cdot \subdual(\mathbf{z}', K)$ and $z'_{ij} \geq 0$.
Together, the two inequalities above give 
\begin{equation}
\label{eqn:key-ineq}
r' \geq \beta \cdot r_{\mathbf{z'}}\ge  \beta \cdot \ref{eq:problem:dual} = \beta \cdot \ref{eq:problem:fair}\,,
\end{equation}
where the last equality follows from strong duality.\footnote{Since both Problem \eqref{eq:problem:fair} and Problem \eqref{eq:problem:dual} are feasible and bounded, they both admit an optimal solution. This allows us to invoke the strong duality theorem (see Theorem 4.4 in \cite{bertsimas-LPbook}), which suggests that $\ref{eq:problem:fair} = \ref{eq:problem:dual}$.} The inequality in Eq. \eqref{eqn:key-ineq} is a key inequality that we will continue to work with in Part 2. 

\textbf{Part 2.} In Part 2, we first create an auxiliary version of the dual problem, defined as follows:
\begin{align}
\begin{aligned}
\textsc{fair-dual-aux}~=~\min_{\rho\ge 0, \mathbf{z}\ge \mathbf{0}} & \, \rho+\sum_{i=1}^{n} \sum_{j=1, j \neq i}^n \delta \cdot  z_{i j} \\
\text { s.t. } & \sum_{i \in S} O_i(S)\cdot \left(\sum_{j=1, j \neq i}^{n} (z_{i j}- z_{ji})\right)+\rho \geq \rev(S), \quad & \forall S \in \mathcal{V}
\end{aligned}
\tag{\textsc{fair-dual-aux}} \label{eq:problem:dual-aux}
\end{align}
Note that the auxiliary dual problem is different from Problem \eqref{eq:problem:dual}, because in the  auxiliary dual problem, we only enforces the constraints on sets $S \in \mathcal{V}$, where $\mathcal{V}$ is the collection sets for which the dual fairness constraint has been violated when we solve Problem \eqref{eq:problem:dual}. 
Since the unviolated constraints did not impact any of the iteration in the Ellipsoid method when we solve Problem \eqref{eq:problem:dual}, if we now apply the Ellipsoid method with the $\beta$-approx. separation oracle to solve Problem \eqref{eq:problem:dual-aux}, the solution we obtain would still be $(\mathbf{z}', \rho')$, which gives objective value $r'$.

Similar to our arguments in Part 1, we note that when we apply the Ellipsoid method using our approximate separation oracle, we essentially increases the feasibility region of the linear program. Hence, the solution $(\mathbf{z}', \rho')$ we found at the end of the Ellipsoid method might not be feasible and the objective we obtain is less than or equal to the actual optimal objective. That is, we must have $r' \leq \ref{eq:problem:dual-aux}$.
This, along with the inequality we established in Eq. \eqref{eqn:key-ineq}, gives
$$
\ref{eq:problem:dual-aux} \geq r' \geq \beta \cdot \ref{eq:problem:fair}\,.
$$

Consider the primal counterpart to the auxiliary dual problem \eqref{eq:problem:dual-aux} defined below:
\begin{align}
    \notag \textsc{fair-aux}~=~\max_{p(S)\ge 0: S \in \mathcal{V}} & \sum_{S: S \in \mathcal{V}} p(S)\cdot \rev(S)\\   \notag
    \text{s.t.} \text{s.t.} \qquad & ~~\sum_{S: i\in S} p(S) \cdot O_i(S) - \sum_{S: j\in S} p(S) \cdot O_j(S) \le \delta \quad\quad~~~~~ \forall~ i,j\in [n], i \neq j\\   \notag
    \quad\quad\quad\quad\quad 
    & \sum_{S:|S|\leq K} p(S) \leq 1 
    \label{eq:problem:fair-aux} \tag{\textsc{fair-aux}}
\end{align}
Essentially, this differs from Problem \eqref{eq:problem:fair} in that here, we set $p(S) = 0$ for all $S \notin \mathcal{V}$. 
This is the primal problem that we solve in Step 2 of the fair Ellipsoid-based algorithm. Let $\widehat{p}(S)$ denote the optimal solution to Problem \eqref{eq:problem:fair-aux}. By the strong duality theorem \footnote{Here, both Problem \eqref{eq:problem:fair-aux} and Problem \eqref{eq:problem:dual-aux} are again feasible and bounded, hence they both admit an optimal solution. For Problem \eqref{eq:problem:fair-aux}, $p(S) = 0$ for all $S \in \mathcal{V}$ is a feasible solution; its objective is upper bounded by $\bar{r}$. For Problem \eqref{eq:problem:dual-aux}, $\rho = \bar{r}, \mathbf{z} = \mathbf{0}$ is a feasible solution; its objective is lower bounded by $0$. This again allows us to invoke the strong duality theorem (see Theorem 4.4 in \cite{bertsimas-LPbook}), which suggests that $\ref{eq:problem:fair-aux} = \ref{eq:problem:dual-aux}$.},  we have $\ref{eq:problem:fair-aux} = \ref{eq:problem:dual-aux} \geq \beta \cdot \ref{eq:problem:fair}$. We have thereby showed that our algorithm returns a solution $\widehat{p}(S)$ that is an $\beta$-approx. feasible solution to Problem \eqref{eq:problem:fair}. 

\textbf{Part 3.} Since we have $\widehat{p}(S) = 0$ for all $S \notin \mathcal{V}$, 
the number of sets that we need to randomize over, i.e. $|\{ S : \widehat{p}(S) > 0\}|$, is bounded by $|\mathcal{V}|$. Recall that in the dual problem \eqref{eq:problem:dual}, we have $\mathcal{O}(n^2)$ variables. By \cite{bertsimas-LPbook}, given a separation oracle, the Ellipsoid method would solve \eqref{eq:problem:dual} in at most $\mathcal{O}(n^{12}\log(n))$ iterations. Within each iteration, one violated constraint is identified. 
Hence, we must have that $|\mathcal{V}|$, i.e., the number of sets for which the dual fairness constraints in \eqref{eq:problem:dual} is violated, is polynomial in $n$. 

That is, after solving our dual problem, we will end up with a polynomial number of sets \( |\{ S : \widehat{p}(S) > 0 \}| \) to consider for the primal problem. We can then solve the primal problem using an algorithm that yields an optimal basic feasible solution. To ensure polynomial runtime for solving the primal problem, we can use a polynomial-time algorithm to solve it and then convert the optimal solution to an optimal basic feasible solution, which can again be done in polynomial time (see \cite{megiddo1991finding}). By the same reasoning as in our proof for Proposition \ref{lem:n_2}, the optimal basic solution randomizes over at most \( \mathcal{O}(n^2) \) sets.

\textbf{Segment 2.} In the second segment of the proof, we remark on the runtime of our fair Ellipsoid-based algorithm. In Step 1 of the algorithm, the Ellipsoid method equipped with a polynomial-time separating oracle runs in polynomial time. In Step 2 of the algorithm, solving a linear program with polynomial number of variables also takes polynomial time. Hence, given a polynomial-time separation oracle, the total time taken by the fair Ellipsoid-based algorithm is polynomial in the size of our input. $\blacksquare$

\section{Proof of Theorem \ref{thm:kp-su-dual}}
Let $S^\star\in\arg\max_{S:|S|\leq K} \revmod(S, \mathbf{z})$ and let $W^\star = \sum_{i \in S^\star} w_i$. We must have
\begin{equation}
\label{eqn:one-direction}
\subdual(\mathbf{z}, K) = \revmod(S^\star, \mathbf{z}) \leq \kpcard{W^\star, K} \leq \max_{W\ge 0}~\kpcard{W, K}\,.
\end{equation}
The first inequality follows from the fact that (i) $S^\star$ is a feasible solution to Problem $\kpcard{W^\star, K}$ since $|S^\star|\le K$ and $\sum_{i \in S^\star} w_i = W^{\star}$ by definition, and (ii)
$$
\revmod(S^\star, \mathbf{z}) 
= \frac{\sum_{i \in S^\star} \r_iw_i}{1+\sum_{i \in S^\star} w_i} - \sum_{i \in S^\star} \c_i 
= \sum_{i \in S^\star} \left(\frac{\r_iw_i}{1+W^\star} - \c_i \right),
$$
where the right-hand side (i.e., $\sum_{i \in S^\star} (\frac{\r_iw_i}{1+W^\star} - \c_i )$) is the objective value of the knapsack problem $(\kpcard{W^{\star}, K})$ at set $S^{\star}$, which is clearly upper bounded by its optimal objective value $\kpcard{W^{\star}, K}$.

Conversely, let $\max_{W\ge 0} \kpcard{W, K}$ be obtained at capacity $\widetilde W$ and with optimal set $\widetilde S$, i.e., $\widetilde S$ is the optimal solution to Problem $\kpcard{\widetilde W, K}$. We then have
\begin{equation}
\label{eqn:reverse-direction}
\begin{aligned}
\max_{W\ge 0} \kpcard{W, K}
= \kpcard{\widetilde W, K} 
& = \sum_{i \in \widetilde S } \left[\frac{\r_iw_i}{1+\widetilde W} - \c_i\right] \\
& \leq \sum_{i \in \widetilde S} \frac{\r_iw_i}{1+\sum_{i \in \widetilde S} w_i} - \sum_{i \in \widetilde S} \c_i = \revmod(\widetilde S, \mathbf{z}) \leq \max_{|S| \leq K} \revmod(S, \mathbf{z})\,,
\end{aligned}
\end{equation}
where the first inequality follows from the feasibility of $\widetilde S$ for Problem $\kpcard{\widetilde W, K}$, which implies that $\sum_{i \in \widetilde S} w_i \leq \widetilde W$, and the second inequality follows from $|\widetilde S| \leq K$.

Overall, Eqs. \eqref{eqn:one-direction} and \eqref{eqn:reverse-direction} together imply that $
\subdual(\mathbf{z}, K) 
= \max_{S:|S| \leq K} \revmod(S, \mathbf{z}) = \max_{W\ge 0} \kpcard{W, K},  
$
as desired. $\blacksquare$

\section{Proof of Lemma \ref{lemma:well-behaving-subinterval}}
{Without loss of generality, we assume the weights of the items satisfy $w_1 \leq w_2 \leq \dots  \le w_n$.} We pre-partition the interval $[0, \infty)$ into sub-intervals $I$, each being the largest well-behaving sub-interval attainable. To identify these $I$'s, we seek the values of $W$ where conditions in Definition \ref{def:well-behaving} cease to hold. 

{For condition (i), we only need to partition the interval $[0, \infty)$ at the values $w_1, w_2, \dots, w_n$. This ensures that within each sub-interval, the set of items that can fit into the knapsacks remains unchanged. Consequently, we need to perform the partition at $\mathcal{O}(n)$ values of $W$ for condition (i) to hold.}

For condition (ii), the sign change in item $i$'s utilities occurs at $u_i(W) = \frac{\r_i w_i}{1+W} - \c_i = 0$, yielding $\mathcal{O}(n)$ such $W$ values at which we need to perform the partition. 

Condition (iii) pertains to utility ordering shifts, happening at $\mathcal{O}(n^2)$ points where $\frac{\r_i w_i}{1+W} - \c_i = \frac{\r_j w_j}{1+W} - \c_j$ for some $i \neq j$. Similarly, condition (iv) involves changes in utility-to-weight ratios, occurring at $\mathcal{O}(n^2)$ $W$ values where $\frac{\r_i}{1+W} - \frac{\c_i}{w_i} = \frac{\r_j}{1+W} - \frac{\c_j}{w_j}$ for some $i \neq j$. 

Lastly, for condition (v), which deals with marginal utility changes upon replacing items, $\mathcal{O}(n^3)$ $W$ values are relevant, marked by $\frac{u_j(W) - u_i(W)}{w_j - w_i} = \frac{u_k(W) - u_i(W)}{w_k - w_i}$.

Overall, we can divide $[0, \infty)$ at at most $\mathcal{O}(n^3)$ values of $W$, which yields at most $\mathcal{O}(n^3)$ well-behaving sub-intervals. $\blacksquare$

\section{Proof of Theorem \ref{thm:half-approx-ratio}}
\label{sec:proof-thm-half-approx-ratio}
The proof consists of two segments. In the first segment, let $\mathcal{C}_I$ be the collection of assortments accumulated by Step 4 of Algorithm \ref{alg:half-approx} and let $S_{I} = \argmax_{S \in \mathcal{C}_I} \revmod(S, \mathbf{z})$ be the returned assortment. We show that for any $W \in I$, there exists a set $S_W \in \mathcal{C}_I$ such that  
\[
\revmod(S_W, \mathbf{z})\ge \frac{1}{2} \kpcard{W,K}\,,
\]
which then implies
\[
\revmod(S_I, \mathbf{z})\ge \frac{1}{2} \max_{W\in I}~ \kpcard{W,K}\,.
\]
In the second segment, we bound the runtime of the algorithm. 

{Recall that with a slight abuse of notation, the items in $[n]$, which are fed into Algorithm \ref{alg:half-approx}, refer to the eligible items of $I$. That is, they can both fit into the knapsack $(i.e., w_i \leq W)$ and have positive utilities $(i.e., u_i(W) > 0)$ for all $W \in I = [W_{\min}, W_{\max})$. 
If, for a given well-behaving interval $I$, the set of eligible items is empty, Algorithm \ref{alg:half-approx} will directly return the trivial solution $\emptyset$, as noted in Footnote \ref{footnote:1/2}.}

\textbf{Segment 1.} We start by considering the LP relaxation of  the original knapsack problem, i.e., Problem \eqref{eqn:relaxed-knapsack}, and 
making the following claim:
\begin{claim}
\label{claim:main}
For any $W\in I$, in the collection of assortments $\mathcal{C}_I$ returned by  Algorithm~\ref{alg:half-approx}, there exists an assortment $S_{W}$ such that $S_{W}$ is an integer $1/2$-approx. feasible solution for Problem $\kprelax{W, K}$.
\end{claim}
Given that Claim~\ref{claim:main} holds, we then have
\begin{equation}
\label{eqn:ms-cost-s-star}
\revmod(S_{W}, \mathbf{z}) = 
\sum_{i \in S_{W}} \big(\frac{\r_i w_i}{1+\sum_{i \in S_{W}}w_i} - \c_i \big) \geq \sum_{i \in S_{W}} \big(\frac{\r_i w_i}{1+W} - \c_i \big) \geq \frac{1}{2} \kprelax{W, K} \geq \frac{1}{2} \kpcard{W, K},
\end{equation}
where the first inequality follows from the feasibility of $S_{W}$ (i.e., $\sum_{i \in S_{W}} w_i \leq W$); the second inequality follows from Claim~\ref{claim:main}; and the third inequality follows from $\kprelax{W, K}$ being an upper bound of $\kpcard{W,K}$. 
Let $W^\star = \argmax_{W\in I} \kpcard{W, K}$. This then further gives 
\begin{equation}
\label{eqn:inequality-half-approx}
\revmod(S_{I}, \mathbf{z}) \geq \revmod(S_{W^\star}, \mathbf{z}) 
\geq \frac{1}{2} \kpcard{W^\star, K}
= \frac{1}{2} \max_{W\in I}~ \kpcard{W,K}\,,
\end{equation}
where the first inequality follows from $S_I = \arg \max_{S \in \mathcal{C}_I} \revmod(S, \mathbf{z})$ (Step 4 of Algorithm~\ref{alg:half-approx}) and that $S_{W^\star} \in \mathcal{C}_I$; and the second inequality follows from Eq. \eqref{eqn:ms-cost-s-star}.
This thus proves the statement of Theorem~\ref{thm:half-approx-ratio}.

In the following, we provide a proof for Claim~\ref{claim:main}. 

\textit{\textbf{Proof of Claim~\ref{claim:main}.}} Our proof consists of five parts. In Part 1, we first consider the capacity level $W_{\text{binding}}$, which is the maximum capacity level under which the capacity constraint is binding for the relaxed knapsack problems on $I$. We show that it suffices to show Claim \ref{claim:main} for $W < W_{\text{binding}}$. This then allows us to justify the stopping rules in Step~\ref{step:2b} and Step~\ref{step:3b_iii} of the algorithm; see Parts 2 and 3 of the proof. Recall that in Algorithm \ref{alg:half-approx}, we consider the two intervals $I_\text{low} = I \cap [0, \th)$ and $I_\text{high}= I \cap [\th, \infty)$ separately. In Part 2, we show that when we hit the stopping rule in Step~\ref{step:2b}, we have reached the upper bound $W_\text{binding}$ and there is no need to consider the interval $I_\text{high}$. Similarly, in Part 3, we show that when we hit stopping rule in Step~\ref{step:3b_iii}, we have reached $W_\text{binding}$ and there is no need to consider larger $W$'s.
In Part 4 of the proof, we show that if $W \in I_\text{low}$, there exists a $1/2$-approx. solution $S_W \in \mathcal{C}_I$. 
In Part 5, which is one of the most challenging parts of the proof, using induction, we show that if $W \in I_\text{high}$, there exists a $1/2$-approx. solution $S_W \in \mathcal{C}_I$. 

\textbf{Part 1: we only need to show Claim \ref{claim:main} for $W \leq W_\text{binding}$.} 
Let $W_\text{binding}$ be the maximum capacity level $W \in I$ for which the capacity constraint is binding for Problem \eqref{eqn:relaxed-knapsack}, i.e., 
\begin{align} \label{eq:w_max} W_\text{binding} := \sup\Big\{W \in I: \sum_{i \in [n]} w_i x_i^\star(W) = W \Big\} ~ \text{and} ~ \sum_{i \in [n]} w_i x_i^\star(W') < W' ~~ \forall ~ W' \in I \cap (W_\text{binding}, \infty) \,.\end{align}
If the capacity constraint is binding for all $W \in I$, we let $W_\text{binding} = W_{\max}$; if the capacity constraint is non-binding for all $W \in I$, we let $W_\text{binding} = W_{\min}$.
Lemma~\ref{lemma:nonbinding} shows that $W_\text{binding}$ is well-defined. The lemma states that if the capacity constraint is non-binding at some capacity level $\widetilde W$, Problem \eqref{eqn:relaxed-knapsack} will have a non-binding capacity constraint for all $W > \widetilde W$. 
\begin{lemma}
\label{lemma:nonbinding}
Consider a capacity level $\widetilde{W} > 0$, and suppose that the capacity constraint is non-binding at the optimal basic solution $\mathbf{x}^\star(\widetilde{W})$ to $\kprelax{\widetilde W, K}$. That is, $\sum_{i \in [n]} w_i x^\star_i(\widetilde{W}) < \widetilde W$. Then, for any $W> \widetilde W$ such that $W \in I$, we have 
$\sum_{i \in [n]} w_i x_i^\star(W) < W$. 
\end{lemma}
We now note that if $W > W_\text{binding}$, then we must have 
\[
\kpcard{W_\text{binding}, K} > \kpcard{W, K}\,.
\]
To see why, we first assume, without loss of generality, that on the well-behaving interval $I$, the items are ranked by the order of their utilities, i.e., $u_1(W) \geq u_2(W) \geq \dots \geq u_n(W)$. Then,
if there exists $W \in I$ such that $W > W_\text{binding}$, we must have that $W_\text{binding} \geq \sum_{i \in [K]} w_i$. That is, the $K$ items with the top utilities can fit into the knapsack of size $W_\text{binding}$. Since the ordering of utilities of items do not change on $I$, we then have
$$
\kpcard{W_\text{binding}, K} = \sum_{i \in [k]} u_i(W_\text{binding}) > \sum_{i \in [k]} u_i(W) = \kpcard{W, K}\,.
$$
Given that, any set $S$ such that 
$\revmod(S, \mathbf{z}) \geq \frac{1}{2}\kpcard{W_\text{binding}, K}$ would also satisfy $\revmod(S, \mathbf{z}) \geq \frac{1}{2}\kpcard{W, K}$ for any $W > W_\text{binding}$. Hence, we only need to show Claim \ref{claim:main} for $W < W_\text{binding}$.

\textbf{Part 2: hitting the stopping rule in Step~\ref{step:2b}.} Assuming $I_\text{low}$ is non-empty (i.e., $W_{\min} < \th$), we would enter Step 2 of Algorithm~\ref{alg:half-approx}. We show that if we hit the stopping rule in Step~\ref{step:2b} of Algorithm \ref{alg:half-approx},  we must have $W_\text{binding} \in I_{\text{low}}$ and hence, we do not need to consider interval  $I_{\text{high}}$, where $I_{\text{low}} = I \cap [0, \th)$ and $I_{\text{high}} = I \cap [\th, \infty)$. Recall that in  Step~\ref{step:2b} of of Algorithm \ref{alg:half-approx}, we stop the algorithm when $u_{h_K}(\th)< 0$, where $h_K$ is the index of the item with the $K$th highest utility-to-weight ratio. 
Note that if $u_{h_K}(\th) < 0$,  then all items with lower utility-to-weight ratios must also have negative utilities, i.e. $u_{h_j}(\th) < 0$ for all $j \geq K$. Hence, the optimal solution $\mathbf{x}^\star(\th)$ for Problem $(\kprelax{\th, K})$ should not include any of the items $h_j$ for all $j \geq K$. This implies that at $\th$, the capacity constraint is no longer binding. (Recall that $\th = w(H_K)$ is the sum of the weights of items $h_1, \ldots, h_K $.) Then, by Lemma~\ref{lemma:nonbinding}, we have $W_{\text{binding}} < \th$. Since in Part 1 of the proof, we show that we only need to work with $W \in I$ such that $W \leq W_{\text{binding}}$, we do not need to consider $I_{\text{high}}$. This justifies the stopping rule in Step \ref{step:2b}.

\textbf{Part 3: hitting the stopping rule in Step~\ref{step:3b}.} 
Assuming that $I_\text{high}$ is non-empty (i.e., $W_{\max} > \th$), we would enter Step 3 of Algorithm~\ref{alg:half-approx}. Here, we show that if we hit the stopping rule in Step~\ref{step:3b}, we must have reached $W_{\text{binding}}$ and we do not need to consider larger $W$'s. 

Before proceeding, we first make the following definitions. In Step~\ref{step:3b} of Algorithm~\ref{alg:half-approx}, we adaptively partition the interval $I_\text{high}$. In particular, at each iteration, the algorithm updates two quantities: (i) the capacity \textit{change point} $W_{\text{next}}$; (ii) two sets of indices that represent the profile at $W_{\text{next}}$: $P_1$ and $P_0$. 
We let $W_{\text{next}}^{(k)}$ be the change point updated in the $k$th iteration of the while loop, and $P_1^{(k)}, P_0^{(k)}$ be the sets of indices updated in the $k$th iteration. Additionally, we let $W_{\text{next}}^{(0)} = \th$ denote the initial change point, 
and $P_1^{(0)} = H_K, P_0^{(k)} = [n]\setminus H_K$ denote the initial sets of indices; 
both are defined in Step~\ref{step:3a}. 

Now, suppose that $W_\text{next}^{(\ell)}$ is the last change point computed in the while loop, before the stopping rule of Step~\ref{step:3b} is invoked. That is, one of the following cases takes place:
\begin{enumerate}[(i)]
\item There does not exist $i \in P_1^{(\ell)}, j \in P_0^{(\ell)}$ such that $w_i < w_j$. In this case, we exit the while loop before the start of the $(\ell+1)$th iteration.
\item 
Within the $(\ell+1)$th iteration, Step \ref{step:3bi} of Algorithm \ref{alg:half-approx}, after making the following update:
\begin{equation}
(\i^\star, \j^\star) \in \argmax_{i \in P_1^{(\ell)}, j \in P_0^{(\ell)} \atop {w_i < w_j}} \left[ \frac{u_j(W_\text{next}^{(\ell)}) - u_i(W_\text{next}^{(\ell)})}{w_j - w_i} \right]\,,
\end{equation}
we have $u_{\j^\star}(W_\text{next}^{(\ell)}) - u_{\i^\star}(W_\text{next}^{(\ell)}) < 0$. That is, we hit the stopping rule in Step~\ref{step:3bii}. 
\end{enumerate}
We show, for each of these cases, that by the time we reach our last change point $W_\text{next}^{(\ell)}$, we have already reached $W_{\text{binding}}$. 
That is, $W_\text{next}^{(\ell)} = W_{\text{binding}}$. 

For Case (i), observe that in Step~\ref{step:I_high} of Algorithm~\ref{alg:half-approx}, 
in every iteration of the while loop, the set $P_1$ always contain $K$ items. 
Hence, if there does not exist $i \in P_1^{(\ell)}, j \in P_0^{(\ell)}$ such that $w_i < w_j$, this means that $P_1^{(\ell)}$ already contains the $K$ items with the highest weights. 
Since we cannot find a set of at most $K$ items with total weight higher than $w(P_1^{(\ell)}) = W_\text{next}^{(\ell)}$, 
the capacity constraint must be non-binding for Problem $\kprelax{W', K}$ for any $W' > W_\text{next}^{(\ell)}$. Hence, $W_\text{next}^{(\ell)} = W_\text{binding}$.

For Case (ii), if we hit the stopping rule, this means that for any $i \in P_1^{(\ell)}$ and any $j \in P_0^{(\ell)}$ such that $w_i < w_j$, we must have $u_j(W_\text{next}^{(\ell)}) < u_i( W_\text{next}^{(\ell)})$. 
We will show, in Part 5 of this proof, that at capacity level $W_{\text{next}}^{(\ell)}$, we have that $\mathcal P(W_\text{next}^{(\ell)})=\{P_1^{(\ell)}, (0, 0), P_0^{(\ell)}\}$ and $w(P_1^{(\ell)}) = W_\text{next}^{(\ell)}$. That is, the capacity constraint is binding for Problem $(\kprelax{W_\text{next}^{(\ell)}, K})$. Based on that, we have that there also does not exist $j \in P_0^{(\ell)}$ such that $w_j \leq w_i$ and $u_j( W_\text{next}^{(\ell)}) > u_i( W_\text{next}^{(\ell)})$; otherwise, the profile $\mathcal{P}(W_\text{next}^{(\ell)})$ no longer represents the optimal solution. Therefore, we must have that 
the set $P_1^{(\ell)}$ contains the $K$ items with the highest utilities. Recall that on the interval $I$, the ordering of utilities of items do not change. This means that for all $W' > W_\text{next}^{(\ell)}$ and $W' \in I$, the profile no longer changes, i.e. $\mathcal{P}(W') = \mathcal P(W_\text{next}^{(\ell)})$, and the capacity constraint becomes non-binding. 
That is, we have 
$W_\text{next}^{(\ell)} = W_{\text{binding}}$.

Hence, if we hit the stopping rule in Step \ref{step:3b}, this implies that our last change point $W_\text{next}^{(\ell)} = W_{\text{binding}}$, and we no longer need to consider larger $W$'s.

\textbf{Part 4: for $W \in I_\text{low}$ such that $W \leq W_\text{binding}$, a $1/2$-approx. solution $S_W$ is found in Step 2 of Algorithm~\ref{alg:half-approx}.} 
Recall that all items in \( [n] \), which we fed into Algorithm \ref{alg:half-approx}, are eligible items of \( I \), where we assume, without loss of generality, that the set of eligible items is non-empty. That is, the items in \( [n] \) can both fit into the knapsack (i.e., \( w_i \leq W \)) and have positive utility (i.e., \( u_i(W) > 0 \)) for all \( W \in I \). 
Also note that on the well-behaving interval $I$, the ordering of the utility-to-weight ratios of items do not change. Hence, the set $H_j$ that includes the $j$ items with the highest utility-to-weight ratios is well-defined for any $j \in [n]$ and does not change for all $W \in I$.
Now, suppose that $W \in I_\text{low}$ and $W \leq W_\text{binding}$. We can use the following lemma:
\begin{lemma}
\label{lemma:interval-low}
Consider the relaxed knapsack problem \eqref{eqn:relaxed-knapsack}. If the capacity level satisfies $W \in I_\text{low} = I \cap [0, w(H_K))$, where $H_K \subset [n]$ is the set of $K$ items in $[n]$ with the highest utility-to-weight ratios, the optimal solution to \eqref{eqn:relaxed-knapsack} is given by filling up the knapsack with items with positive utilities in the descending order of the utility-to-weight ratio until the capacity is reached.
\end{lemma}
Since $W \in I_\text{low}$, Lemma~\ref{lemma:interval-low} shows that an optimal basic feasible solution $\mathbf{x}^\star(W)$ to the relaxed knapsack problem $(\kprelax{W, K})$ is simply filling up the knapsack with the items with the highest utility-to-weight ratios until we reach capacity $W$. 
This means that the profile $\mathcal{P}(W)$ of the optimal solution must take one of the following forms:
\begin{enumerate}[(i)]
    \item 
    $
    P_1(W) = H_j$ for some $j < K$, and $P_f(W) = (0, 0).
    $ 
    That is, the knapsack is precisely filled by the $j$ items in $[n]$ with the highest utility-to-weight ratios. 
    \item 
    $P_1(W) = H_{j-1}$ for some $j \in [K]$, and $P_f(W) = (h_j,0)$.  This means that the knapsack is filled by the $(j-1)$ items in $[n]$ with the highest utility-to-space ratios, and partially filled by the item $h_j$, i.e., the item with the $j$-th highest utility-to-space ratio.
\end{enumerate}
   Given the form of the profile $\mathcal{P}(W)$, we know from  Lemma~\ref{lemma:1/2-approx} that there must exist a $1/2$-approx. feasible assortment $S_{W}$ to Problem $(\kprelax{W, K})$ in the following collection:
   $
   \big\{\{j\} : j \in [n] \big\} \cup \big\{H_j : j \in [K-1] \big\}.
   $ 
Observe that in Step 1 of Algorithm~\ref{alg:half-approx}, we add $\{j\}$ to $\mathcal{C}_I$ for all $j \in [n]$. In Step 2 of Algorithm~\ref{alg:half-approx}, we add $\{H_j : j \in [K-1] \}$ to $\mathcal{C}_I$. 
Therefore, 
$$
S_{W} \in \big\{\{j\} : j \in [n] \big\} \cup \big\{H_j : j \in [K-1] \big\} \subseteq \mathcal{C}_I\,,
$$
where $S_{W}$ is a $1/2$-approx. solution for Problem $(\kprelax{W, K})$. 

\textbf{Part 5: for $W \in I_{\text{high}}$ such that $W \leq W_\text{binding}$, a $1/2$-approx. solution $S_W$ is found in Step 3 of Algorithm~\ref{alg:half-approx}.} 
Similar to Part 3 of the proof, we define $W_{\text{next}}^{(k)}$ as the change point updated in the $k$th iteration of the while loop, and $P_1^{(k)}, P_0^{(k)}$ as the sets of indices updated in the $k$th iteration. 
Let $W_{\text{next}}^{(0)}, P_1^{(0)}, P_0^{(0)}$ be the initial change points and sets of indices we have in the initialization step (see Step~\ref{step:3a}). 

Let us start by showing the following statement: suppose that at the $k$th iteration of Step~\ref{step:3b}, for some $k \geq 0$, Algorithm~\ref{alg:half-approx} has not hit the stopping rule in Step~\ref{step:3bii} and updated the change point to be $W_{\text{next}}^{(k)}$ and the two sets to be $P_1^{(k)}, P_0^{(k)}$, then, as long as $W_{\text{next}}^{(k)} \in I_\text{high}$,
the optimal basic solution to Problem $\kprelax{W_{\text{next}}^{(k)}, K}$ must have profile $\mathcal P(W_{\text{next}}^{(k)}) = \{P_1^{(k)}, (0, 0),  P_0^{(k)}\}$, and $W_{\text{next}}^{(k)} = w(P_1^{(k)})$.
We will prove this statement via induction:
\begin{itemize} [leftmargin=*]
    \item \textit{Base step} ($k = 0$): For the base step, we would like to show that $\mathcal P(W_{\text{next}}^{(0)}) = \{P_1^{(0)}, (0, 0),  P_0^{(0)}\}$ and $W_{\text{next}}^{(0)} = w(P_1^{(0)})$, where $W_{\text{next}}^{(0)}, P_1^{(0)}, P_0^{(0)}$ are the initial change points and sets of indices we have in the initialization step. 
    That is, the profile for Problem $\kprelax{W_{\text{next}}^{(0)}, K}$ is represented by the two sets $P_1^{(0)}$ and $P_0^{(0)}$, and the capacity constraint is binding.
    Since we have two different ways of initialization based on whether $W_{\min} < \th$ or $W_{\min} \geq \th$, we divide our base step into two cases.
     \begin{enumerate}[(i)]
     \item \textbf{Case (i): $W_{\min} <\th$.}
     In this case, we perform the following initialization: $W_{\text{next}}^{(0)} = \th$ is the initial change point, and the initial sets of indices are $P_1^{(0)} = H_K, P_0^{(0)} = [n]\setminus H_K$. 
     Since $W_{\min} < \th$, $I_\text{low}$ must be non-empty and we have already entered Step 2 of Algorithm \ref{alg:half-approx}. 
     Given that Algorithm \ref{alg:half-approx} has not terminated at Step~\ref{step:2b}, we must have that $u_j(\th) > 0$ for all $j \in H_K$. 
    
     Then, by Lemma \ref{lemma:interval-low}, we know that the optimal basic solution $\mathbf{x}^\star(\th)$ for Problem $\kprelax{\th, K}$ has the profile $P_1(\th) = H_K$, $P_f(\th) = (0, 0)$, $P_0(\th) = [n] \setminus H_K$, where $\th = w(H_K)$. Given our definitions of the initial change points and sets of indices, this is precisely $\mathcal P(W_{\text{next}}^{(0)}) = \{P_1^{(0)}, (0, 0),  P_0^{(0)}\}$ and $W_{\text{next}}^{(0)} = w(P_1^{(0)})$. Hence, the base step holds under Case (i).

    \item \textbf{Case (ii): $W_{\min} \geq \th$.} In this case, we solve Problem $(\kprelax{W_{\min}, K})$ and obtain an optimal basic solution, which has the profile $\mathcal{P}(W_{\min}) = \{P_1(W_{\min}), (\i, \j), P_0(W_{\min})\}$. We set
    $P_1^{(0)} = P_1(W_{\min}) \cup \{\j\}, P_0^{(0)} = [n] \setminus P_1^{(0)}, W_{\text{next}}^{(0)} = w(P_1^{(0)})$ as our initial sets of indices and change point. 
    To see why this initialization satisfies the base case, 
    we can invoke Lemma~\ref{lemma:feasibility} stated below.
    \begin{lemma}
    \label{lemma:feasibility}
    Suppose that at capacity level $\widehat{W} \in I_\text{high}$, an optimal basic solution $\mathbf{x}^\star(\widehat{W})$ to Problem $(\kprelax{\widehat{W}, K})$ has profile $\mathcal{P}(\widehat{W}) = \{ P_1 \setminus \{\i\}, (\i, \j), P_0 \setminus \{\j\} \}$. 
    Then, for any $\widetilde{W} \in [w(P_1), w(P_1) - w_{\i} + w_{\j}]$ such that $\widetilde{W} \in I_\text{high}$,
    we have an optimal basic solution $\mathbf x$ to Problem $(\kprelax{\widetilde{W}, K})$ such that $x_k=1$ for any $k \in P_1 \setminus \{\i\} $,  $x_k =0$
    for any $k \in P_0 \setminus \{\j\}$, and 
    \begin{equation}
    \label{eqn:form-of-BFS-2}
    x_{k} = 
    \begin{cases}
    \dfrac{\widetilde{W} - w(P_1)+ w_{\i} - w_{\j}}{w_{\i} - w_{\j}} & \text{$k = \i$} \\
    \dfrac{\widetilde{W} - w(P_1)}{w_{\j} - w_{\i}} & \text{$k = \j$} 
    \end{cases}\,\,.
    \end{equation}
    \end{lemma} 
    By invoking Lemma~\ref{lemma:feasibility} with 
    $\widehat{W} = W_{\min}$ and $\widetilde{W} = w(P_1(W_{\min})) + w_{\j} = W^{(0)}_\text{next}$, 
    we have that as long as $W^{(0)}_\text{next} \in I_\text{high}$, the optimal basic solution $\mathbf{x}$ to Problem $(\kprelax{W^{(0)}_\text{next}})$ takes the form that $x_k = 1$ for any $k \in P_1(W_{\min}) \cup \{\j\}$ and $x_k = 0$ for any $k \in P_0(W_{\min}) \cup \{\i\}$. 
    That is, we have that the profile $\mathcal{P}(\kprelax{W^{(0)}_\text{next}}) = \{P_1(W_{\min}) \cup \{\j\}, (0, 0), P_0(W_{\min}) \cup \{\i\}\} = \{P_1^{(0)}, (0, 0), P_0^{(0)}\}$ and $W^{(0)}_\text{next} = w(P_1(W_{\min}) \cup \{\j\}) = w(P_1^{(0)})$, which shows that the base case also holds for Case (ii). 
    \end{enumerate}
    \item \textit{Inductive step}: 
    Suppose that $W_{\text{next}}^{(k)}, P_1^{(k)}, P_0^{(k)}$ are the change point and the sets of indices at the $k$th iteration. 
    By our inductive assumption, 
    assume that the optimal solution to Problem $(\kprelax{W_{\text{next}}^{(k)}, K})$ indeed has profile $\mathcal P(W_{\text{next}}^{(k)}) = \{P_1^{(k)}, (0, 0),  P_0^{(k)}\}$ and $w(P_1^{(k)}) = W_{\text{next}}^{(k)}$. 
    Now, assume that at the $(k+1)$th iteration of Step \ref{step:3b}, Algorithm~\ref{alg:half-approx} has not hit the stopping rule in Step~\ref{step:3bii} and updated the change point to be $W_{\text{next}}^{(k+1)}$ and the two sets to be $P_1^{(k+1)}, P_0^{(k+1)}$.
    We would like to show that as long as $W_{\text{next}}^{(k+1)} \in I_\text{high}$,
    the optimal basic solution to Problem $(\kprelax{W_{\text{next}}^{(k+1)}, K})$ must have profile 
    $\mathcal P(W_{\text{next}}^{(k+1)}) = \{P_1^{(k+1)}, (0, 0),  P_0^{(k+1)}\}$, and $W_{\text{next}}^{(k+1)} = w(P_1^{(k+1)})$.

    Recall that within the $(k+1)$th iteration of Step~\ref{step:3b} of Algorithm~\ref{alg:half-approx}, we first solve the following optimization problem $(\i^\star, \j^\star) \in \argmax_{i \in P_1^{(k)}, j \in P_0^{(k)} \atop {w_i < w_j}} \left[ \frac{u_j(W_{\text{next}}^{(k)}) - u_i(W_{\text{next}}^{(k)})}{w_j -w_i} \right]$, and then, given that $W_{\text{next}}^{(k)} < W_{\max}$ and $u_{\j^\star}(W_{\text{next}}^{(k)}) > u_{\i^\star}(W_{\text{next}}^{(k)})$, make the following updates: $ W_{\text{next}}^{(k+1)} = W_{\text{next}}^{(k)} - w_{\i^\star} + w_{\j^\star}$, $P_1^{(k+1)} = P_1^{(k)} \cup \{\j^\star\} \setminus \{\i^\star\}$ and $P_0^{(k+1)} = P_0^{(k)} \cup \{\i^\star\} \setminus \{\j^\star\}$. To show the result, we use the following lemma:
    \begin{lemma}
\label{corollary:form-sol}
Suppose that at capacity level $W \in I_\text{high}$, the capacity constraint is binding and the relaxed knapsack problem $(\kprelax{W, K})$ has a degenerate, integer optimal basic solution $\mathbf{x}^\star$, with profile $\mathcal{P}(W) = \{P_1, (0, 0), P_0\}$. Let $(\i^\star, \j^\star) \in \argmax_{i \in P_1, j \in P_0 \atop {w_i < w_j}} \left[\frac{u_j(W) - u_i(W)}{w_j - w_i}\right]$. If $u_{\j^\star}(W) > u_{\i^\star}(W)$, for any $\eta \in [0, w_{\j^\star} - w_{\i^\star}]$ such that $W + \eta \in I_\text{high}$, Problem $(\kprelax{W + \eta, K})$ has an optimal basic solution $\mathbf x$ such that  $x_k =1$ for any $k \in P_1 \setminus \{\i^\star\}$, $x_k=0$ for 
any $k \in P_0 \setminus \{\j^\star\}$, and   \begin{equation}
x_k = 
\begin{cases}
1 - \frac{\eta}{w_{\j^\star} - w_{\i^\star}} & \text{$k = \i^\star$} \\ \frac{\eta}{w_{\j^\star} - w_{\i^\star}} & \text{$k = \j^\star$} 
\end{cases}.
\end{equation}
\end{lemma}
The proof of the inductive step is then completed by invoking Lemma \ref{corollary:form-sol} with $W = W_{\text{next}}^{(k)}$ and $\eta = w_{\j^{\star}}- w_{\i^{\star}}$. 
We indeed have that the optimal basic solution to Problem $(\kprelax{W_{\text{next}}^{(k+1)}, K})$
has the profile $\mathcal{P}(W_{\text{next}}^{(k+1)}) = \{P_1^{(k)} \cup \{\j^\star\} \setminus \{\i^\star\}, (0, 0),  P_0^{(k+1)} \cup \{\i^\star\} \setminus \{\j^\star\}\} = \{P_1^{(k+1)}, (0, 0),  P_0^{(k+1)}\}$, 
and $w(P_1^{(k+1)}) = w(P_1^{(k)}) - w_{\i^\star} + w_{\j^\star} = W_{\text{next}}^{(k+1)}$, as long as $W_{\text{next}}^{(k+1)} \in I_\text{high}$.
    
\end{itemize}
The base step and the inductive step together prove the statement that at the $k$th iteration of Step~\ref{step:3b} for any given $k \geq 0$, 
as long as the change point $W_{\text{next}}^{(k)} \in I_\text{high}$, the optimal basic solution to Problem $(\kprelax{W_\text{next}^{(k)}, K})$ has profile $\mathcal P(W_{\text{next}}^{(k)}) = \{P_1^{(k)}, (0, 0),  P_0^{(k)}\}$ and the capacity constraint is binding, i.e., $w(P_1^{(k)}) = W_{\text{next}}^{(k)}$. 

Now, suppose that throughout Algorithm~\ref{alg:half-approx}, we have the following change points: $W_{\text{next}}^{(0)} < W_{\text{next}}^{(1)} < \dots < W_{\text{next}}^{(\ell)}$, where $W_{\text{next}}^{(\ell)}$ is the last change point computed by Algorithm~\ref{alg:half-approx} before the stopping rule in Step~\ref{step:3b} is invoked. 
We have shown, in part 3 of the proof, that $W_{\text{next}}^{(\ell)} = W_\text{binding}$. Now, consider any $W \in I_\text{high}$ such that $W \leq W_\text{binding}$. We must have one of the following two cases: (1) $W \in [\th, W_{\text{next}}^{(0)}]$; note that this interval is non-empty only when $W_{\min} > \th$; or (2) $W$ lies between two consecutive change points, i.e. $W \in [W_{\text{next}}^{(k)}, W_{\text{next}}^{(k+1)})$ for some $0 \leq k < l$ (or, $W = W_{\text{next}}^{(\ell)}$). Let us discuss these two cases separately:
\begin{enumerate}
\item \textbf{Case 1: $W \in [\th, W_{\text{next}}^{(0)})$.}\footnote{Recall from Part 1 that if the capacity constraint is non-binding for all $W \in I$, we have defined $W_\text{binding} = W_{\min}$. We remark that in this case, our statement of Part 5 trivially holds for $W =  W_{\min}$. In the initialization step (Step \ref{step:3a}), we compute an optimal basic solution for Problem $(\kprelax{W_{\min}, K})$. Since the capacity constraint is non-binding at $W_{\min}$, the profile $\mathcal{P}(W_{\min})$ will take the form $\mathcal{P}(W_{\min}) = \{P_1(W_{\min}), (0, 0), P_0(W_{\min})\}$. where $P_1(W_{\min})$ includes the $K$ items with the highest positive utilities. By design of Algorithm \ref{alg:half-approx}, we will add $P_1(W_{\min})$ to the collection $\mathcal{C}_I$, and $P_1(W_{\min})$ serves as 
a $1/2$-approx. solution for the knapsack problem $(\kprelax{W_{\min}, K})$.} 
Since this interval is only non-empty when $W_{\min} > \th$, in the initialization step (Step \ref{step:3a}), we will compute the optimal basic solution for Problem $(\kprelax{W_{\min}, K})$, which has the profile $\mathcal{P}(W_{\min}) = \{P_1(W_{\min}), (\i, \j), P_0(W_{\min})\}$. By invoking Lemma \ref{lemma:feasibility} with $\widehat{W} = W_{\min}$ and $\widetilde{W} = W$, we know that the optimal basic solution $x^\star(W)$ has profile $\mathcal{P}(W) = \{P_1(W_{\min}), (\i, \j), P_0(W_{\min})\}$. Given the form of the profile $\mathcal{P}(W)$, we know from Lemma \ref{lemma:1/2-approx} that either $P_1(W_{\min}) \cup \{\i\}$ or $\{\j\}$ is a $1/2$-approx. feasible solution $S_{W}$ to Problem $(\kprelax{W, K})$. Observe that in Step \ref{step:3a}, we have added $P_1(W_{\min}) \cup \{\i\}$ to the collection $\mathcal{C}_I$. In Step \ref{step:initialization}, we have added $\{\j\}$ to $\mathcal{C}_I$. Therefore, if $W \in [\th, W_{\text{next}}^{(0)})$, we must have $S_{W} \in \mathcal{C}_I$.

\item \textbf{Case 2: $W \in [W_{\text{next}}^{(k)}, W_{\text{next}}^{(k+1)})$ for some $0 \leq k < l$ (or, $W = W_{\text{next}}^{(\ell)}$).} Since $W \leq W_{\text{binding}}$, we must also have $W_{\text{next}}^{(k)} \leq W_{\text{binding}}$; hence, our inductive statement holds for the $k$th iteration.
At the $k$th iteration of Step~\ref{step:3b}, we have updated the change point to be $W_{\text{next}}^{(k)}$, and the two sets to be $P_1^{(k)}, P_0^{(k)}$. By the inductive statement, we have that $\mathcal P(W_{\text{next}}^{(k)})=\{P_1^{(k)}, (0,0), P_0^{(k)}\}$. 
Since $W \in [W_{\text{next}}^{(k)}, W_{\text{next}}^{(k+1)})$ and that the capacity constraint is binding at $W$, by Lemma~\ref{corollary:form-sol}, we have: 
\begin{enumerate}[(i)]
    \item if $W = W_{\text{next}}^{(k)}$ for some $k\in \{0, \ldots, l\}$, the profile of $\mathbf{x}^\star(W )$ is $\mathcal{P}(W )=  \{P_1^{(k)}, (0, 0), P_0^{(k)}\}$.
    \item if $W \in (W_{\text{next}}^{(k)}, W_{\text{next}}^{(k+1)})$ for some $k\in \{0, \ldots, l-1\}$, the profile of $\mathbf{x}^\star(W)$ is $\mathcal{P}(W)=  \{P_1^{(k)} \setminus \{\i^\star\},  (\i^\star, \j^\star),  P_0^{(k)} \setminus \{\j^\star\}\}$.
\end{enumerate}
Given the form of the profile $\mathcal{P}(W)$, we know from Lemma~\ref{lemma:1/2-approx} that either $P_1^{(k)}$ or $\{\j^\star\}$ is an $1/2$-approx. feasible solution $S_{W}$ to Problem $\kprelax{W, K}$. Observe that in Step~\ref{step:3b_iii} of the $k$th iteration, we have added $P_1^{(k)}$ to the collection $\mathcal{C}_I$. In Step~\ref{step:initialization}, we have added $\{j\}$ to $\mathcal{C}_I$ for all $j \in [n]$. Therefore, in this case, we again have $S_{W} \in \mathcal{C}_I$.
\end{enumerate}
This concludes Part 5 of this proof, which shows that for any $W \in I_\text{high}$ such that $W \leq W_\text{binding}$, we must also have $S_{W} \in \mathcal{C}_I$, where $S_{W}$ is a $1/2$-approx. solution for Problem $(\kprelax{W, K})$.

\vspace{2mm}

Overall, Parts 2--5 imply that for any $W \in I$ such that $W \leq W_\text{binding}$, Algorithm~\ref{alg:half-approx} would add a $1/2$-approx. feasible solution $S_{W}$ to the collection $\mathcal{C}_I$. Together with Part 1 of the proof, we show that Claim~\ref{claim:main} holds.

\medskip

\textbf{Segment 2.} We now comment on the runtime of Algorithm~\ref{alg:half-approx}. 
\begin{enumerate}
    \item \emph{Step 1.} Adding $n$ singletons to $\mathcal{C}_I$ takes $\mathcal{O}(n)$, and ranking the items by their utility-to-weight ratios takes $\mathcal{O}(n\log n)$. 
    \item \emph{Step 2.} Adding $n$ assortments to $\mathcal{C}_I$ and checking the signs of the utilities takes $\mathcal{O}(n)$. 
    \item \emph{Step 3.} The initialization step (Step~\ref{step:3a}) requires us to solve a single relaxed knapsack problem with $n$ items, which takes at most $\mathcal{O}(n\log(n))$. 
    
    In Step \ref{step:3bi}, we need to find the pair of items $(\i^\star, \j^\star)$ to be swapped at each iteration. Instead of solving the problem directly at every iteration, we can do the following to improve the time complexity.
    At the start of Step~\ref{step:3b}, we first pre-compute and sort the array $\mathcal{A}_i = \{\frac{u_j(W) - u_i(W)}{w_j - w_i} : j \in [n], w_j > w_i \}$ for each $i \in [n]$, for any given $W \in I_\text{high}$. Note that since interval $I$ is well-behaving, by Definition \ref{def:well-behaving}, the ordering of terms in $\mathcal{A}_i$ for any fixed $i \in [n]$ does not change as long as $W \in I$. This pre-computing step takes $\mathcal{O}(n^2)$, and sorting these $n$ arrays takes $\mathcal{O}(n^2\log n)$ in total. 
    The first iteration of Step~\ref{step:3b} will take $\mathcal{O}(K^2 + K)$ time. This is because we will restrict our attention to the $K$ sorted arrays $\{\mathcal{A}_i : i \in P_1\}$ that correspond to elements $i \in P_1$, and for each of these arrays, finding the optimal index $j_{\max}(i) := \argmax \{\frac{u_j(W_\text{next}) - u_i(W_\text{next})}{w_j - w_i} : j \in P_0, w_j > w_i\}$ takes $\mathcal{O}(K)$ given that $\mathcal{A}_i$ has been sorted. 
    We can now determine $(\i^\star, \j^\star) = \argmax \{\frac{u_j(W_\text{next}) - u_i(W_\text{next})}{w_j - w_i} : i \in P_1, j \in P_0, w_i < w_j\}$ by simply choosing the maximizing index $\i^\star = \argmax\{\frac{u_{j_{\max}(i)}(W_\text{next}) - u_i(W_\text{next})}{w_{j_{\max}(i)} - w_i} : i \in P_1\}$, which takes $\mathcal{O}(K)$. 
    
    Each time after we determine $(\i^\star, \j^\star)$ of the current iteration, checking the stopping rule and updating $P_1, P_0, W_\text{next}$ take $\mathcal{O}(1)$ time. Then, we can perform $\mathcal{O}(K)$ updates to find the pair of items to be swapped in the next iteration. 
    For each sorted array $\mathcal{A}_i$ where $i \in P_1 \setminus \{\i^\star\}$, we will now consider $\i^\star$ as an element in $P_0$ and no longer consider $\j^\star$ when we search for the next optimal index $\j_{\max}(i)$. 
    In addition, we will now consider $\mathcal{A}_{\j^\star}$ as one of the $K$ sorted arrays that correspond to items in $P_1$, and no longer consider $\mathcal{A}_{\i^\star}$. 
    Therefore, the time taken in each iteration of Step~\ref{step:3b} would be $\mathcal{O}(K)$.

    We further remark that the adaptive partition of $I_\text{high}$ would divide $I_\text{high}$ into at most $\mathcal{O}(nK)$ sub-intervals. This is because at each iteration of Step~\ref{step:3b}, we need to swap some item $i \in P_1$ and $j \in P_0$ such that $w_i < w_j$; that is, at each iteration, we always replace one item in $P_1$ with another item of higher weight. 
    Let us think of $P_1$ as a set of $K$ spots, where each spot can hold one item. Since the weight of item in one spot in $P_1$ is strictly increasing, each spot in $P_1$ can only undergo $n$ replacement of items. Hence, in total, we can perform at most $\mathcal{O}(nK)$ such replacements for the items in $P_1$.
    
    Overall, Step 3 takes $\mathcal{O}(n\log{n} + n^2 + n^2\log{n} + (K^2 + K) + nK \cdot K) = \mathcal{O}(n^2\log{n} + nK^2)$. 
    
    \item \emph{Step 4.} Finally, since $\mathcal{O}(n)$ assortments are added in Step 1, $\mathcal{O}(n)$ assortments are added in Step 2, and $\mathcal{O}(nK)$ assortments are added in Step 3, there are at most $\mathcal{O}(nK)$ assortments in the collection $\mathcal{C}_I$. The time taken to find the assortment that maximizes the cost-adjusted revenue is thus $\mathcal{O}(nK)$. 
\end{enumerate}
The overall runtime of Algorithm~\ref{alg:half-approx} is at most
$\mathcal{O}(n\log n + n + n^2\log{n} + nK^2 + nK) = \mathcal{O}(n^2\log{n} + nK^2)$. $\blacksquare$

\subsection{Proof of Corollary \ref{corollary:half-approx-ratio}}

Corollary \ref{corollary:half-approx-ratio} immediately follows from the proof of Theorem \ref{thm:half-approx-ratio}. Given any $W \subseteq I$, we have:

\begin{enumerate}
    \item If $W \in I_\text{low} = I \cap [0, \th)$, by Part 4 of the proof of Theorem \ref{thm:half-approx-ratio}, an $1/2$-approx. solution for Problem $\kprelax{W, K}$ is found in Step 2 of Algorithm \ref{alg:half-approx-modified}. In particular, there exists $j \in [K-1]$ such that $W \in [w(H_j), w(H_{j+1}))$. By Lemma \ref{lemma:1/2-approx}, we know that for any $W \in I' = [w(H_j), w(H_{j+1})) \cap I$, the profile $\mathcal{P}(W) = \{H_j, (h_{j+1}, 0), [n]\setminus H_{j+1}\}$ remains the same and the $1/2$-approx. set should be either $S_0 = H_j$ or $S_1 = \{h_{j+1}\}$. We thus inputs $I', S_0, S_1$ into Algorithm \ref{alg:subroutine} in Step 2(a) of Algorithm \ref{alg:half-approx-modified}.
    \item If $W \in I_\text{high} = I \cap [\th, \infty)$, by Part 5 of the proof of Theorem \ref{thm:half-approx-ratio}, an $1/2$-approx. solution for Problem $\kprelax{W, K}$ is found in Step 3 of Algorithm \ref{alg:half-approx-modified}. In particular,
    \begin{itemize}
    \item If $W \in [\th, W^{(0)}_\text{next})$, where $W_{\text{next}}^{(0)} = w(P_1(W_{\min})) + w_{\j}$, by Lemma \ref{lemma:1/2-approx}, we know that for any $W \in I' = [\th, W^{(0)}_\text{next}) \cap I$, the profile $\mathcal{P}(W) = \{P_1(W_{\min}), (\i, \j), P_0(W_{\min}) \}$ remains the same and the $1/2$-approx. set should be either $S_0 = P_1(W_{\min}) \cup \{\i\}$ and $S_1 = \{\j\}$. We thus inputs $I', S_0, S_1$ into Algorithm \ref{alg:subroutine} in Step 3(a) of Algorithm \ref{alg:half-approx-modified}.
    \item If $W \in [W_{\text{next}}^{(k)}, W_{\text{next}}^{(k+1)})$ for some $k \ge 0$, by Lemma \ref{lemma:1/2-approx}, we know that for any $W \in I' = [W_{\text{next}}^{(k)}, W_{\text{next}}^{(k+1)}) \cap I$, the profile $\mathcal{P}(W)=  \{P_1^{(k)} \setminus \{\i^\star\},  (\i^\star, \j^\star),  P_0^{(k)} \setminus \{\j^\star\}\}$ remains the same and the $1/2$-approx. set should be either $S_0 = P_1^{(k)}$ or $S_1 = \{\j^\star\}$. We thus inputs $I', S_0, S_1$ into Algorithm \ref{alg:subroutine} in Step 3(b) of Algorithm \ref{alg:half-approx-modified}.
    \end{itemize}
\end{enumerate}

Given a well-behaving interval $I'$, on which the profile does not change, and the two candidates for its $1/2$-approx. solutions $S_0, S_1$,
Algorithm \ref{alg:subroutine} (i) determines whether we need to further partition $I'$ once and (ii) maps each sub-interval $I_\ell$ to its exact $1/2$-approx. solution $\mathcal{D}(I_\ell)$. Given that the ordering of the total utilities produced by our two candidate sets---$\sum_{i \in S_0} u_i(W)$ and $\sum_{i \in S_1} u_i(W)$---would change at most once at value $\widehat{W}$ such that $\sum_{i \in S_0} u_i(\widehat{W}) = \sum_{i \in S_1} u_i(\widehat{W})$ (defined in Eq.\eqref{eqn:dividing-pt}), we can simply compare the utilities generated by $S_0$ and $S_1$ at the two end points of $I'$ to determine the exact identities of the $1/2$-approx sets. The details are outlined in Step 2 of Algorithm \ref{alg:subroutine}.

The design of Algorithm \ref{alg:half-approx-modified} ensures that $\Pi_I$ forms a partition of interval $I$.  
Given the results shown above, we also have that for any $I_\ell \in \Pi_I$, $\mathcal{D}(I_\ell)$ is a feasible $1/2$-approx. set for any $W \in I_\ell$. $\blacksquare$

\subsection{Proof of Lemma~\ref{lemma:nonbinding}}
Consider a capacity level $\widetilde{W} > 0$, and suppose that the capacity constraint is non-binding at the optimal basic solution $\mathbf{x}^\star(\widetilde{W})$ to $\kprelax{\widetilde W, K}$. That is, $\sum_{i \in [n]} w_i x^\star_i(\widetilde{W}) < \widetilde W$. We would like to show that for any $W> \widetilde W$, we have 
$\sum_{i \in [n]} w_i x_i^\star(W) < W$. To do so, for any $\eta > 0$, we  show that the optimal solution $x^\star(\widetilde W)$ is in fact also the optimal solution for Problem $(\kprelax{W^\star + \eta, K})$, and hence the capacity constraint remains non-binding at capacity level $\widetilde W + \eta$ for any $\eta > 0$.

To do that, we first consider the following auxiliary relaxed problem:
\begin{equation}
\label{eqn:relaxed-knapsack-2}
\max_{\mathbf{x} \in [0,1]^n} 
\sum_{i \in [n]} u_i(\widetilde{W}) x_{i} \quad
\mathrm{s.t.} \quad \sum_{i \in [n]} w_{i} x_{i} \leq \widetilde{W} + \eta \quad \text{and} \quad \sum_{i \in [n]} {x_i} \leq K
\end{equation}
Note that the above relaxed problem only differs from $\kprelax{\widetilde{W}, K}$ in the capacity constraint. Since the capacity constraint is non-binding for Problem $\kprelax{\widetilde{W}, K}$, 
we have that $\mathbf{x}^\star(\widetilde{W})$ is also the optimal solution to Problem  \eqref{eqn:relaxed-knapsack-2}. This is because both feasibility and optimality conditions remain unchanged.

We claim that the optimal solution $\mathbf{x}^\star(\widetilde{W})$ must fully contain the $K$ items with the highest utilities (here, we assume that at least $K$ items have positive utilities at capacity level $\widetilde{W}$; otherwise, this Lemma holds trivially). Without loss of generality, let the items be ranked based on their utilities $u_i(\widetilde{W})$; that is, $u_1(\widetilde{W}) \geq u_2(\widetilde{W}) \geq \dots \geq u_n(\widetilde{W})$. (Recall that on $I$, the ordering of utilities of items does not change.) We will show that 
(i) $x^\star_k(\widetilde{W}) = 0$ for $k \geq K+1$ and 
(ii) $x^\star_k(\widetilde{W}) = 1$ for $k \leq K$. 
To show the first statement, suppose by contradiction that there exists $i \geq K+1$ such that $x^\star_i(\widetilde{W}) > 0$. Then, there must exist some $j \in [K]$ such that $x^\star_j(\widetilde{W}) < 1$.
Suppose $w_i \geq w_j$, then if we set $\mathbf{x}$ 
to be 
$x_{i} = x^\star_i(\widetilde{W}) + x^\star_j(\widetilde{W}) - x_j,
x_j = \min\{1, x^\star_i(\widetilde{W}) + x^\star_j(\widetilde{W})\}, \text{and}
~
x_k = x^\star_k(\widetilde{W}) ~
\text{for any}
~k \neq i, j
$, $\mathbf{x}$ is a better, feasible solution to Problem~\eqref{eqn:relaxed-knapsack-2}. 
On the other hand, suppose $w_i < w_j$, then if we set $\mathbf{x}$ to be $
x_{i} = x^\star_i(\widetilde{W}) + x^\star_j(\widetilde{W}) - x_j,
x_j = 
\min\{1, x^\star_j(\widetilde{W}) + 
\frac{\eta}{w_j - w_i}\}, \text{and}
~
x_k = x^\star_k(\widetilde{W}) ~
\text{for any}
~k \neq i, j
$, $\mathbf{x}$ is a better, feasible solution to Problem~\eqref{eqn:relaxed-knapsack-2}. The above arguments thus raise a contradiction. Therefore, we must have $\mathbf{x}^\star_k(\widetilde{W}) = 0$ for all $k \geq K+1$. To show the second statement, suppose by contradiction that $x^\star_i(\widetilde{W}) < 1$ for some $i \in [K]$, then if we set $\mathbf{x}$ to be $x_i = \min \{1, x^\star_i(\widetilde{W}) + \frac{\eta}{w_i}\}, \text{and}
~
x_k = x^\star_k(\widetilde{W}) ~
\text{for any}
~k \neq i$, $\mathbf{x}$ is a better, feasible solution to Problem~\eqref{eqn:relaxed-knapsack-2}. Therefore, we must have $\mathbf{x}^\star_k(\widetilde{W}) = 1$ for all $k \leq K$.

Since the ordering of the utilities of items do not change on $I$, when we consider Problem $(\kprelax{W^\star + \eta, K})$, the optimal solution $\mathbf{x}^\star(\widetilde{W})$ is again a feasible solution to Problem $(\kprelax{W^\star + \eta, K})$ that fully contains the $K$ items with the highest utilities. Hence, $\mathbf{x}^\star(\widetilde{W})$ again serves as the optimal solution to Problem $(\kprelax{W^\star + \eta, K})$. Since $\sum_{i \in [n]} w_i x_i^\star(\widetilde{W}) < \widetilde{W} < \widetilde{W} + \eta$, we thus have that the capacity constraint is non-binding for Problem $(\kprelax{W^\star + \eta, K})$. $\blacksquare$

\subsection{Proof of Lemma~\ref{lemma:interval-low}}
Consider the following relaxed knapsack problem without the cardinality constraint:
\begin{equation}
\label{eqn:knapsack-without-card}
\begin{aligned}
\max_{\mathbf{x}\in [0,1]^n} & 
\sum_{i=1}^{n} u_i(W) x_{i} \quad \mathrm{s.t.} & \sum_{i=1}^{n} w_{i} x_{i} \leq W 
\end{aligned}
\end{equation}
It is known that the optimal solution to Problem \eqref{eqn:knapsack-without-card} is given by filling up the knapsack with items in descending order of the utility-to-weight ratio until the capacity is reached. Recall that $H_K$ is the set of $K$ items with the highest utility-to-weight ratios. Since $W \leq w(H_K)$, this optimal solution also satisfies $|\sum_{i \in [n]} {x_i}| \leq K$ and is thus feasible for Problem \eqref{eqn:relaxed-knapsack} as well. Since $\kprelax{W, K}$ is always upper bounded by the optimal objective of \eqref{eqn:knapsack-without-card}, 
we have the optimal solution to Problem \eqref{eqn:knapsack-without-card} is also optimal for Problem \eqref{eqn:relaxed-knapsack}. $\blacksquare$

\subsection{Proof of Lemma \ref{lemma:feasibility}}
Suppose that at capacity level $\widehat{W} \in I_\text{high}$, an optimal basic solution $\mathbf{x}^\star(\widehat{W})$ to Problem $(\kprelax{\widehat{W}, K})$ has profile $\mathcal{P}(\widehat{W}) = \{ P_1 \setminus \{\i\}, (\i, \j), P_0 \setminus \{\j\} \}$. By Lemma \ref{lemma:other-lemma-1}, we know that $x^\star(\widehat{W})$ either has two fractional variables, or no fractional variables. If $(\i, \j) = (0, 0)$ (i.e., $x^\star(\widehat{W})$ has no fractional variables), the statement of Lemma \ref{lemma:feasibility} is trivially satisfied. In the following of this proof, we focus on the case when $x^\star(\widehat{W})$ has two fractional variables, i.e., $\i \neq 0, \j \neq 0$.

Fix $\widetilde{W} \in [w(P_1), w(P_1) - w_{\i} + w_{\j}]$ such that $\widetilde{W} \in I_\text{high}$. For simplicity of notation, we let $\mathbf{x}^\star = \mathbf{x}^\star(\widehat{W})$ and $\mathbf{x} = \mathbf{x}^\star(\widetilde{W})$. As shown in \cite{caprara2000approximation}, for any $W\ge 0$, in the optimal basic solution of a relaxed knapsack problem $\kprelax{W,K}$, we have at most two basic variables. Here, for Problem $\kprelax{\widehat{W}, K}$, $x^\star_{\i}$ and $x^\star_{\j}$ are the two fractional variables, hence they also serve as the two basic variables in the optimal basic solution $\mathbf{x}^\star$. 
In the following proof, we will show that if we move from Problem $(\kprelax{\widehat{W}, K})$ to Problem $(\kprelax{\widetilde{W}, K})$, $x_{\i}$ and $x_{\j}$ will continue to be the two basic variables. To show that, recall that a basic solution that satisfies (i) the optimality condition and (ii) the feasibility condition is an optimal basic solution (see Section \ref{sec:LP} for formal definitions of both conditions). In the following, we will show that both conditions continue to hold with the same basis when we transition to Problem $(\kprelax{\widetilde{W}, K})$. This then allows us to solve for $\mathbf{x} = \mathbf{x}^\star(\widetilde{W})$ using a linear system defined by capacity and cardinality constraints.

\textbf{Optimality condition.} We first consider the optimality condition, which requires that the reduced costs of non-basic variables that attain their lower bounds to be non-positive, and the reduced costs of non-basic variables that attain their upper bounds to be non-negative (see Section \ref{sec:LP} for formal definition of \emph{reduced cost}). Since $x^\star$ is a nondegenerate, optimal basic solution to Problem $(\kprelax{\widehat{W}, K})$, we know that it satisfies the optimality condition. Our main idea here is to show that if we now transition to Problem $(\kprelax{\widetilde{W}, K})$, the signs of the reduced costs of all variables remain unaffected.

To see that, we consider $(\kprelax{\widehat W, K})$ in the following equivalent form:
\begin{equation}
\label{eqn:relaxed-knapsack-eq}
\tag{$\textsc{kp-relax-eq}(\widehat{W}, K)$}
\max_{\mathbf {x}\in [0,1]^n, s_1, s_2\ge 0 } 
\sum_{i \in [n]} u_i(\widehat{W}) x_{i} \quad
\mathrm{s.t.} \quad  \sum_{i \in [n]} w_{i} x_{i} + s_1 = \widehat{W} \quad \text{and} \quad \sum_{i \in [n]} {x_i} + s_2 = K 
\end{equation}
Since $x^\star_{\i}$ and $x^\star_{\j}$ are the two basic variables in the optimal solution for the above problem, we know that for any non-basic variable $x^\star_k$, where $k \neq \i, \j$, the reduced cost $\overline{C}_k \leq 0$ if $x^\star_k = 0$ and $\overline{C}_k \geq 0$ if $x^\star_k = 1$ (intuitively, we can think of the reduced cost as the change in our total utility due to including non-basic variables into our basis. These conditions ensure that changing the values of any of the non-basic variables above will only decrease our objective function; see Section \ref{sec:LP} for more details). 
The non-basic variables $s_1, s_2$, also have non-positive reduced costs.

Let us first focus on any non-basic variables with  $x^\star_k = 0$. The corresponding reduced cost of this variable is as follows:
$$
\begin{aligned}
& \overline{C}_k = u_k(\widehat{W}) + u_\i(\widehat{W})\frac{w_k - w_{\j}}{w_{\j}- w_{\i}} - u_\j(\widehat{W})\frac{w_k - w_{\i}}{w_{\j} - w_{\i}} \leq 0 \\
\Longleftrightarrow & (w_{\j} - w_{\i})u_k(\widehat{W}) + (w_k - w_{\j})u_\i(\widehat{W}) + (w_\i 
- w_k)u_\j(\widehat{W}) \leq 0 
\\
\Longleftrightarrow & 
(u_k(\widehat W) - u_\i(\widehat W))(w_\j - w_\i) \leq 
(u_\j(\widehat W) - u_\i(\widehat W))(w_k - w_\i)
\end{aligned}
$$
Note that since $\widehat{W}, \widetilde{W} \in I$, based on Definition \ref{def:well-behaving} of the well-behaving interval, we must have the ordering of $\frac{u_k(W) - u_\i(W)}{w_k - w_\i}$ and $\frac{u_\j(W) - u_\i(W)}{w_\j - w_\i}$ for any $W \in I$. 
This implies that we must also have
$$
(w_{\j} - w_{\i})u_k(\widetilde{W}) + (w_k - w_{\j})u_\i(\widetilde{W}) + 
(w_\i - w_k)u_k(\widetilde{W}) \leq 0\,.
$$
A similar argument applies to non-basic variables $x^\star_k$ that takes value $1$. 
That is, the reduced cost associated with any item $k$ such that $k \neq \i, k \neq \j$ does not change sign as we transition from Problem $(\kprelax{\widehat{W}, K})$ to Problem $(\kprelax{\widetilde{W}, K})$. Let us now consider the reduced cost of $s_1$, which is non-positive when we consider Problem $(\kprelaxeq{\widehat{W}, K})$ 
$$
\overline{C}_{s_1} = \frac{u_\j(\widehat{W}) - u_\i(\widehat{W})}{w_{\i} - w_{\j}} \leq 0.
$$
As $w_{\i} < w_{\j}$, this condition essentially implies that $u_\j(\widehat{W}) \geq u_\i(\widehat{W})$. Recall that on interval $I$, the ordering of utilities of items do not change, so we must also have 
$$
\frac{u_\j(\widetilde{W}) - u_\i(\widetilde{W})}{w_\i - w_\j} \leq 0\,.
$$
Hence, the optimality condition associated with $s_1$ continues to hold under Problem $(\kprelaxeq{\widetilde{W}, K})$. 
Finally, we compute the reduced cost of $s_2$ when we consider Problem $(\kprelaxeq{\widehat{W}, K})$:
$$
\overline{C}_{s_2} 
= \frac{w_\j}{w_\i - w_\j}u_\i(\widehat W) - \frac{w_\i}{w_\i - w_\j}u_\j(\widehat W) \leq 0  \quad 
\Leftrightarrow \quad  \frac{u_\i(\widehat W)}{w_\i} \geq \frac{u_\j(\widehat W)}{w_\j} \quad  \text{since $w_\i < w_\j$}\,.
$$
Recall that the ordering of utility-to-weight ratios of items do not change on $I$. Hence, replacing $\widehat{W}$ with $\widetilde{W}$ again does not change the sign of the reduced cost of $s_2$. Overall, we have showed that optimality conditions are unaffected when we move from Problem $(\kprelax{\widehat{W}, K})$ to $(\kprelax{\widetilde{W}, K})$. 

\textbf{Feasibility condition.} We then check whether the feasibility condition still holds for Problem $(\kprelax{\widetilde{W}, K})$. To do that, we simply let $\mathbf{x}$ have the same basis as $\mathbf{x}^\star$ (that is, $x_\i$ and $x_\j$ are the two basic variables), and see if $\mathbf{x}$ is a feasible solution for Problem $(\kprelax{\widetilde{W}, K})$. To compute $\mathbf{x}$, we solve the following linear system:
$$
w_{\i} x_{\i} +  w_{\j} x_{\j} 
= \widetilde{W} - \sum_{k \in P_1 \setminus \{\i\}} w_k \quad \text{and} \quad
x_{\i} + x_{\j} = 1\,,
$$
where the first linear equation ensures that the capacity constraint is binding at solution $\mathbf x$, and the second linear equation ensures that the cardinality constraint is binding  at solution $\mathbf x$.
The aforementioned linear system has the following solution
\begin{equation}
\label{eqn:sol-linear-sys}
x_{\i} = \frac{\widetilde{W} - \sum_{k \in P_1} w_k + w_{\i} - w_{\j}}{w_{\i} - w_{\j}} \quad 
\text{and} \quad 
x_{\j} = \frac{\widetilde{W} - \sum_{k \in P_1} w_k}{w_{\j} - w_{\i}}.
\end{equation}
It is then easy to verify that 
for any $\widetilde{W} \in [ \sum_{k \in P_1} w_k, \sum_{k \in P_1} w_k - w_{\i} + w_{\j}]$, we have $x_{\i}, x_{\j} \in [0, 1]$. That is, the basic solution $\mathbf{x}$ with $x_\i$ and $x_\j$ as the two basic variables is always feasible for any $\widetilde{W} \in [ \sum_{k \in P_1} w_k, \sum_{k \in P_1} w_k - w_{i} + w_{j}]$. In particular, the solution takes the form $x_k = x^\star_k$ for any $k \neq \i, \j$ and $x_\i, x_\j$ are defined as in Eq. \eqref{eqn:sol-linear-sys}.
Hence, the feasibility condition also holds.

To summarize, when we transition from Problem $(\kprelax{\widehat W, K})$ to Problem $(\kprelax{\widetilde W, K}$) and keep $x_\i, x_\j$ as our two basic variables, both the optimality and feasibility conditions continue to hold. Hence, the optimal solution to Problem $(\kprelax{\widetilde W, K}$) satisfies that $x_k = x^\star_k$ for any $k \neq \i, \j$, while $x_\i$ and $x_\j$ takes the form in Eq. \eqref{eqn:sol-linear-sys}. We thus show the desired result. $\blacksquare$

\subsection{Proof of Lemma~\ref{corollary:form-sol}}
The result trivially holds when $\eta = 0$. For any fixed $\eta \in (0, w_{\j^\star} - w_{\i^\star}]$, we can first choose some $\eta' \in (0, \min\{\eta, \min_{w_i \neq w_j} |w_i - w_j |\})$ such that $W + \eta' \in I_\text{high}$ and $W + \eta' \neq \sum_{i \in S} w_i$ for any $S \subset [n], |S| = K$. Note that such $\eta'$ always exists. We will see shortly in the following discussion why we consider such a value of $\eta'$.

Our proof consists of two parts: (i) we first solve for the optimal solution to Problem $(\kprelax{W+\eta', K})$,
(ii) we next show that the optimal solutions to Problem $(\kprelax{W+\eta', K})$ and $(\kprelax{W+\eta, K})$ share the same profile, which allows us to derive $\mathbf{x}^\star(W + \eta)$. 

First, let us solve for the optimal solution to Problem $(\kprelax{W+\eta', K})$ using the following lemma:
\begin{lemma}
\label{lemma:form-of-BFS}
Suppose that at capacity level $W \in I_\text{high}$, the capacity constraint is binding and the relaxed knapsack problem $(\kprelax{W, K})$ has a degenerate, integer optimal basic solution $\mathbf{x}^\star$, with profile $\mathcal{P}(W) = \{P_1, (0, 0), P_0\}$. 
Let $(\i^\star, \j^\star) \in \argmax_{i \in P_1, j \in P_0 \atop {w_i < w_j}} \left[\frac{u_j(W) - u_i(W)}{w_j - w_i} \right]$, and assume $u_{\j^\star}(W) > u_{\i^\star}(W)$. 
For any $0 < \eta' \leq \min_{w_i \neq w_j} |w_i - w_j|$ such that $W + \eta' \in I_\text{high}$ and $W + \eta' \neq \sum_{i \in S} w_i$ 
for any $S \subset [n], |S| = K$,  Problem $(\kprelax{W + \eta', K})$ has an optimal basic solution, where $x_k =1$ for any $k\in P_1 \backslash\{\i^\star\}$, $x_k =0$ for any $k\in P_0 \backslash\{\j^\star\}$, and 
\begin{equation}
\label{eqn:optimal-BFS-form}
x_k = 
\begin{cases}
1 - \frac{\eta'}{w_{\j^\star} - w_{\i^\star}} & \text{$k = \i^\star$} \\
\frac{\eta'}{w_{\j^\star} - w_{\i^\star}} & \text{$k = \j^\star$} 
\end{cases}\,.
\end{equation}
\end{lemma}
Using Lemma~\ref{lemma:form-of-BFS}, we can solve for the optimal basic solution $\mathbf{x}^\star(W+\eta')$, which takes the form in Eq. \eqref{eqn:optimal-BFS-form}. 
The solution has the profile $\mathcal{P} (W+\eta') =\{ P_1 \setminus \{\i^\star\}, (\i^\star, \j^\star), P_0 \setminus \{\j^\star\}\}$. 

Now, consider $\eta \ge \eta'$, which falls into the interval $(0, w_{\j^\star} - w_{\i^\star}]$. We can now invoke Lemma \ref{lemma:feasibility} by taking $\widehat W = W + \eta'$ and $\widetilde{W} = W + \eta \in [w(P_1), w(P_1) + w_{\j^\star} - w_{\i^\star}]$. Recall that we have assumed that $\widetilde{W} \in I_\text{high}$. Then, given the profile $\mathcal{P}(W + \eta')=\{ P_1 \setminus \{\i^\star\}, (\i^\star, \j^\star), P_0 \setminus \{\j^\star\}\}$, by Lemma~\ref{lemma:feasibility}, we have that the optimal basic solution $\mathbf x$ to Problem $(\kprelax{W+\eta, K})$ takes the following form: $x_k=1$ for any $k \in P_1 \setminus \{\i^\star\}$,  $x_k =0$ for any $k \in P_0 \setminus \{\j^\star\}$,
$$
x_{\i^\star}=\dfrac{W + \eta - W + w_{\i^\star} - w_{\j^\star}}{w_{\i^\star} - w_{\j^\star}} = 1 - \dfrac{\eta}{w_{\j^\star} - w_{\i^\star}}
\quad \text{and} \quad 
x_{\j^\star} = \dfrac{W + \eta - W}{w_{\j^\star} - w_{\i^\star}} = \dfrac{\eta}{w_{\j^\star} - w_{\i^\star}}\,,
$$
which is the desired form. $\blacksquare$

\subsection{Proof of Lemma \ref{lemma:form-of-BFS}}
Suppose that at capacity level $W$, the capacity constraint is binding and the relaxed knapsack problem $(\kprelax{W, K})$ has a degenerate, integer optimal basic solution $\mathbf{x}^\star$, with profile $\mathcal{P}(W) = \{P_1, (0, 0), P_0\}$. 
Fix $0 < \eta' \leq \min_{i \neq j} |w_i - w_j|$ such that $W + \eta' \in I_\text{high}$ and $W + \eta' \neq \sum_{i \in S} w_i$ for any $S \subset[n], |S| = K$. 
Let us first consider the following auxiliary relaxed problem, in which we only increase the capacity level from $W$ to $W+\eta'$, without changing the utility of each item: 
\begin{equation}
\label{eqn:relaxed-knapsack-3}
\max_{\mathbf {x}\in [0,1]^n, s_1, s_2\ge 0 } 
\sum_{i \in [n]} u_i(W) x_{i} \quad 
\mathrm{s.t.} \quad  \sum_{i \in [n]} w_{i} x_{i} + s_1 = W + \eta' \quad \text{and} \quad 
\sum_{i \in [n]} {x_i} + s_2 = K 
\end{equation}
Note that in the above optimization problem,  we introduce $s_1, s_2$ to turn inequality constraints into equality constraints. Our proof consists of two parts. In Part 1, we find an optimal basic solution $\widehat{\mathbf{x}}$ to the auxiliary problem in \eqref{eqn:relaxed-knapsack-3}. To do that, we apply ideas from the dual simplex method, and show that the optimal basis $(i^\star, j^\star)$ are the two indices that satisfy: (i) $i^\star \in P_1, j^\star \in P_0$ and $w_{i^\star} < w_{j^\star}$, and (ii) $(i^\star, j^\star) \in \argmax_{i \in P_1, j \in P_0 \atop {w_i < w_j}} \left[\frac{u_j(W) - u_i(W)}{w_j - w_i}\right]$. That is, $(i^\star, j^\star) =(\i^\star, \j^\star)$. In Part 2, we show that such an optimal solution $\widehat{\mathbf{x}}$ is also an optimal basic solution to Problem $(\kprelax{W+\eta', K})$. Our proof again relies on notions such as optimality and feasibility conditions, reduced costs, the simplex and dual simplex methods, which are defined and described in more details in Section~\ref{sec:LP}. 

\textbf{Part 1: Find an optimal basic solution $\widehat{\mathbf{x}}$ to the auxiliary problem.}
Observe that when we increase the capacity level from $W$ to $W + \eta'$, the optimality condition still holds, since the reduced cost associated with each variable is unaffected.
However, the feasibility condition does not necessarily hold for the original basic variables.
To look for a new optimal basis for the auxiliary problem, we can apply the dual simplex method.
By design of the dual simplex method, throughout its execution,
the optimality condition always hold (i.e., the reduced costs for all non-basic variables at their lower bounds are non-positive; the reduced costs for all non-basic variables at their upper bounds are non-negative; the reduced costs for all basic variables are zero).
The dual simplex method terminates when the feasibility condition is also satisfied; see Section~\ref{sec:LP} and \cite{bertsimas-LPbook} for details.

Before applying the dual simplex method to find a new optimal basis, we first remark that an optimal basic solution to Problem \eqref{eqn:relaxed-knapsack-3}, which we denote as $\widehat{\mathbf{x}}$, must have two fractional components. To see that, we can invoke Lemma~\ref{lemma:other-lemma-1} as follows. Since we have $\i^\star \in P_1$ and $\j^\star \in P_0$ such that $u_{\i^\star}(W) < u_{\j^\star}(W)$, there must exist some $\i \in P_1$ such that $\i \notin G_K$, where $G_K$ is the set of items with top $K$ utilities (recall that on interval $I$, the ordering of utilities of items does not change). Then, since we also have $\eta' < \min_{w_i \neq w_j} |w_i - w_j|$, by Part (iii) of Lemma \ref{lemma:other-lemma-1}, we must have that the capacity constraint is binding at $\widehat{\mathbf{x}}$. By Part (ii) of Lemma \ref{lemma:other-lemma-1}, this further implies that the cardinality constraint is also binding at $\widehat{\mathbf{x}}$. That is, we must have $s_1 = 0$ and $s_2 = 0$ at the optimal basic solution to Problem \eqref{eqn:relaxed-knapsack-3}. By Part (i) of Lemma \ref{lemma:other-lemma-1}, we know that $\widehat{\mathbf{x}}$ either has two fractional variables, or all variables are integers. Since we have assumed that $W + \eta' \neq \sum_{i \in S} w_i$ for any $S \subset [n], |S| = K$, we know that $\widehat{\mathbf{x}}$ cannot be integer. Hence, the optimal basic solution $\widehat{\mathbf{x}}$ must have two fractional variables, which we denote as $\widehat{\mathbf{x}}_{i^\star}$ and $\widehat{\mathbf{x}}_{j^\star}$, that also serve as the two basic variables. In the following, we look for this optimal basis $(i^\star, j^\star)$ with the help of the dual simplex method.

Suppose that we apply the dual simplex method to solving Problem \eqref{eqn:relaxed-knapsack-3}. The dual simplex method works in the following way.
At each iteration, it updates the basis $(i, j)$ by removing one index out of the basis and adding a different index into the basis. It then computes the values of the basic variables $x_i, x_j$ to ensure that the capacity and cardinality constraints are tight. The non-basic variables $\{x_k: k \neq i, k \neq j\}$ would remain at their respective lower/upper bounds. The dual simplex method then checks whether the feasibility condition is satisfied, i.e., $x_i, x_j \in [0, 1]$. If it is satisfied, the dual simplex terminates; otherwise, it starts a new iteration.
(For more details on the dual simplex method, please see Section~\ref{sec:LP} or \cite{bertsimas-LPbook}.) 
Our first claim is that when we apply the dual simplex method to solving Problem \eqref{eqn:relaxed-knapsack-3} and it terminates at the optimal basis $\{i^\star, j^\star\}$, we must have $i^\star \in P_1, j^\star \in P_0$ and $w_{i^\star} < w_{j^\star}$. Recall that we assumed that $(\kprelax{W, K})$ has a degenerate, integer optimal basic solution $\mathbf{x}^\star$, with profile $\mathcal{P}(W) = \{P_1, (0, 0), P_0\}$. Further recall that the dual simplex method terminates when the feasibility conditions hold for our chosen basis. 
We can check the feasibility conditions for the following three cases:
\begin{enumerate} [leftmargin=*]
    \item Case 1: Both $i^\star$ and $j^\star$ are from $P_0$. If we solve for $\widehat{x}_{i^\star}, \widehat{x}_{j^\star}$ using the binding capacity and cardinality constraints, i.e.
    $$
    w_{i^\star}\widehat{x}_{i^\star} + w_{j^\star}\widehat{x}_{j^\star} = \eta' \quad \text{and} \quad 
    \widehat{x}_{i^\star} + \widehat{x}_{j^\star} = 0  \,,
    $$
    we have
    $$
    \widehat{x}_{i^\star} = - \frac{\eta'}{w_{j^\star}-w_{i^\star}} \quad \text{and} \quad \widehat{x}_{j^\star} = \frac{\eta'}{w_{j^\star}-w_{i^\star}} \,.
    $$
    Clearly, here the feasibility condition is not satisfied because one of the variables would take a negative value. Hence, $\widehat{\mathbf{x}}$ can not be the optimal solution. 
    \item Case 2: Both $i^\star$ and $j^\star$ are from $P_1$. If we solve for $\widehat{x}_{i^\star}, \widehat{x}_{j^\star}$ using the binding capacity and cardinality constraints, i.e.
    $$
    w_{i^\star}\widehat{x}_{i^\star} + w_{j^\star}\widehat{x}_{j^\star} = w_{i^\star} + w_{j^\star} + \eta' \quad \text{and} \quad 
    \widehat{x}_{i^\star} + \widehat{x}_{j^\star} = 2  \,,
    $$
 we have
    $$
    \widehat{x}_{i^\star} = 1 - \frac{\eta'}{w_{j^\star} - w_{i^\star}} \quad \text{and} \quad \widehat{x}_{j^\star} = 1 + \frac{\eta'}{w_{j^\star}-w_{i^\star}} \,.
    $$
    Clearly, here the feasibility condition is again not satisfied because one of the variables would exceed one. Hence, $\widehat{\mathbf{x}}$ cannot be the optimal solution.
    \item Case 3: One variable is from $P_1$ and the other is from $P_0$. (WLOG, let $i^\star \in P_1$ and $j^\star \in P_0$.) 
    If we solve for $\widehat{x}_{i^\star}, \widehat{x}_{j^\star}$ using the binding capacity and cardinality constraints, i.e.
    $$
    w_{i^\star}\widehat{x}_{i^\star} + w_{j^\star}\widehat{x}_{j^\star} = w_{i^\star} + \eta' \quad \text{and} \quad 
    \widehat{x}_{i^\star} + \widehat{x}_{j^\star} = 1  \,,
    $$
 we get
    $$
    \widehat{x}_{i^\star} = 1 - \frac{\eta'}{w_{j^\star} - w_{i^\star}} \quad \text{and} \quad \widehat{x}_{j^\star} = \frac{\eta'}{w_{j^\star}-w_{i^\star}} \,.
    $$
    Here, the solution $\widehat{\mathbf{x}}$ would be feasible if and only if $w_{j^\star} - w_{i^\star} > 0$.
\end{enumerate}
From the case discussion above, we have seen that
the dual simplex method must terminate at an optimal basis $(i^\star, j^\star)$ that satisfies $i^\star \in P_1, j^\star \in P_0$ and $w_{i^\star} < w_{j^\star}$. In particular, in the optimal basic solution $\widehat{\mathbf{x}}$ returned by the dual simplex method, the two basic variables would take the value $\widehat{x}_{i^\star} = 1 - \frac{\eta'}{w_{j^\star} - w_{i^\star}}$ and 
$\widehat{x}_{j^\star} = \frac{\eta'}{w_{j^\star}-w_{i^\star}}$, while the other non-basic variables would remain unchanged form the optimal solution to problem $(\kprelax{W, K})$, i.e., $\widehat{x}_k = x^\star(k)$. 

To determine $(i^\star, j^\star)$, we further note that if we choose $i \in P_1, j \in P_0$ where $w_i < w_j$ to be in our basis, and let $x_i = 1- \frac{\eta'}{w_j - w_i}, x_j = \frac{\eta'}{w_j - w_i}$, the objective of \eqref{eqn:relaxed-knapsack-3} would be
$$
\begin{aligned}
\sum_{k \in [n]}u_k(W)x_k 
& = \sum_{k \in P_1 \setminus \{i\}}u_k(W)x_k  + (1-\frac{\eta'}{w_j - w_i})u_i(W) + \frac{\eta'}{w_j - w_i}u_j(W) \\
& = \sum_{k \in P_1}u_k(W)x_k + \eta' \frac{u_j(W) - u_i(W)}{w_j - w_i}
\end{aligned}
$$
Therefore, a basis $(i^\star, j^\star)$ that maximizies $\dfrac{u_j(W) - u_i(W)}{w_j - w_i}$ would also maximize the objective. 

Our discussion above shows that $(\i^\star, \j^\star) \in \argmax_{i \in P_1, j \in P_0 \atop {w_i < w_j}} \left[\frac{u_j(W) - u_i(W)}{w_j - w_i} \right]$ is the optimal basis for the optimal basic solution $\widehat{\mathbf{x}}$. That is, $(i^\star, j^\star) = (\i^\star, \j^\star)$. Further, $\widehat{\mathbf{x}}$ takes the following form:
\begin{equation*}
\widehat{x}_k = 
\begin{cases}
1 - \frac{\eta'}{w_{\j^\star} - w_{\i^\star}} & \text{$k = \i^\star$} \\
\frac{\eta'}{w_{\j^\star} - w_{\i^\star}} & \text{$k = \j^\star$} 
\end{cases}\,.
\end{equation*}
and $\widehat{x}_k = 1$ for any $k \in P_1 \setminus \{\i^\star\}$, $\widehat{x}_k = 1$ for any $k \in P_0 \setminus \{\j^\star\}$. This is the desired form given in Eq. \eqref{eqn:optimal-BFS-form}.

\textbf{Part 2: $\widehat{\mathbf{x}}$ is also an optimal basic solution to Problem ($\kprelax{W+\eta', K}$).}
To see this, we again consider an equivalent form of Problem $(\kprelax{W + \eta', K})$ stated below:
\begin{equation*}
\max_{\mathbf {x}\in [0,1]^n, s_1, s_2\ge 0 } 
\sum_{i \in [n]} u_i(W + \eta') \quad 
\mathrm{s.t.} \quad  \sum_{i \in [n]} w_{i} x_{i} + s_1 = W + \eta' \quad \text{and} \quad \sum_{i \in [n]} {x_i} + s_2 = K 
\end{equation*}
where we introduced slack variables $s_1, s_2$ to turn inequality constraints into equalities.
Note that when we transition from Problem \eqref{eqn:relaxed-knapsack-3} to $(\kprelax{W + \eta', K})$, we only change the utility of each item. Hence, the feasibility condition still hold. It suffices to check whether the optimality condition hold for each variable. Applying the same argument as in the proof of Lemma~\ref{lemma:feasibility}, 
we have that the signs of reduced costs associated with the non-basic variables $\{x_k : k \neq \i^\star, k \neq \j^\star\}$, $s_1$ and $s_2$ do not change as we change the utility of each item $k$ from $u_k(W)$ to $u_k(W + \eta')$, since $W, W + \eta' \in I$. Hence, the optimality conditions related to these variables continue to hold.
Since both feasibility and optimality conditions hold for $\widehat{\mathbf{x}}$, 
it is indeed an optimal basic solution to Problem $(\kprelax{W + \eta, K})$, and it takes the desired form. $\blacksquare$

\subsection{Other Lemmas}
\begin{lemma}
Suppose that $W \in I_\text{high} = I \cap [\th, \infty)$. For any fixed $\eta' \geq 0$ such that $W + \eta' \in I_\text{high}$, consider the knapsack problem of the following form:
\begin{equation}
\label{eqn:relaxed-knapsack-5}
\max_{\mathbf {x}\in [0,1]^n, s_1, s_2\ge 0 } 
\sum_{i \in [n]} u_i(W) x_{i} \quad 
\mathrm{s.t.} \quad \sum_{i \in [n]} w_{i} x_{i} + s_1 = W + \eta' \quad \text{and} \quad 
\sum_{i \in [n]} {x_i} + s_2 = K 
\end{equation}
where $u_i(W) = \frac{\r_i w_i}{1+W} - \c_i$ is the utility of item $i$. 
Assume that $\mathbf{x}^\star$ is an optimal basic solution to this knapsack problem. We must have the following:
\begin{enumerate}[(i)]
\item $\mathbf{x}^\star$ either has exactly two fractional components, or all variables take integer values.
\item If the capacity constraint is binding at $\mathbf{x}^\star$, the cardinality constraint must also be binding at $\mathbf{x}^\star$. 
\item 
If Problem $(\kprelax{W, K})$ has a degenerate, integer optimal basic solution $\mathbf{x}^\star(W)$ with profile $\mathcal{P}(W) = \{P_1, (0, 0), P_0\}$, and there exists $\i \in P_1$ such that $\i \notin G_K$, where $G_K$ is the set of items with top $K$ utilities, we must have that the capacity constraint is binding at $\mathbf{x}^\star$ for any $\eta' \leq \min_{w_i \neq w_j}|w_i - w_j|$.
\end{enumerate}
\label{lemma:other-lemma-1}
\end{lemma}

\textit{Proof of Lemma~\ref{lemma:other-lemma-1}.} For fixed $W \in I_\text{high}$ and $\eta' \geq 0$ such that $W + \eta' \in I_\text{high}$, let $\mathbf{x}^\star$ be an optimal basic solution to the knapsack problem in \eqref{eqn:relaxed-knapsack-5}. We will now prove each of the statements.
\begin{enumerate}[(i)]
\item Recall from Lemma 1 in \cite{caprara2000approximation} that an optimal basic solution $\mathbf{x}^\star$ has at most two fractional variables. 
Hence, it suffices to show that we cannot have an optimal basic solution $\mathbf{x}^\star$ that has only one fractional component.
Suppose by contradiction that $\mathbf{x}^\star$ is an optimal basic solution to Problem \eqref{eqn:relaxed-knapsack-5} that has one fractional component $x^\star_\i$, which implies that the cardinality constraint $\sum_{i \in [n]} x^\star_i \leq K$ is not tight. That is, the slack variable $s_2 > 0$.
In this case, $x_\i$ and $s_2$ are the two basic variables that take fractional values. Let $P_1$ denote the indices of variables with value $1$, and let $m := |P_1| \leq K - 1$. 
Now, suppose that we decrease the capacity limit $W + \eta'$ to some $W' = \sum_{j \in P_1} w_j$, without changing the utilities of items. That is, we consider the following relaxed knapsack problem:
\begin{equation}
\label{eqn:relaxed-knapsack-4}
\max_{\mathbf {x}\in [0,1]^n, s_1, s_2\ge 0 } \sum_{i \in [n]} u_i(W) x_{i} \quad 
\mathrm{s.t.} \sum_{i \in [n]} w_{i} x_{i} + s_1 = W' \quad \text{and} \quad \sum_{i \in [n]} {x_i} + s_2 = K 
\end{equation}
Let $\mathbf{x}$ be an optimal basic solution to Problem \eqref{eqn:relaxed-knapsack-4}. Observe that Problem \eqref{eqn:relaxed-knapsack-4} only differs from Problem \eqref{eqn:relaxed-knapsack-5} in the capacity limit. We note that if we keep $x_\i$ and $s_2$ as our basic variables in a basic solution to Problem \eqref{eqn:relaxed-knapsack-4},
the optimality condition does not change (since the utility of the items are not changed). Then, if we solve for $x_\i$ and $s_2$ using the capacity/cardinality constraints in \eqref{eqn:relaxed-knapsack-4}, 
we have the linear system
$$
\sum_{j \in P_1} w_j + w_\i x_\i = W' \quad \text{and} \quad
\sum_{j \in P_1} x_j + x_\i + s_2 = K
$$
which gives $x_\i = 0, s_2 = K - m$, where $m = |P_1| \leq K-1$. 
Here, the feasibility condition holds because $x_i \in [0, 1]$ and $s_2 \geq 0$. Since both optimality and feasibility conditions hold, the solution $\mathbf{x}$ where $x_\i = 0$ and $x_k = x^\star_k$ for all $k \neq \i$ is an optimal solution to Problem \eqref{eqn:relaxed-knapsack-4}.

Recall that the ordering of utility-to-weight ratios of items does not change on the interval $I$. We will now derive a contradiction by dividing our analysis into two cases: (1) if $P_1$ contains the $m$ items with the highest utility-to-weight ratios, we cannot have $W \geq w(H_K) = \th$, which violates our assumption in the lemma; (2) if $P_1$ contains an item that is not among the items with top $m$ utility-to-weight ratios, the solution $\mathbf{x}$ cannot be an optimal solution to Problem \eqref{eqn:relaxed-knapsack-4}. 
We now discuss the two possibilities:
\begin{itemize}
    \item \textbf{Case 1:} $P_1$ contains the $m$ items with the highest utility-to-weight ratios. In this case, we have that $W' = \sum_{j \in P_1} w_j = w(H_m) < w(H_K)$. By Lemma~\ref{lemma:interval-low}, we must have that $\i$ is the item with the $(m+1)$th highest utility-to-weight ratio. 
    However, this implies that $W + \eta' \leq w(H_K)$, and hence $W \leq w(H_K) = \th$, which leads to a contradiction. 
    \item \textbf{Case 2:} There exists an item $\j \in P_1$ and another item $\k \in [n]\setminus P_1$ 
    such that the utility-to-weight ratio of item $\k$ is higher than that of item $\j$, i.e. $\frac{u_\k(W)}{w_\k} > \frac{u_\j(W)}{w_\j}$. 
    We will show, in this case, that $\mathbf{x}$ cannot be optimal for Problem \eqref{eqn:relaxed-knapsack-4}. 
    In particular, we will find another solution $\mathbf{x}'$ that yields a higher objective than $\mathbf{x}$ in Problem \eqref{eqn:relaxed-knapsack-4}.
    Let us define $\mathbf{x}'$ as follows:
    \begin{itemize}
        \item if $w_\k \geq w_\j$, we define $\mathbf{x}'$ as 
        $$
        x'_i = \begin{cases} 
        1 & \text{if $i \in P_1$ and $i \neq \j$} \\
        \frac{w_\j}{w_\k} & \text{if $i = \k$} \\
        0 & \text{otherwise}
        \end{cases}
        $$
        The solution $\mathbf{x}'$ is feasible for Problem \eqref{eqn:relaxed-knapsack-4} since $\sum_{i \in [n]} w_i x_i' = \sum_{j \in P_1} w_j = W'$ and 
        $\sum_{i \in [n]} x_i' = (m-1) + \frac{w_\j}{w_\k} < K$. It yields a higher objective than $\mathbf{x}$ in Problem \eqref{eqn:relaxed-knapsack-4}, since $\frac{w_\j}{w_\k} u_\k(W) > u_\j(W)$.
    \item if $w_\k < w_\j$, we define $\mathbf{x}'$ as: 
    $$
    x'_i = 
    \begin{cases} 
    1 & \text{if $i \in P_1 \setminus \{\j\}$ or $i = \k$} \\
    \frac{w_\j-w_\k}{w_\j} & \text{if $i=\j$} \\
    0 & \text{otherwise}
    \end{cases}
    $$
    The solution $\mathbf{x}'$ is feasible for Problem \eqref{eqn:relaxed-knapsack-4} since $\sum_{i \in [n]} w_ix_i' = \sum_{j\in P_1} w_j = W'$ and $\sum_{i \in [n]} x_i' = m + \frac{w_\j - w_\k}{w_\j} < K$. It again yields a higher objective than $\mathbf{x}$ in Problem \eqref{eqn:relaxed-knapsack-4}, since $u_\k(W) + \frac{w_\j-w_\k}{w_\j} u_\j(W) > u_\j(W)$.
    \end{itemize}
    To summarize, we can always find a feasible solution $\mathbf{x}'$ to Problem \eqref{eqn:relaxed-knapsack-4} that gives a higher objective value than $\mathbf{x}$; hence, $\mathbf{x}$ cannot be optimal for Problem \eqref{eqn:relaxed-knapsack-4}. This again leads to a contradiction.
\end{itemize}
Overall, we have showed that it is not possible for $\mathbf{x}^\star$ to have just one fractional component. Hence, it either has exactly two fractional components, or all variables take integer values. 
\item Assume that the capacity constraint is binding at $\mathbf{x}^\star$, and suppose by contradiction that the cardinality constraint is non-binding at $\mathbf{x}^\star$.
That is, 
$s_1 = 0$ and $s_2 > 0$ in Problem \eqref{eqn:relaxed-knapsack-5}. 
We have already showed, in Part (i) of this proof, that $\mathbf{x}^\star$ cannot have just one fractional component. Hence, 
$\mathbf{x}^\star$ must be an integer solution with $\sum_{i \in [n]} x_i^\star = m \leq K - 1$. 
However, using the same argument as in Part (i) of this proof, we can show that such $\mathbf{x}^\star$ cannot be an optimal solution to Problem \eqref{eqn:relaxed-knapsack-5}. That is, if the optimal solution contains the $m$ items with highest utility-to-weight ratios, we would have $W + \eta' = w(H_m) < w(H_K) = \th$, which leads to a contradiction; otherwise, we can find another solution $\mathbf{x}'$ which yields higher utilities, which contradicts the optimality of $\mathbf{x}'$.

Via proof of contradiction, we have showed that if the capacity constraint is binding at $\mathbf{x}^\star$, the cardinality constraint must also be binding at $\mathbf{x}^\star$. 

\item First, note that 
if $W + \eta' \leq w(G_K)$, where $G_K$ is the set of items with the top $K$ utilities, the capacity constraint must be binding at $\mathbf{x}^\star$. 
Now, suppose Problem $(\kprelax{W, K})$ has a degenerate, integer optimal basic solution $\mathbf{x}^\star(W)$ with profile $\mathcal{P}(W) = \{P_1, (0, 0), P_0\}$, and there exists $\i \in P_1$ such that $\i \notin G_K$. This means that there must exist $\j \in P_0$ such that $\j \in G_K$. 

We must also have $w_\i < w_\j$. To see this, suppose $w_\i \geq w_\j$ by contradiction. We can define $\mathbf{x}'$ such that $x'_k = 1$ if $k \in P_1 \cup \{\j\} \setminus \{\i\}$ and $x'_k = 1$ if $k \in P_0 \cup \{\i\} \setminus \{\j\}$. This $\mathbf{x}'$ is feasible for Problem $(\kprelax{W, K})$ and yields higher utilities, which contradicts the optimality of $\mathbf{x}^\star(W)$). 

Therefore, we must have $w(G_K) - W \geq w_\j - w_\i \geq \min_{w_i \neq w_j} |w_i - w_j|$. 
That is, if $\eta' \leq \min_{w_i \neq w_j} |w_i - w_j|$, we must have that the capacity constraint is binding at $\mathbf{x}^\star$. $\blacksquare$
\end{enumerate} 
\endproof

\section{Proof of Theorem \ref{thm:fptas-optimality}}
\label{sec:proof-thm-fptas}
The proof consists of two segments. In the first segment, we show that 
\[\revmod(S_I, \mathbf{z})\ge (1-\epsilon) \max_{W\in I}~ \kpcard{W,K},\]
where $S_I$ is the set returned by Algorithm \ref{alg:fptas}. In the second segment, we bound the runtime of the algorithm.

\textbf{Segment 1.} Suppose the maximum $\max_{W \in I} \kpcard{W,K}$ is achieved at $W^\star \in I$ with assortment $S^\star$. Given our partition rules, there exists a sub-interval $I'_\ell \subseteq I_\ell \subseteq I$ such that $W^\star \in I'_\ell$. Let us focus on the interval $I'_\ell = [W_{\min}, W_{\max})$. 

Let $(\chi^\star, \kappa^\star) = \argmax_{\chi\in \{0,\ldots,  \chi_{\max}\}, \kappa\in \{0, \ldots, K\}} \{ \chi : \minwt_n(\chi, \kappa) \leq W^\star \}$. Consider the set 
$\tilde{S} := \minwtset_n(\chi^\star, \kappa^\star) \in \mathcal{C}_I$. We must have 
\begin{equation}
\label{eqn:fptas-eqn-1}
\sum_{i \in \Tilde{S}} \tilde{u}_i \geq \sum_{i \in S^\star} \tilde{u}_i 
\end{equation}
since $\Tilde{S}$ is the set, among all sets with weights bounded by $W^\star$, that yields that highest total scaled utility.
Now, given the way we rescale the utilities in Eq. \eqref{eqn:rescaled-util}, and since for any $W \in I'_\ell$, the rescaled utility $\tilde{u}_i(W) = \tilde{u}_i$ takes the same value, we have that for all $i \in [n]$,
$$
\tilde{u}_i = \Big\lceil \dfrac{u_{i}(W^\star)}{\sum_{i' \in \mathcal{D}(I_\ell)} u_{i'}(W^\star)} \cdot \frac{K}{\epsilon} \Big\rceil\,.
$$
This gives 
\begin{equation}
\label{eqn:fptas-left-right-bound}
(\tilde{u}_i - 1) \cdot \frac{\epsilon}{K} \sum_{i' \in \mathcal{D}(I_\ell)} u_{i'}(W^\star)
< u_i(W^\star) \leq 
\tilde{u}_i\cdot \frac{\epsilon}{K} \sum_{i' \in \mathcal{D}(I_\ell)} u_{i'}(W^\star)\,.
\end{equation}
Hence, we have 
$$
\begin{aligned}
\sum_{i \in S^\star} u_i(W^\star) - 
\sum_{i \in \tilde{S}} u_i(W^\star) 
& \overset{(a)}\leq \frac{\epsilon}{K} \sum_{i' \in \mathcal{D}(I_\ell)} u_{i'}(W^\star)
\cdot \left[ \sum_{i \in S^\star} \Tilde{u}_i + \kappa^\star - \sum_{i \in \Tilde{S}} \Tilde{u}_i \right] \\
& \overset{(b)}\leq \frac{\epsilon}{K} \sum_{i' \in \mathcal{D}(I_\ell)} u_{i'}(W^\star)
\cdot \kappa^\star \\
& \overset{(c)}\leq \epsilon \cdot \sum_{i' \in \mathcal{D}(I_\ell)} u_{i'}(W^\star) \\
& \overset{(d)}\leq \epsilon \cdot \sum_{i' \in S^{\star}} u_{i'}(W^\star) \,.
\end{aligned}
$$
where (a) follows from \eqref{eqn:fptas-left-right-bound}, (b) follows from \eqref{eqn:fptas-eqn-1}, (c) follows from $\kappa^\star \leq K$ and (d) follows from $S^\star$ being the utility-maximizing set. Hence,
$$
\sum_{i \in \tilde{S}} u_i(W^\star) \geq (1-\epsilon) \sum_{i \in S^\star} u_i(W^\star) = (1-\epsilon) \max_{W \in I} \kpcard{W,K}\,.
$$
Finally, since $\tilde{S} \in \mathcal{C}_I$ and $w(\tilde{S}) \leq W^\star$, we have
$$
\begin{aligned}
\revmod(S_I, \mathbf{z}) \geq \revmod(\tilde{S}, \mathbf{z}) & = \sum_{i \in \tilde{S}} \frac{\r_i w_i}{1 + \sum_{i \in \tilde{S}} w_i} - \sum_{i \in \tilde{S}} \c_i \\
& = \sum_{i \in \tilde{S}} u_i(w(\tilde{S})) \geq \sum_{i \in \tilde{S}} u_i(W^\star)  \geq (1 - \epsilon) \max_{W\in I} \kpcard{W, K}\,.
\end{aligned}
$$

\medskip

{\textbf{Segment 2.}} 
We now comment on the runtime of our FPTAS. 
Step 2 of Algorithm \ref{alg:fptas} takes $\mathcal{O}(n^2\log{n} + nK^2)$, which is the runtime of our $1/2$-approx. algorithm established in Theorem \ref{thm:half-approx-ratio}. In our proof of Theorem \ref{thm:half-approx-ratio}, we have shown that $I$ would get adaptively partitioned into at most $\mathcal{O}(nK)$ sub-intervals, denoted by $\{I_\ell\}_{\ell \in [L]}$, such that each $I_\ell$ admits a common $1/2$-approx. solution. 
Step 3 of Algorithm \ref{alg:fptas}, by Lemma \ref{lemma:monotonic}, would further partition each sub-interval $I_\ell$ into at most $\mathcal{O}(nK/\epsilon)$ sub-intervals $I'_\ell$ such that on each $I'_\ell$ the rescaled utility takes the same value for all $i \in [n]$, which takes at most $\mathcal{O}(nK/\epsilon)$ time.  
The total runtime of Steps 2 and 3 is thus $\mathcal{O}(n^2\log{n} + nK^2 + nK/\epsilon)$. 
Note that after Steps 2 and 3, we have partitioned our well-behaving interval $I$ into at most $\mathcal{O}(nK \cdot nK/\epsilon) = \mathcal{O}(n^2K^2/\epsilon)$ sub-intervals $I'_\ell$.

In Step 4, for each sub-interval $I'_\ell$ we perform a dynamic programming scheme. Step 4(a) takes $O(n)$ since it simply rescales the utilities of each item. 
In Steps 4(b) and 4(c), we  operate on a matrix of size $\mathcal{O}(n \cdot \chi_{\max} \cdot K) = \mathcal{O}(nK^2/\epsilon)$. Computing each entry of the matrix ($\minwt_i(\chi, \kappa)$) takes $\mathcal{O}(1)$. For each entry of the matrix, updating its corresponding assortment $\minwtset_i(\chi, \kappa)$ also takes $\mathcal{O}(1)$. Hence the overall complexity of Step 4(b) and 4(c) is $\mathcal{O}(nK^2/\epsilon)$. 
In Step 4(d), we collect at most $\mathcal{O}(\chi_{\max} \cdot K) = \mathcal{O}(K^2/\epsilon)$ sets $\minwt_n(\chi, \kappa)$. 
Overall, for each sub-interval $I'_\ell$, our DP scheme takes runtime of at most $\mathcal{O}(nK^2/\epsilon)$, and collects at most $\mathcal{O}(K^2/\epsilon)$ sets. The total runtime of Step 4 is thus $\mathcal{O}(n^2K^2/\epsilon \cdot nK^2/\epsilon) = \mathcal{O}(n^3K^4/\epsilon^2)$.

In Step 5, the collection $\mathcal{C}_I$ contains at most $\mathcal{O}(n^2K^2/\epsilon \cdot K^2/\epsilon) = \mathcal{O}(n^2K^4/\epsilon^2)$ sets. Finding the set that maximizes $\revmod(S, \mathbf{z})$ hence takes $\mathcal{O}(n^2K^4/\epsilon^2)$.

In summary, the overall runtime complexity of Algorithm~\ref{alg:fptas} is  dominated by Step 4, which is at most $\mathcal{O}(n^3K^4/\epsilon^2)$. 
$\blacksquare$

\subsection{Proof of Lemma \ref{lemma:monotonic}}
Consider a well-behaving interval $I_{\ell}$ that admits a common $1/2$-approx. solution $\mathcal{D}(I_{\ell})$ for all $W \in I_{\ell}$. Recall from Eq. \eqref{eqn:rescaled-util} that the rescaled utility $\tilde{u}_i(W)$ is defined as follow:
$$
\tilde{u}_i(W) = \Big\lceil \dfrac{u_{i}(W)}{\sum_{i' \in \mathcal{D}(I_{\ell})} u_{i'}(W)} \cdot \frac{K}{\epsilon} \Big\rceil \,.
$$
For simplicity, let us perform the following changes of variables: $\gamma = \frac{1}{1+W}, \alpha_i = \r_iw_i, A_i = \sum_{i' \in \mathcal{D}(I_{\ell})} \r_i w_i, C_i = \sum_{i' \in \mathcal{D}(I_{\ell})} \c_i$. These allows us to write $u_i(W) = \alpha_i \gamma - \c_i$ and 
$\sum_{i' \in \mathcal{D}(I_{\ell})} u_{i'}(W) = \sum_{i' \in \mathcal{D}(I_{\ell})} A_i \gamma - C_i$.
To show the monotonicity of $\tilde{u}_i(W)$, it suffices to show the monotonicity of the following function $f_i$:
$$
f_i(\gamma(W)) =  \dfrac{\alpha_i \gamma - \c_i}{A_i \gamma - C_i} = \dfrac{u_{i}(W)}{\sum_{i' \in \mathcal{D}(I_{\ell})} u_{i'}(W)} \,.
$$
We have that
$$
f_i'(W) = f_i'(\gamma) \cdot \gamma'(W) = \dfrac{A_i\c_i - \alpha_i C_i}{(A_i \gamma - C_i)^2} \cdot \dfrac{-1}{(1+W)^2}\,.
$$
Since the sign of $f_i'(W)$ does not change with respect to $W$, we thus have $f_i(W)$ is monotonic in terms of $W$. This thus implies that the rescaled utility $\tilde{u}_i(W)$ also changes monotonically with respect to $W$ on interval $I_\ell$.
$\blacksquare$

\section{Backgrounds on Linear Programming}
\label{sec:LP}
In this section, we provide an brief overview of the key terminologies in linear programming that we used throughout the paper. For an more detailed discussion, see \cite{bertsimas-LPbook}.

In our overview, we consider a linear program in its standard form, i.e. 
\begin{equation*}
\max_{\mathbf{x}} \quad 
\mathbf{c}^\top \mathbf{x} \quad \quad
\mathrm{s.t.} \quad  \mathbf{A}\mathbf{x} = \mathbf{b} \quad \text{and} \quad \mathbf{x} \geq \mathbf{0}\,,
\end{equation*}
where the dimension of $\mathbf{A}$ is $m \times n$, and its rows are linearly independent; $\mathbf{x}, \mathbf{c}, \mathbf{z}$ are vectors of size $n$. 
We remark that an LP with inequality constraints can be transformed into the standard form by introducing extra slack variables (see \cite{bertsimas-LPbook}). 

\textbf{Basic solution.} 
We first provide the formal definition of a basic solution to the LP. We say that a solution $\mathbf{x} \in \mathbb{R}^n$ is a \emph{basic solution} to the LP if and only if we have $\mathbf{A}\mathbf{x} = \mathbf{b}$ and there exist indices $B(1), \dots, B(m)$ such that:
\begin{enumerate}
    \item The columns of matrix $\mathbf{A}$ indexed by $B(1), \dots, B(m)$, which we denote as $\mathbf{A}_{B(1)}, \dots, \mathbf{A}_{B(m)}$, are linearly independent;
    \item If $i \notin \{ B(1), \dots, B(m)\}$, then $x_i = 0$.
\end{enumerate}
In addition, we call $x_{B(1)}, \dots, x_{B(m)}$ the \emph{basic variables} and $x_i$ for $i \notin \{B(1), \dots, B(m)\}$ the non-basic variables.
We say that a basic solution $\mathbf{x} \in \mathbb{R}^n$ is a \emph{basic feasible solution} if it is also feasible (i.e., $\mathbf{x} \geq \mathbf{0}$). We say that a basic solution $\mathbf{x} \in \mathbb{R}^n$ is an \emph{optimal basic solution} if it is both feasible and optimal to the LP.

\textbf{Simplex/dual simplex method}. 
It is known that if an LP in the standard form has an optimal solution, there must exist an optimal basic solution. The simplex and dual simplex method are both algorithms designed to find the \emph{optimal basic solution} $\mathbf{x}^\star$ to the LP. In the implementation of the simplex/dual simplex method, the algorithm keeps track of two sets of conditions:
\begin{enumerate}
    \item \textbf{Optimality condition.} Let $\mathbf{x}$ be a basic solution and let $\mathbf{B} = \left[\mathbf{A}_{B(1)}, \dots, \mathbf{A}_{B(m)}\right]$ be its associated basis matrix. Let $\mathbf{c}_B = [c_{B(1)}, \dots, c_{B(m)}]^\top$ be the vector of costs of the basic variables. For each $j \in [n]$, we define the \emph{reduced cost} $\overline{C}_j$ of the variable $x_j$ as:
    \begin{align} \label{eq:reduced_cost} \overline{C}_j = c_j - \mathbf{c}_B^\top \mathbf{B}^{-1}\mathbf{A}_j \,.\end{align}
    At a high level, for a non-basic variable $x_j$, the reduced cost of $\overline{C}_j$ measures the rate of cost change if we bring $x_j$ into the basis and move along the direction of $x_j$. Intuitively, if there exists an non-basic variable with positive reduced-cost, we should put it into our basis. Our \emph{optimality condition} is thus defined as :
    \begin{equation*}
    \label{eqn:optimality-condition-def}
    \overline{\mathbf{C}} \leq 0 \,.
    \end{equation*}
    Intuitively, if the optimality condition is satisfied at some solution $\mathbf{x}$, then moving along any direction would only decrease our objective function.
    \item \textbf{Feasibility condition.} Let $\mathbf{x}$ and $\mathbf{B}$ be its corresponding basis matrix; let $\mathbf{x}_B = [x_{B(1)}, \dots, x_{B(m)}]$. Since the columns of $\mathbf{B}$ are independent and $\mathbf{A}\mathbf{x} = \mathbf{B}\mathbf{x}_B = \mathbf{b}$, the \emph{feasibility condition} is defined as:
    $$
    \mathbf{x}_B = \mathbf{B}^{-1}\mathbf{b} \geq \mathbf{0}.
    $$
    This ensures that $\mathbf{x} \geq 0$.
\end{enumerate}
In Theorem 3.1 of \cite{bertsimas-LPbook}, it is shown that a basic solution that satisfies both the optimality and feasibility conditions is an optimal solution to the LP. Reversely, we also have that a nondegenerate, optimal basic solution to the LP must satisfy both the optimality and feasibility conditions. 

Both the simplex method and the dual simplex methods are designed based on the optimality and feasibility conditions above. In the simplex method, the algorithm keeps a basis $\mathbf{B}$ at each iteration, which corresponds to $m$ basic variables. At every iteration, the simplex method makes sure that the feasibility condition is always satisfied, and it swaps a basic variable with a non-basic variable that can increase the objective function. In the dual simplex method, the algorithm also keeps a basis $\mathbf{B}$ at each iteration, which corresponds to a basic solution that might not be feasible. At every iteration, the dual simplex method makes sure that the optimality condition is always satisfied for its basic solution, and it swaps a basic variable with a non-basic variable such that the basic solution gets closer to the feasibility region. The simplex/dual simplex method terminates when both optimality and feasibility conditions are satisfied, which implies that given the current choice of basis $\mathbf{B}$, an optimal basic solution has been found. See \cite{bertsimas-LPbook} for an extended discussion of simplex/dual-simplex method.

\end{APPENDICES}

\end{document}